\documentclass[14pt,preprint]{aastex}

\usepackage{natbib}
\usepackage{epsfig}
\usepackage{graphicx}
\usepackage{longtable}
\usepackage{multirow}
\usepackage{epstopdf}
\slugcomment{Accepted for publication in the ApJ}

\shorttitle{Fundamental Parameters of Carbon Stars}

\shortauthors{van Belle \& Paladini et al.}

\begin{document}

\title{The {\it PTI} Carbon Star Angular Size Survey:\\Effective Temperatures and Non-Sphericity}

\author{Gerard T. van Belle\altaffilmark{1}, Claudia Paladini\altaffilmark{2}, Bernhard Aringer\altaffilmark{3}, Josef Hron\altaffilmark{2}, 
David Ciardi\altaffilmark{4}}

\altaffiltext{1}{Lowell Observatory, 1400 W. Mars Hill Rd., Flagstaff, Arizona, USA; gerard@lowell.edu}
\altaffiltext{2}{Institut f. Astronomie, T\"urkenschanzstr. 17, A-1180 Vienna, Austria; claudia.paladini@univie.ac.at, josef.hron@univie.ac.at, thomas.lebzelter@univie.ac.at}
\altaffiltext{3}{INAF-OAPD, Vicolo dell'Osservatorio 5, 35122 Padova, Italy, bernhard.aringer@oapd.inaf.it}
\altaffiltext{4}{California Institute of Technology, 770 South Wilson Avenue, Pasadena, California, USA; ciardi@ipac.caltech.edu}

\begin{abstract}
We report new interferometric angular diameter observations of 41 carbon stars observed with the {\it Palomar Testbed Interferometer} (PTI).  Two of these stars are CH carbon stars and represent the first such measurements for this subtype. Of these, 39 have \citet{Yamashita1972AnTok..13..169Y,Yamashita1975AnTok..15...47Y} spectral classes and are of sufficiently high quality that we may determine the dependence of effective temperature on spectral type.  We find that there is a tendency for the effective temperature to increase with increasing temperature index by $\sim 120$K per step, starting at $T_{\rm EFF} \simeq 2500$K for $C3,y$, although there is a large amount of scatter about this relationship. Overall, the median effective temperature for the carbon star sample is found to be $2800\pm270$K, and the median linear radius is $360 \pm 100 R_\odot$.  We also find agreement on average within 15K with the $T_{\rm EFF}$ determinations of \citet{Bergeat2001A_A...369..178B,Bergeat2002A&A...390..987B,Bergeat2002A_A...390..967B}, and a refinement of carbon star angular size prediction based on $V$ \& $K$ magnitudes is presented that is good to an rms of 12\%. A subsample of our stars have sufficient $\{u,v\}$ coverage to permit non-spherical modeling of their photospheres, and a general tendency for detection of statistically significant departures from sphericity with increasing signal-to-noise of the interferometric data is seen.  The implications of most --- and potentially all --- carbon stars being non-spherical is considered in the context of surface inhomogeneities and a rotation-mass loss connection.
\end{abstract}

%





\section{Introduction}\label{sec_introduction}

General knowledge of the fundamental parameters of carbon stars is an important foundation for understanding the properties of these core constituents of the galactic population of stars.  This is particularly true given that carbon stars contribute 10-50\% of the carbon enrichment of the interstellar medium \citep{Thronson1987ApJ...322..770T}.  These `classical' carbon stars have spectra that are dominated by carbon species features such as C$_2$, C$_2$H$_2$, C$_3$, CN, and HCN \citep{Goebel1981ApJ...246..455G,Joyce1998AJ....115.2059J,Lancon2000A_AS..146..217L,Loidl2001A_A...371.1065L}.  Evolutionarily speaking, this high carbon abundance is due to the 3rd dredge-up phenomenon in the AGB stars that become these carbon stars \citep{Iben1983ARA_A..21..271I,Herwig2005ARA_A..43..435H}, where the convective envelope penetrates the inter-shell region between the He- and H-burning shells, leaving these objects with C/O $> 1$.

Yet it remains true that very little directly obtained fundamental information exists for the carbon stars. From \citet{Dyck1996AJ....112..294D} and \citet{Bergeat2001A_A...369..178B}, it is apparent that only roughly a few dozen objects have had angular diameters and effective temperatures ($T_{\rm EFF}$) directly determined through high-resolution techniques such as interferometry and lunar occultations, from a wide variety of disparate sources.  As such, the effective temperature scale for carbon stars has not been previously well established owing to these small numbers from inhomogenous sources.  Additionally, these measurements to date all represent diameters along a single measurement axis across these complex objects.

For early studies such as \citet{Mendoza1965ApJ...141..161M}, $T_{\rm EFF}$ estimates range from 2270K to 5500K; later studies such as  \citet{Cohen1979MNRAS.186..837C} produced a $T_{\rm EFF}$ scale that ranged from 3240K to 2230K as a function of \citet{Yamashita1972AnTok..13..169Y,Yamashita1975AnTok..15...47Y} spectral class, largely confirmed by the interferometric measurements in \citet{Dyck1996AJ....112..294D}, who found a mean measured $T_{\rm EFF}$ of $3000\pm200$ for their sample, and a slight indication that $T_{\rm EFF}$ increased with Yamashita spectral type.

Early model spectra in \citet{Rowan-Robinson1983MNRAS.202..797R} indicated that the atmospheric temperatures for the earlier types are around 2500K and for the later types around 2000K; later models \citep{Jorgensen2000A_A...356..253J,Loidl2001A_A...371.1065L} appear to point to higher temperatures, with a mean around 3000K. A $T_{\rm EFF}$ calibration by \citet{Bergeat2001A_A...369..178B} attempted to extend this result, primarily through leveraging a preliminary set of Palomar Testbed Interferometer (PTI) results found in \citet{vanBelle1999AAS...195.4501V}.  However, those results represent an incomplete sample with preliminary data reduction techniques; one of the intents of this manuscript is to present a finalized set of PTI data, both in terms of scope and processing.

Recently we published a study of 5 carbon stars \citep{Paladini2011A_A...533A..27P}, which combined infrared spectra with interferometric determinations of angular size to constrain the model atmospheres of \citet{Aringer2009A_A...503..913A}; which resulted in successful reproduction of first time spectra between 0.9 and 4 $\mu$m for these objects.   $T_{\rm EFF}$, $M$, $\log(g)$, C/O, and distances were derived for these objects. Our aim in this study is to complement that earlier comprehensive work with a broader survey of a large sample of stars with interferometric angular sizes.

Of special note is that, while this study does not have the comprehensive spectroscopic component of our  \citet{Paladini2011A_A...533A..27P} study, a number of the stars in this sample set have considerably more dense angular size data sets.  These data sets allow us to establish that some, and potentially all, carbon stars present non-spherical photospheres.


The observations from the PTI facility are presented in \S \ref{sec_observations}, along with the specific carbon star models, and a discussion of the details of the measured angular sizes for both the carbon stars and the giant `check' stars.  Derived quantities are found in \S \ref{sec_TEFF}, including bolometric fluxes, effective temperatures, and distances.  Finally, in \S \ref{sec_discussion} is a discussion of the implications of these measures, including comparisons to the works of \citet{Yamashita1972AnTok..13..169Y,Yamashita1975AnTok..15...47Y} and \citet{Bergeat2001A_A...369..178B,Bergeat2002A&A...390..987B,Bergeat2002A_A...390..967B}, a calibration of the `reference' angular sizes for carbon stars, and a discussion of the possible origin of the non-spherical appearance of these stars.


\section{Observations and Data Reduction}\label{sec_observations}

\subsection{Target Selection}

The primary source for targets were Yamashita's investigations of carbon stars \citep{Yamashita1972AnTok..13..169Y,Yamashita1975AnTok..15...47Y}, as already preliminarily examined in \citet{Dyck1996AJ....112..294D}\footnote{The original {\it Annals of the Tokyo Astronomical Observatory} are not available electronically, but the Yamashita spectral types are present in the catalog of \citet{Skiff2010yCat....102023S}.}.  Yamashita spectral typing was considered by Keenan himself to be an improvement over the original MK carbon star typing \citep{Keenan1994ASPC...60...15K}, although with still some limitations \citep{Keenan1993PASP..105..905K}.  Supplementing those sources was the most recent edition of Stephenson's {\it General Catalog of Galatic Carbon Stars} \citep{Alksnis2001BaltA..10....1A}.  From these sources, an observation list was culled, largely based on the rough observational constraints of $m_{\rm R}<8$ (necessary for tip-tilt tracking of the individual interferometer telescopes), $m_{\rm K}<3$ (for sufficiently large angular size to be resolved by the interferometer), and $m_{\rm K}>1$ (for sufficiently small angular size to be fringe tracked by the interferometer). Roughly a dozen related S-type stars were also observed during this program and will be addressed in a separate publication.
(An additional comprehensive source of well-typed carbon stars, \citet{Barnbaum1996ApJS..105..419B}, which builds on the typing improvements suggested in \citet{Keenan1993PASP..105..905K}, also matches well to the obesrvational constraints of PTI but unfortunately was not used in the target selection process.)

\subsection{Carbon Star Models}\label{sec_carbonmodels}

For comparison to the interferometric observations, we utilized the hydrostatic carbon star models of \citet{Aringer2009A_A...503..913A}, which describe the observed properties of such stars quite well.  These models, improve relative to the earlier generation of models \citep{Jorgensen2000A_A...356..253J, Loidl2001A_A...371.1065L} through the inclusion of atomic opacities and updated molecular opacities.  A grid of models was generated, covering the following steps\footnote{The lower limit of $\log g$ varies with $T_{\rm EFF}$ and is detailed in Table 1 of \citet{Aringer2009A_A...503..913A}.}: model effective temperature $2400 \leq T_{\rm EFF}^M \leq 4000 {\rm K}$ in steps of 100 K, surface gravity $-1.0 \leq \log g [{\rm cm~s}^{-2} ] \leq +0.0 $ in steps of 0.2dex, with $Z/Z_\odot=1$, $M = 2 M_\odot$ and C/O $= 1.05$.  Although additional steps of grid parameters were available ($M = 1 M_\odot$; C/O $= 1.10, 1.40, 2.00$), these options were not considered be well-constrained by our data and were not exercised; in particular, the higher mass range is slightly favored by our earlier study \citep{Paladini2011A_A...533A..27P}.
A more detailed description of our approach with the \citet{Aringer2009A_A...503..913A} models can be found in \citet{Paladini2009A&A...501.1073P,Paladini2011A_A...533A..27P}.

For each of the models considered herein, two complementary data sets were derived: (1) a synthetic spectral energy distribution (SED), covering the range from 0.45$\mu$m to 25$\mu$m, for use with determination of bolometric flux estimation (\S \ref{sec_SED_fitting}), and (2) a set of center-to-limb variation (CLV) brightness profiles specific to the PTI bandpasses, for use with determination of limb-darkend angular sizes (\S \ref{sec_LDandUD}).  Although these models are hydrostatic and not dynamic, our expectation was that these models were sufficient given the low amplitude photometric variability seen for these stars across the peak of the SED (discussed in detail in \S \ref{sec_SED_fitting}) and thereby were sufficient for $F_{\rm BOL}$ determination.

\subsection{PTI Observations}\label{sec_PTIobservations}

PTI was an 85 to 110 m $H$- and $K$-band 1.6 $\mu$m and 2.2 $\mu$m) interferometer located at Palomar Observatory in San Diego County, California,
and is described in detail in \citet{Colavita1999ApJ...510..505C}.  It had three 40-cm apertures used in pairwise combination for measurement of stellar fringe visibility on sources that range in angular size from 0.05 to 5.0 milliarcseconds (mas), being able to resolve individual sources with angular diameter $(\theta)$ generally greater than 1.4 mas in size \citep[e.g.][]{vanBelle1999AJ....117..521V,vanBelle2009MNRAS.394.1925V} and in certain cases to diameters roughly half that \citep[e.g.][]{vanBelle2009ApJ...694.1085V}.  The three baselines of PTI allowed diameter measurements across stellar photospheres at a variety of projection angles, allowing for detection of oblateness in those photospheres \citep{vanbelle2001ApJ...559.1155V}.  PTI had been in nightly operation from 1997 until early 2009, with minimum downtime throughout the intervening years; in 2009 it was decommissioned and removed from the mountaintop.

The data from PTI considered herein cover the range from the beginning of 1998 (when the standardized data collection and pipeline reduction went into place) until the cessation of operations in 2009.  In addition to the target stars discussed herein, appropriate calibration sources were observed as well and can be found {\it en masse} in \citet{vanBelle2008ApJS..176..276V}.

\subsection{Visibility and Uniform Disk Angular Sizes}\label{sec_angularsizes}

The `canonical' calibration of the target star visibility $(V^2)$ data is performed by estimating the interferometer system visibility ($V^2_{\textrm{\tiny SYS}}$) using the calibration sources with model angular diameters and then normalizing the raw target star visibility by $V^2_{\textrm{\tiny SYS}}$ to estimate the $V^2$ measured by an ideal interferometer at that epoch
\citep{Mozurkewich1991AJ....101.2207M,Boden1998SPIE.3350..872B,vanBelle2005PASP..117.1263V}. Uncertainties in the system visibility and the calibrated target visibility are inferred from internal scatter among the data in an observation using standard error-propagation calculations \citep{Boden1999ApJ...515..356B}. Calibrating our point-like calibration objects against each other produced no evidence of
systematics, with all objects delivering reduced $V^2 = 1$.

Visibility and uniform disk (UD) angular size $(\theta_{\rm UD})$ are related using the first Bessel function $(J_1)$:  $V^2 =[2 J_1 (x) / x]^2$, where spatial frequency $x = \pi B \theta_{UD} \lambda^{-1}$. We may establish uniform disk angular sizes for the target stars observed by the interferometer since the accompanying parameters (projected telescope-to-telescope separation, or baseline, $B$ and wavelength of observation $\lambda$) are well-characterized during the observation.  Below, the UD angular size will be connected to a more physical limb darkened angular size.

\subsection{Limb Darkened Angular Sizes}\label{sec_LDandUD}

Limb darkened (LD) angular sizes $(\theta_{\rm LD})$ are typically utilized as a reasonable proxy for the Rosseland angular diameter, which corresponds to the surface where the Rosseland mean optical depth equals unity, as advocated by \citet{Scholz1987A_A...186..200S} as the most appropriate surface for computing an effective temperature.  To properly consider the `true' extended atmospheres of the carbons stars observed in this investigation, we used the model CLV profiles presented in \S \ref{sec_carbonmodels} that were  specific to the PTI filter sets.

\begin{figure}
\includegraphics[scale=0.9,angle=0]{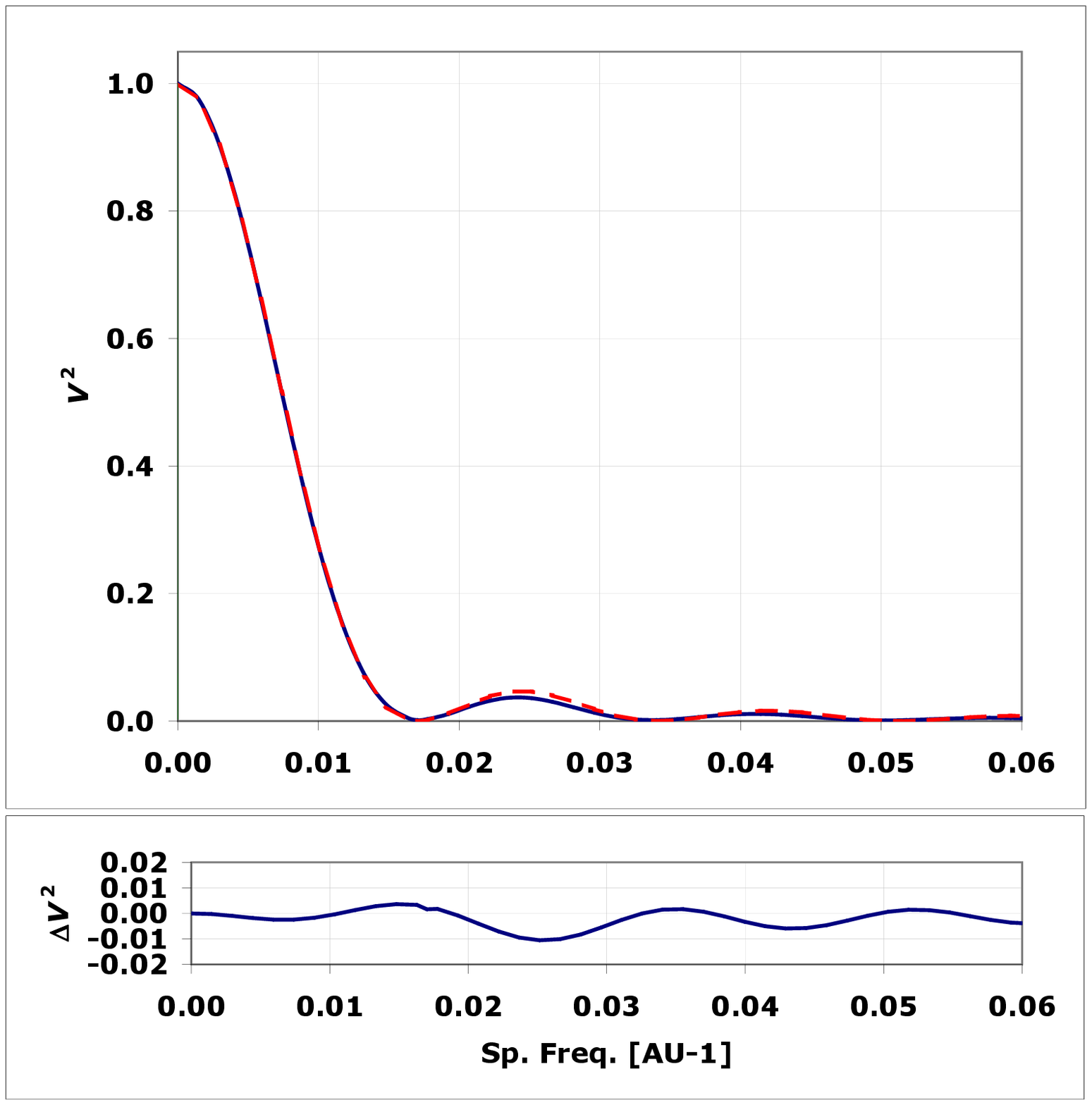}
\caption{\label{fig_LD2UD} Based on \citet{Aringer2009A_A...503..913A}, a plot of a representative visibility-squared curve indicated for the PTI $K$-band filter for a limb-darkened model atmosphere with $T_{\rm EFF}=2900$K, $\log g=-0.40$, $M = 2 M_\odot$ (solid blue line) and a uniform disk toy model (red dashed line).  The lower panel shows the LD-to-UD residuals and has a vertical scale magnified by 5$\times$.}
\end{figure}

\begin{figure}
\includegraphics[scale=0.9,angle=0]{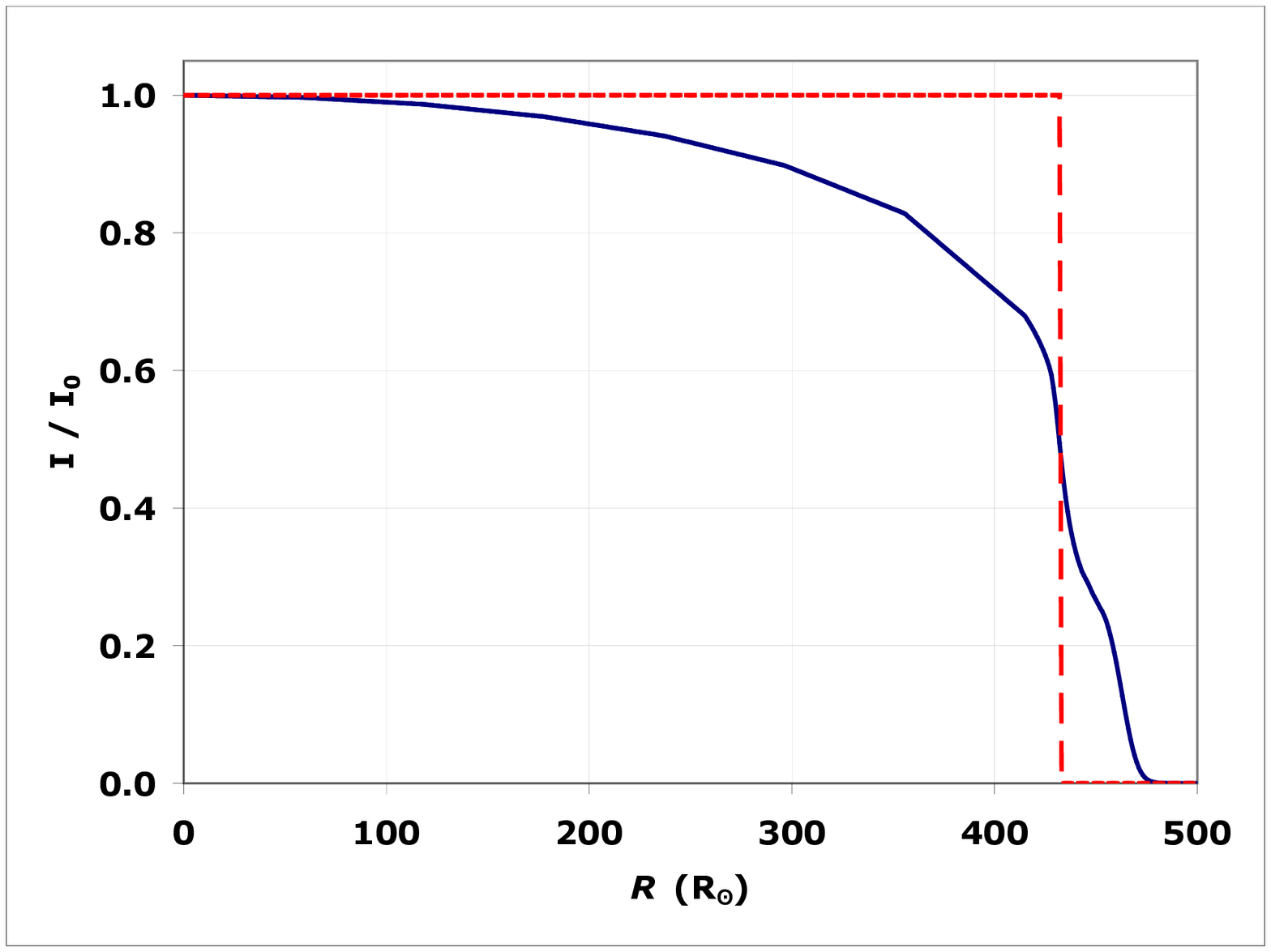}
\caption{\label{fig_LD} Based on \citet{Aringer2009A_A...503..913A}, a plot of a representative CLV curve indicated for the PTI $K$-band filter for a limb-darkened model atmosphere with $T_{\rm EFF}=2900$K, $\log g=-0.40$, $M = 2 M_\odot$ (solid blue line) and a uniform disk toy model (red dashed line).}
\end{figure}

In general, for interferometric data taken in the central lobe (as is the case for all of our data points), the difference in visibility between LD disks and a simple UD fit is on the order of $\Delta V < 1$\%.  The reason for this extremely small offset --- in contrast to earlier studies that had factors of $1.022\times$ or more \citep[e.g.,][]{Dyck1996AJ....112..294D} --- is in the close match between the central lobes of a UD visibility curve and the LD visibility curves from the center-limb-variation (CLV) brightness profiles of the \citet{Aringer2009A_A...503..913A} models (Figure \ref{fig_LD2UD}).  Comparison of the source UD CLV to the LD CLV  shows the two intersect at the $I/I(0)=\sim 50\%$ point  (Figure \ref{fig_LD}), and matches the linear `radius' defined for these models (which fundamentally is set by the models' Rosseland radii, where $\tau_{\rm ROSS}=2/3$).  A net result of this definition is that, for the central lobe visibility data, there is significant correspondence between the `toy' UD model and the visibility data from the model CLV.  As such, a LD-to-UD conversion factor will be not be implemented here.




\begin{deluxetable}{llllcccccccccc}
\tabletypesize{\scriptsize}

\rotate

\tablewidth{650pt}

\tablecaption{Observed targets, coordinates, number of visibility points, resultant uniform disk angular size, and visible variability information.\label{tab_Targets}}


\tablehead{
\colhead{} & \colhead{} & \colhead{} & \colhead{} & \colhead{}  & \colhead{}  & \colhead{} & \colhead{} & \colhead{} & \multicolumn{3}{c}{GCVS\tablenotemark{d}} & \colhead{AFOEV} & \colhead{HIP} \\
\cline{10-12}
\colhead{} & \colhead{} & \colhead{} & \colhead{} & \colhead{V2}  & \colhead{$\theta$\tablenotemark{a}}  & \colhead{$\chi^2$/} & \colhead{Average} & \colhead{} & \colhead{VarType} & \colhead{Amp} & \colhead{Band} & \colhead{Amp} & \colhead{Amp} \\
\colhead{Target ID} & \colhead{Alt.} & \colhead{RA (J2000)} & \colhead{DE (J2000)} & \colhead{Points\tablenotemark{c}}  & \colhead{[mas]}  & \colhead{DOF} & \colhead{Residual} & \colhead{Group\tablenotemark{b}} & \colhead{Type} & \colhead{[mag]} & \colhead{} & \colhead{[mag]} & \colhead{[mag]}
}
\startdata
HD225217 & SU And & 00 04 36.4 & +43 33 04.7 & 10 / 10 & $2.577 \pm 0.021$ & 0.855 & 0.016 & CV3 & LC & 0.50 & V & 0.80 & 0.23\\
HD2342 & AQ And & 00 27 31.6 & +35 35 14.4 & 2 / 2 & $3.867 \pm 0.015$ & 0.217 & 0.001 & CV5 & SR & 1.90 & p & 1.80 & 0.36\\
HD19881 & V410 Per & 03 13 38.6 & +47 49 33.8 & 19 / 21 & $2.357 \pm 0.011$ & 0.649 & 0.017 & CV3 & SRB: & 0.60 & V & 1.10 & 0.28\\
IRC+40067 & AC Per & 03 45 03.3 & +44 46 51.0 & 11 / 12 & $3.281 \pm 0.018$ & 0.797 & 0.010 & CV5 & LB: & 0.60 & p &  & \\
HD30443 &  & 04 49 16.0 & +35 00 06.4 & 3 / 3 & $1.430 \pm 0.094$ & 0.226 & 0.020 & HC4 &  &  &  &  & \\
HD280188 & V346 Aur & 04 52 34.8 & +38 30 19.9 & 1 / 1 & $3.399 \pm 0.028$ & inf & N/A & SCV & SRA: & 1.01 & B & 1.10 & 0.26\\
HD33016 & TX Aur & 05 09 05.4 & +39 00 08.4 & 3 / 3 & $2.945 \pm 0.036$ & 0.701 & 0.012 & CV4 & LB & 0.70 & V & 0.90 & 0.28\\
HD34467 & V348 Aur & 05 19 10.2 & +35 47 32.4 & 2 / 2 & $2.237 \pm 0.049$ & 0.005 & 0.001 & CV3 & LB & 1.05 & B & 0.60 & 0.15\\
HIP25004 & V1368 Ori & 05 21 13.3 & +07 21 19.3 & 2 / 2 & $3.283 \pm 0.042$ & 0.230 & 0.005 & CV2 & SRA & 0.35 & Hp &  & 0.35\\
HD38218 & TU Tau & 05 45 13.7 & +24 25 12.4 & 2 / 4 & $3.894 \pm 0.011$ & 2.050 & 0.002 & CV3 & SRB & 3.30 & V & 2.20 & 0.46\\
HD247224 & CP Tau & 05 45 26.5 & +15 30 45.3 & 2 / 2 & $1.602 \pm 0.069$ & 0.276 & 0.011 & CV3 & LB & 1.90 & p & 1.40 & \\
HD38572 & FU Aur & 05 48 08.1 & +30 37 51.8 & 2 / 2 & $2.664 \pm 0.027$ & 0.009 & 0.001 & CV2 & LB & 1.20 & B & 1.00 & 0.23\\
HD38521 & AF Aur & 05 48 44.7 & +44 54 36.0 & 2 / 2 & $2.331 \pm 0.046$ & 0.912 & 0.015 &  & SR: & 3.20 & p & 1.70 & \\
HIP29896 & GK Ori & 06 17 42.0 & +08 31 11.3 & 5 / 5 & $3.500 \pm 0.022$ & 0.580 & 0.008 & CV6 & SR & 1.50 & V & 1.50 & 0.31\\
HD45087 & AB Gem & 06 26 14.0 & +19 04 23.0 & 4 / 4 & $2.755 \pm 0.031$ & 0.501 & 0.010 & CV6 & LB & 2.20 & p & 2.00 & \\
HIP31349 & CR Gem & 06 34 23.9 & +16 04 30.3 & 1 / 1 & $3.864 \pm 0.015$ & inf & N/A & CV2 & LB & 1.20 & B & 1.00 & 0.54\\
HD46321 & RV Aur & 06 34 44.6 & +42 30 12.7 & 4 / 4 & $2.126 \pm 0.064$ & 2.196 & 0.044 & CV3 & SRB & 1.30 & p & 1.50 & 0.16\\
HD47883 & VW Gem & 06 42 08.5 & +31 27 17.5 & 8 / 8 & $2.187 \pm 0.028$ & 0.512 & 0.020 & CV2 & LB & 0.38 & V & 1.30 & 0.23\\
HD48664 & CZ Mon & 06 44 40.8 & +03 19 00.0 & 5 / 5 & $2.883 \pm 0.040$ & 0.503 & 0.015 & CV5 & LB & 2.00 & p &  & \\
HD51620 & RV Mon & 06 58 21.4 & +06 10 01.5 & 1 / 1 & $3.421 \pm 0.062$ & inf & N/A & CV3 & SRB & 2.19 & B & 2.40 & 0.39\\
HD54361 & W CMa & 07 08 03.4 & -11 55 23.7 & 7 / 7 & $4.890 \pm 0.016$ & 0.471 & 0.003 & CV3 & LB & 1.55 & V & 2.00 & 0.45\\
IRC+10158 & BK CMi & 07 15 38.8 & +05 03 39.7 & 6 / 6 & $2.946 \pm 0.046$ & 0.745 & 0.033 & CV5 &  &  &  & 1.50 & \\
HD57160 & BM Gem & 07 20 59.0 & +24 59 58.0 & 1 / 1 & $2.213 \pm 0.028$ & inf & N/A & CV1 & SRB & 0.60 & p & 0.93 & 0.29\\
HD59643 & NQ Gem & 07 31 54.5 & +24 30 12.5 & 41 / 41 & $1.499 \pm 0.066$ & 2.580 & 0.101 & HC5 & SR+ZAND & 0.59 & V & 1.50 & 0.18\\
HD60826 & BE CMi & 07 36 29.1 & +02 04 44.1 & 3 / 3 & $2.396 \pm 0.070$ & 0.079 & 0.009 & CV3 & LB & 0.50 & p & 0.55 & \\
HD70072 & RY Hya & 08 20 06.3 & +02 45 56.0 & 11 / 13 & $4.112 \pm 0.011$ & 1.854 & 0.011 & CV6 & SRB & 2.80 & p & 2.75 & \\
HD144578 & RR Her & 16 04 13.4 & +50 29 56.9 & 58 / 64 & $1.582 \pm 0.015$ & 1.633 & 0.047 & CV2 & SRB & 4.70 & B & 2.60 & 1.61\\
HD173291 & HK Lyr & 18 42 50.0 & +36 57 30.8 & 6 / 9 & $3.945 \pm 0.006$ & 2.924 & 0.003 & CV3 & LB & 1.80 & V & 1.60 & 0.32\\
HIP92194 & DR Ser & 18 47 21.0 & +05 27 18.6 & 54 / 75 & $3.447 \pm 0.005$ & 2.375 & 0.017 & CV5 & LB & 2.99 & B & 0.40 & 0.27\\
HIP95024 & U Lyr & 19 20 03.9 & +37 52 27.4 & 5 / 7 & $3.933 \pm 0.024$ & 0.161 & 0.003 & CV5 &  &  &  & 2.80 & \\
HIP95777 & AW Cyg & 19 28 47.5 & +46 02 38.1 & 8 / 9 & $3.387 \pm 0.010$ & 6.115 & 0.016 & CV4 & SRB & 3.50 & p & 2.50 & 0.29\\
HD186047 & TT Cyg & 19 40 57.0 & +32 37 05.7 & 1 / 1 & $3.305 \pm 0.024$ & inf & N/A & CV4 & SRB & 1.70 & B & 1.70 & 0.26\\
HIP99336 & AY Cyg & 20 09 44.2 & +41 29 36.4 & 2 / 2 & $2.156 \pm 0.128$ & 4.568 & 0.075 & CV5 & LB & 2.20 & B & 0.80 & 0.32\\
HD191783 & RY Cyg & 20 10 23.0 & +35 56 50.1 & 17 / 17 & $2.382 \pm 0.017$ & 1.283 & 0.033 & CV4 & LB & 1.80 & V & 2.00 & \\
HIP102764 & V1862 Cyg & 20 49 16.2 & +33 13 47.1 & 17 / 20 & $2.902 \pm 0.010$ & 3.899 & 0.031 & CV6 & SR & 2.60 & B &  & 0.33\\
HIP105539 & YY Cyg & 21 22 28.7 & +42 23 46.4 & 42 / 49 & $2.576 \pm 0.007$ & 1.288 & 0.016 & CV4 & SRA & 1.10 & p & 0.80 & 0.51\\
IRC+50399 & LQ Cyg & 21 46 39.1 & +52 34 00.0 & 1 / 1 & $2.069 \pm 0.098$ & inf & N/A &  & LB: & 1.10 & B &  & \\
HD208512 & LW Cyg & 21 55 13.7 & +50 29 49.7 & 2 / 2 & $3.954 \pm 0.025$ & 1.045 & 0.003 & CV5 & LB & 2.20 & B & 1.50 & 0.26\\
HD208526 & RX Peg & 21 56 22.3 & +22 51 39.9 & 1 / 1 & $2.988 \pm 0.127$ & inf & N/A & CV3 & SRB & 1.90 & p & 1.90 & \\
HIP113715 & VY And & 23 01 49.4 & +45 53 09.1 & 96 / 116 & $2.456 \pm 0.005$ & 1.758 & 0.021 & CV5 & SRB & 2.20 & V & 1.40 & 0.64\\
HD220870 & EW And & 23 26 57.3 & +49 30 58.9 & 71 / 73 & $1.904 \pm 0.008$ & 1.217 & 0.023 & CV3 & LB: & 1.00 & p & 1.30 & 0.26\\
\enddata
\tablenotetext{a}{Error bar is the formal fitting error.  For ensuing computations that make use of $\theta_{\rm UD}$, an error floor of $\pm 4.6\%$ will be used, as per the discussion at the end of \S \ref{sec_PA}.}
\tablenotetext{b}{From the group designations in \citet{Bergeat2001A_A...369..178B}, as discussed in \S \ref{sec_BergeatCompare}.}
\tablenotetext{c}{Visibility points used in fitting $\theta_{\rm UD}$, versus number available; 3-$\sigma$ outliers were discarded.}
\tablenotetext{d}{Variability type and amplitude from the AAVSO Variability Index \citep{Watson2012yCat....102027W}; amplitude indicated as $V$-band, $B$-band, photographic (pg), or Hipparcos (Hp).}
\tablecomments{No data available for HD30443, IRC+10158, HIP 95024.
AFOEV data has been selected for the available V-band data.
}



\end{deluxetable}

Table \ref{tab_Targets} presents our observed angular sizes, including target IDs, coordinates, the number of visibility points $N(V^2)$ (available and used), derived $\theta_{\rm UD}$ sizes, associated goodness-of-fit as characterized by a reduced chi-squared ($\chi^2_\nu$) and average squared visibility residuals.

\subsection{Variability of Angular Sizes}

\subsubsection{Temporal}\label{sec_temporal}

All of these carbon stars exhibit photometric variability, particularly in the visible, where $\Delta V\simeq$1 is not uncommon \citep{Watson2006SASS...25...47W}, though in the near-IR (where the bulk of the spectral energy is emitted), the variability is significantly less (\S \ref{sec_SED_fitting}).  For HIP92194 and HIP113715, with $N>50$ data points over $\Delta t > 400$ days, each had a Lomb-Scargle periodgram analysis was examined \citep{Lomb1976Ap&SS..39..447L,Scargle1982ApJ...263..835S}, to see if any periodicity was detectable in the angular size data.  None was detected, unlike the $\sim\pm 20\%$ angular diameter pulsations seen for Miras \citep{Thompson2002ApJ...577..447T} which have far greater variability with $\Delta V \gtrsim3$.  Examination of the visible light curves for these objects show behavior that markedly more stochastic than is the case for Miras.

\subsubsection{Size versus Position Angle for Carbon Stars}\label{sec_PA}

Recent studies have shown that certain stars can take on a non-spherical appearance due to the effects of surface inhomogeneities \citep{Chiavassa2011A&A...535A..22C}, rotation \citep{vanbelle2001ApJ...559.1155V,vanBelle2012A&ARv..20...51V} or --- of particular interest in the case of carbon stars --- for stars undergoing extreme mass loss \citep[as illustrated for IRC+10216 in Figure 1 of][]{Tuthill2005ApJ...624..352T}.  For interferometry data of this nature  --- namely, very high resolution but not particularly well sampled with regards to full image reconstruction --- other phenomena can take on the appearance of a varying uniform disk size with changes in position angle.  Specifically, a circular stellar disk with prominent, non-central surface inhomogeneities is one solution, as well as a circular stellar disk with an encircling disk.  To inform our discussion of the possible sources of in \S \ref{sec_departures_from_spherical}, we will first establish for which stars we can confidently measure departures from on-sky spherical symmetry.

For most of our sample, there is insufficient $\{u,v\}$ plane sampling to definitively establish for each individual object whether or not they are non-spherical.  However, for 12 of our stars, we managed to obtain visibility measurements for both the North-South (NS) and North-West (NW) baselines of PTI, permitting examination of these stars as non-spherical objects.




\begin{deluxetable}{lcccccccccc}
\tabletypesize{\scriptsize}											

\rotate



\tablecaption{Parameters of ellipse fits to angular size data for those stars with semi-orthogonal NS and NW baseline data.\label{tab_ellipseFits}}


\tablehead{
Star ID & $\theta_{\rm circ}$ & a & b & $\phi$ & $\chi^2_{\nu,{\rm ellipse}}$ & $\chi^2_{\nu,{\rm circle}}$ &$\Delta \chi^2_\nu$ & $\theta_{\rm equiv}$ & $\theta_{\rm equiv} \over \theta_{\rm circ}$ & oblateness\\
 & (mas) & (mas) & (mas) & (deg)\tablenotemark{a} & & & & (mas) &  & $o_{\rm ab}=a/b-1$
}
\startdata
HD19881 & $2.371 \pm 0.010$ & $2.441 \pm 0.085$ & $2.165 \pm 0.096$ & $-50 \pm 8$ & 1.1 & 1.6 & 0.4 & 2.298 & 0.969 & $0.128^{+0.093}_{-0.128}$ \\
IRC+40067 & $3.297 \pm 0.018$ & $3.396 \pm 0.079$ & $3.005 \pm 0.160$ & $-47 \pm 6$ & 0.8 & 1.7 & 0.8 & 3.194 & 0.969 & $0.130^{+0.087}_{-0.073}$ \\
HD59643 & $1.499 \pm 0.024$ & $1.766$ & $1.476$ & $57$ & 1.2 & 1.2 & 0.0 & 1.614 & 1.077 & 0.197\\
HD144578 & $1.642 \pm 0.015$ & $1.658 \pm 0.036$ & $1.612 \pm 0.080$ & $-78$ & 2.2 & 2.1 & -0.1 & 1.635 & 0.995 & $0.029^{+0.096}_{-0.029}$ \\
HIP92194 & $3.459 \pm 0.004$ & $3.917 \pm 0.086$ & $3.365 \pm 0.016$ & $28 \pm 1$ & 2.0 & 4.8 & 2.7 & 3.630 & 1.049 & $0.164^{+0.032}_{-0.037}$ \\
HIP95024 & $3.808 \pm 0.018$ & $3.951 \pm 0.048$ & $3.588 \pm 0.053$ & $-76 \pm 12$ & 0.1 & 14.1 & 14.1 & 3.765 & 0.989 & $0.101^{+0.038}_{-0.030}$ \\
HIP95777 & $3.394 \pm 0.010$ & $3.634 \pm 0.076$ & $3.152 \pm 0.062$ & $-54 \pm 11$ & 2.6 & 7.8 & 5.2 & 3.385 & 0.997 & $0.153^{+0.042}_{-0.049}$ \\
HD191783 & $2.391 \pm 0.018$ & $2.432 \pm 0.048$ & $2.174 \pm 0.170$ & $-62 \pm 12$ & 1.2 & 1.5 & 0.3 & 2.300 & 0.962 & $0.118^{+0.065}_{-0.118}$ \\
HIP102764 & $2.938 \pm 0.010$ & $3.191 \pm 0.068$ & $2.846 \pm 0.024$ & $76 \pm 7$ & 1.3 & 5.7 & 4.4 & 3.013 & 1.026 & $0.121^{+0.027}_{-0.039}$ \\
HIP105539 & $2.619 \pm 0.005$ & $2.776 \pm 0.047$ & $2.539 \pm 0.022$ & $42 \pm 1$ & 2.2 & 3.2 & 1.0 & 2.655 & 1.014 & $0.093^{+0.026}_{-0.032}$ \\
HIP113715 & $2.450 \pm 0.004$ & $2.530 \pm 0.030$ & $2.429 \pm 0.010$ & $66 \pm 11$ & 3.8 & 4.0 & 0.2 & 2.479 & 1.012 & $0.042^{+0.014}_{-0.042}$ \\
HD220870 & $1.901 \pm 0.007$ & $1.940 \pm 0.020$ & $1.839 \pm 0.027$ & $-75 \pm 18$ & 0.9 & 1.3 & 0.4 & 1.889 & 0.993 & $0.055^{+0.049}_{-0.055}$ \\
\multicolumn{8}{l}{Check stars (NS, NW baselines only)} \\
HD113226 & $3.062 \pm 0.009$ & $3.073 \pm 0.096$ & $3.001 \pm 0.073$ & $-7 \pm 42$ & 2.1 & 1.9 & -0.2 & 3.037 & 0.992 & $0.024^{+0.033}_{-0.024}$ \\
HD216131 & $2.366 \pm 0.011$ & $2.388 \pm 0.159$ & $2.344 \pm 0.056$ & $29$ & 0.3 & 0.3 & 0.0 & 2.366 & 1.000 & $0.019^{+0.071}_{-0.019}$ \\
\multicolumn{8}{l}{Check stars (all baselines)} \\
HD113226 & $3.066 \pm 0.008$ & $3.088 \pm 0.030$ & $3.018 \pm 0.054$ & $10 \pm 9$ & 1.8 & 1.8 & 0.0 & 3.053 & 0.996 & $0.023^{+0.028}_{-0.023}$ \\
HD216131 & $2.369 \pm 0.010$ & $2.395 \pm 0.047$ & $2.333 \pm 0.047$ & $27$ & 0.3 & 0.5 & 0.1 & 2.364 & 0.998 & $0.027^{+0.036}_{-0.027}$
\enddata
\tablenotetext{a}{No error given for $\{a,b,\phi\}$ or oblateness $o_{\rm ab}=a/b-1$ indicates that the error term was not constrained by examination of the $\Delta \chi^2_{\rm ellipse}$ surface.}




\end{deluxetable}

\begin{figure*}
\begin{center}
\includegraphics[scale=0.9,angle=0]{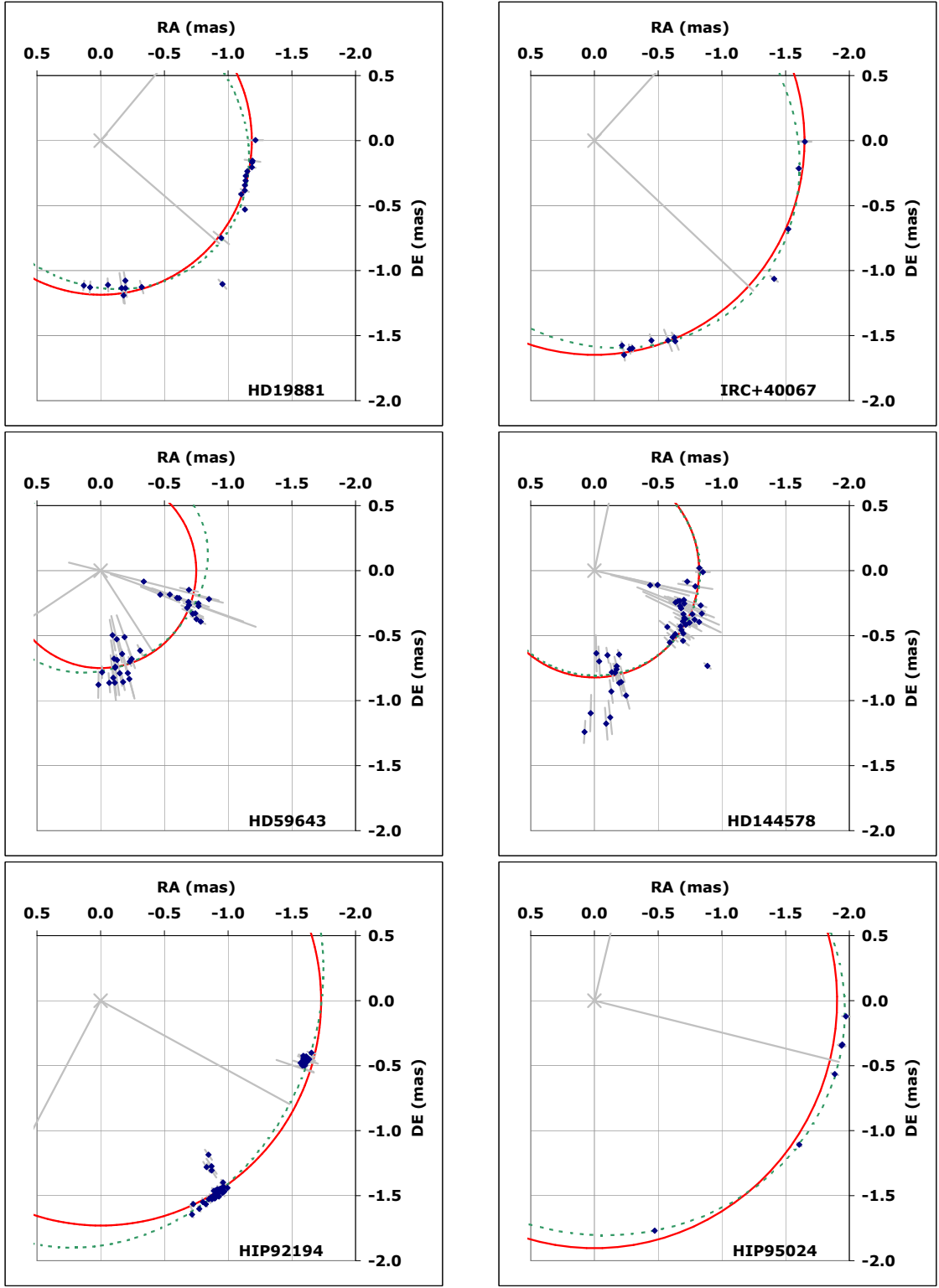}
\caption{\label{fig_star-limbs-1} Stellar limb fits for the carbon stars in this study, including circular (red solid line) and elliptical (green dotted line).  Outlier data point at $\{-0.9,-1.1\}$mas for HD 19881, $\{-0.9,-0.7\}$mas for HD144578 not included in fit data of Table \ref{tab_ellipseFits}; outliers at $\{-0.8,-1.3\}$mas for HIP92194 also discarded.}
\end{center}
\end{figure*}

\begin{figure*}
\begin{center}
\includegraphics[scale=0.9,angle=0]{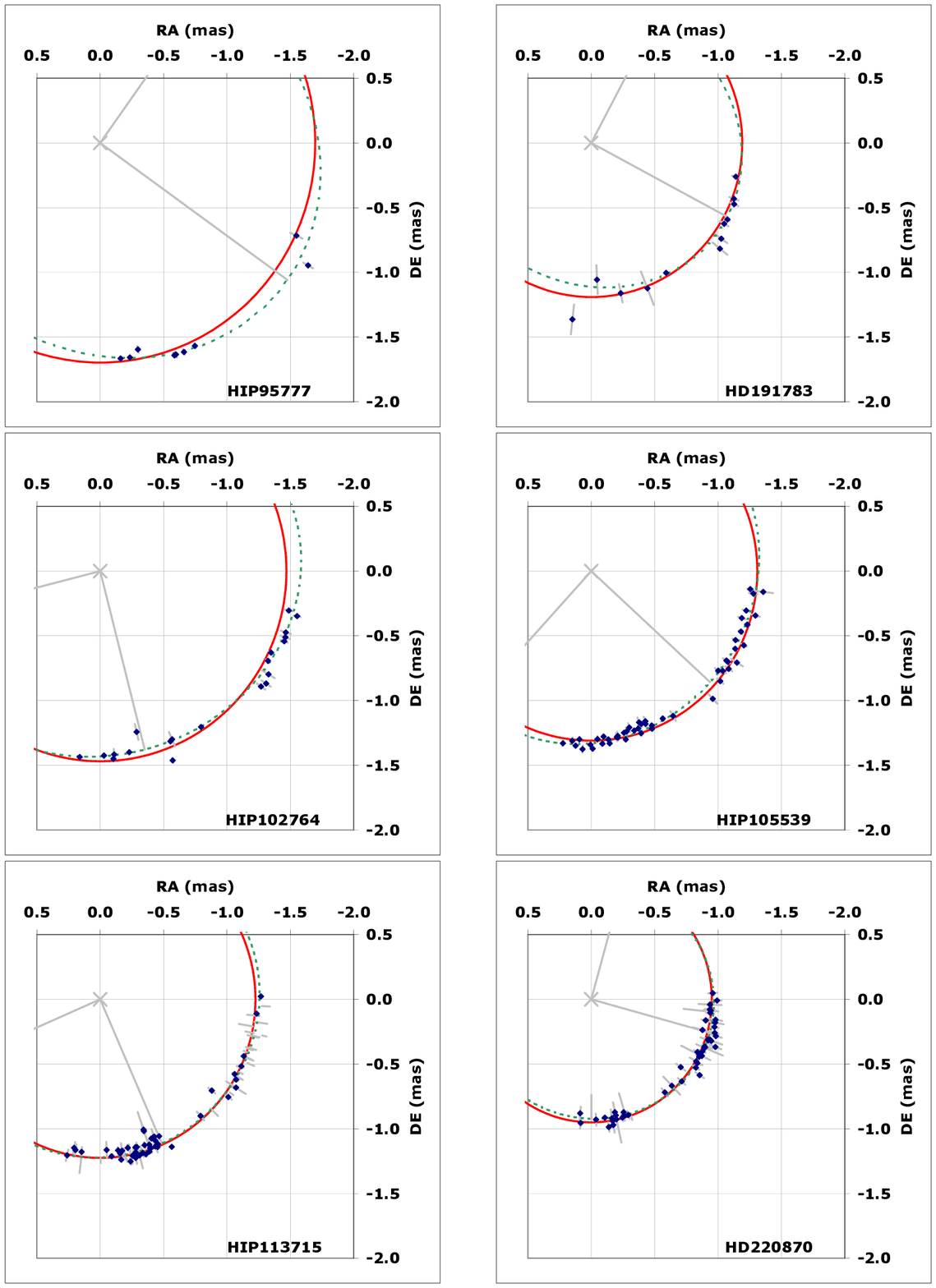}
\caption{\label{fig_star-limbs-2} Stellar limb fits for the carbon stars in this study (continued), including circular (red solid line) and elliptical (green dotted line).}
\end{center}
\end{figure*}

For each of these 12 stars, each individual visibility point was fit with a corresponding uniform disk diameter.  These UD diameters were then colllectively fit with (1) a simple single-parameter circle ($\theta_{\rm circ}$), and (2) an ellipsoidal fit, with a major axis, minor axis, and position angle $(a, b, \phi)$  (Table \ref{tab_ellipseFits}; Figures \ref{fig_star-limbs-1} and \ref{fig_star-limbs-2})\footnote{Individual calibrated visibility points and associated metadata (observation MJD, baseline data including length and apparent position angle) are available upon request for any star in our sample (1,305 in total).  All raw visibility data is also available online at the PTI Archive, http://nexsci.caltech.edu.}.  A few of the visibility points are clear outliers and are discarded (as noted in the captions of Figures \ref{fig_star-limbs-1} and \ref{fig_star-limbs-2}).  Visibility measurements were made at a variety of baselines corresponding to the central lobe of the star's visibility curve.  Slight differences can be expected between the UD visibility function and the `true' stellar visibility curve can be expected, but as seen in Figure \ref{fig_LD2UD}, this will be at the $\ll 1\%$ level - well below the threshold of our measurement error.

Overall, the ellipsoidal fit is an improved characterization of the angular size data; for one object (HD59643) the data are insufficient to properly constrain the elliptical fit; for a second (HD144578) the elliptical fit does not improve upon the circular fit.
For the rest of this subset of the target sample, an elliptical fit is an improvement, though statistically significant one (with a improvement in reduced chi-squared of $\Delta \chi^2_\nu>1$) for only 5 of the objects: HIP 92194, HIP 95024, HIP 95777, HIP 102764, HIP 105539.  For our ensuing analyses, it is useful to note three statistics that we can establish with these 5 stars:  First, we can compute the {\it equivalent} angular size by examining the area associated with an ellipse of dimensions $\{a,b\}$, where $\theta_{\rm equiv}=\sqrt{ab}$.  
This parameter will allow us to proceed with the effective temperature analysis in \S \ref{sec_TEFF}.

Second, we can examine the ratio between the angular size indicated by $\theta_{\rm circ}$ and the ellipse axes $\{a,b\}$: $a / \theta_{\rm circ} = 0.954 \pm 0.041$ and $b / \theta_{\rm circ} = 1.046 \pm 0.021$ --- overall, an average difference of $\sim 4.6\%$.  Thus, for those stars presented that have insufficient $\{u,v\}$ coverage (the 7 other stars in Table \ref{tab_ellipseFits}, plus all other objects found in Table \ref{tab_Targets}), we can establish an angular size error floor for circular uniform disk fits of $\pm4.6\%$: the formal $\theta_{\rm UD}$ fitting error to the visibility data see in Table \ref{tab_Targets} might be well below this, but without better $\{u,v\}$ coverage, it is unclear where on the stellar photosphere the angular size data is being sampled.

Finally, the average oblateness for these 5 stars is $\overline{o_{\rm ab}}=\overline{a/b-1} = 0.127 \pm 0.031$, which we will examine in \S \ref{sec_departures_from_spherical} in considering origins of non-spherical symmetry for these objects.

\subsubsection{Check Stars: Spherical Giants}

The indication that many of carbon stars we have observed are non-spherical in a statistically significant manner (and that more may be, beyond of the statistical significance of our measurements) is a remarkable finding.  One necessary check of consistency is to examine similar data for stars that are, {\it a priori},  {\it not} expected to show any departures from spherical symmetry.  Towards that end, we examined the PTI archive for giant stars that had sufficiently complete data sets for similar such analysis as in \S \ref{sec_PA}.  These objects were not necessarily observed on the same nights as the carbon stars in this study, but data were collected in an identical queue-scheduled manner and reduced following an identical data reduction prescription.  Additionally, we cross-referenced this list against the catalog of \citet{deMedeiros1999A&AS..139..433D} to select stars with small $v \sin i$ values.

For the purposes of demonstrating the integrity of our approach, we selected HD113226 and HD216131, which are G8III, G8+III giants \citep{Gray2003AJ....126.2048G,Keenan1989ApJS...71..245K} with rotational velocities of $v \sin i = 1.2$ and $2.3$ km s$^{-1}$, respectively.  These stars were selected purely on the criteria that (1) a proxy estimate \citep{vanBelle1999PASP..111.1515V} of their angular size indicated $3.0< \theta_{\rm EST} < 4.0$mas, and (2) of all the giants in the PTI Archive, these two objects had the greatest number of visibility points available in all three PTI configurations: NS, NW, and South-West (SW).

\begin{figure*}
\begin{center}
\includegraphics[scale=0.9,angle=0]{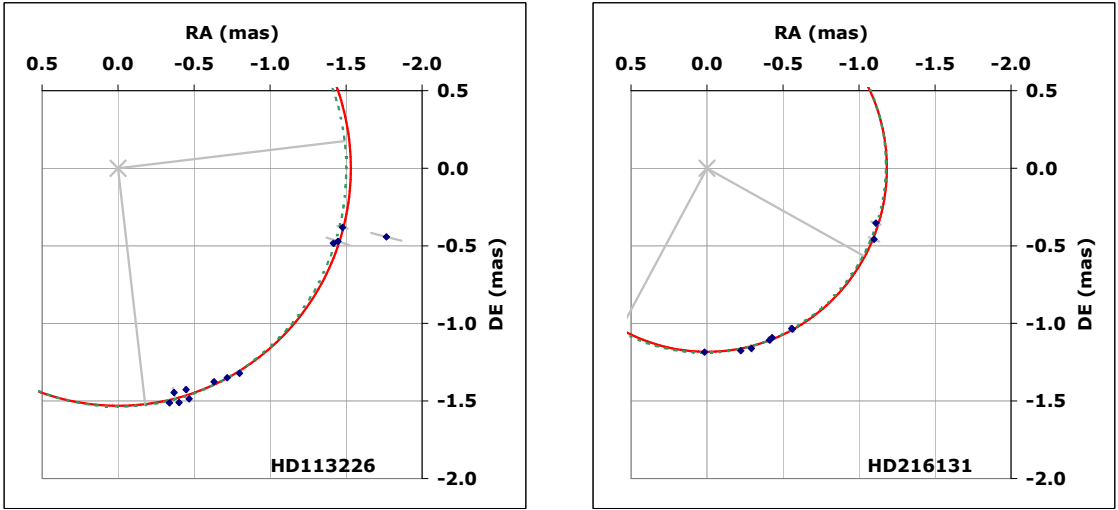}
\caption{\label{fig_star-limbs-check-NS-NW} Angular size data for the slow-rotating giant check stars HD113226 (G8III) and HD216131 (G8+III) from the PTI NS and NW baselines; these stars have $v \sin i=1.2$ and $2.3$ km s$^{-1}$, respectively \citep{deMedeiros1999A&AS..139..433D}, and, as such, are expected to present circularly symmetric geometry upon the sky.  This expectation is substantiated by the fit data in Table \ref{tab_ellipseFits}.}
\end{center}
\end{figure*}

Using the same analysis as described for the carbon stars in \S \ref{sec_PA}, for just the NS, NW interferometry data, we fit the angular sizes for these two giants using both circular and elliptical fits (bottom of Table \ref{tab_ellipseFits}, and Figure \ref{fig_star-limbs-check-NS-NW}).    HD113226 exhibits one clear outlier in the angular size data that is discarded.  The expectation here is that, if there was a systematic (and non-astrophysical) offset between the calibrated visibility points of the NS versus the NW baselines (which would manifest itself as artificial ellipticity), we would see elliptical fits for these two objects appearing (improperly) as statistically significant.  As seen in Table \ref{tab_ellipseFits}, this is not the case: HD113226 is robustly $\theta_{\rm circ}=3.062 \pm 0.030$, and HD216131 is $\theta_{\rm circ}=2.366 \pm 0.031$, with no statistical significance to an elliptical fit in either case.

\begin{figure*}
\begin{center}
\includegraphics[scale=0.9,angle=0]{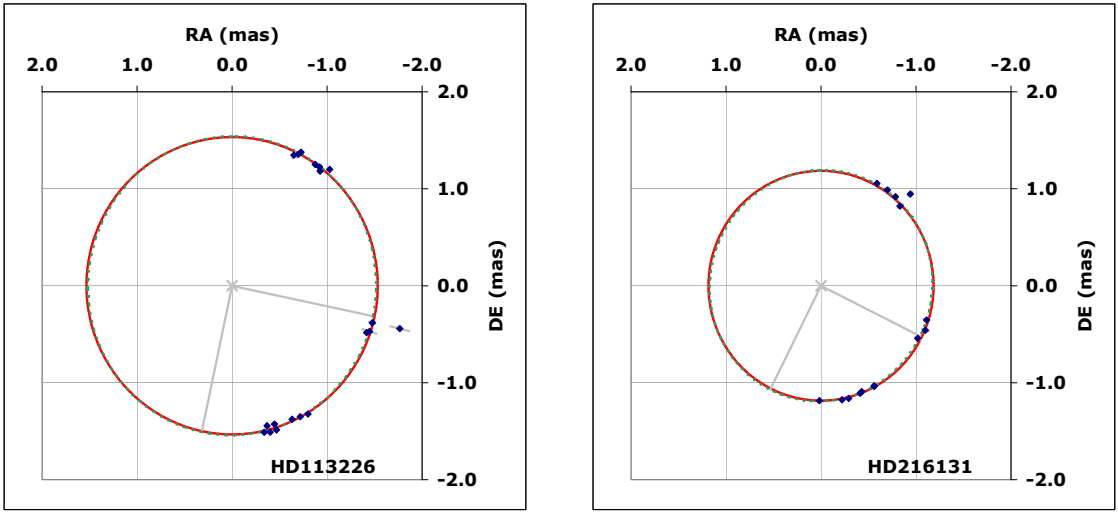}
\caption{\label{fig_star-limbs-check} As Figure \ref{fig_star-limbs-check-NS-NW} but including the additional SW baseline of PTI.  The outlier at $\{-1.0,+1.0\}$mas for HD216131 was discarded.}
\end{center}
\end{figure*}

Additionally, these objects have the benefit of further $\{u,v\}$ coverage in the form of SW baseline data (Figure~\ref{fig_star-limbs-check}), which permit us one further level of consistency check.  For this case, HD216131 also exhibits one clear outlier in the angular size data that is discarded.  The additional constraint of increased $\{u,v\}$ coverage improves the major axis $a$ in the elliptical fit for both objects, but in both cases, the elliptical fit remains as no improvement over a simple circular fit, and the fit for a simple circle is statistically unchanged with HD113226 at $\theta_{\rm circ}=3.066 \pm 0.034$ and HD216131 is $\theta_{\rm circ}=2.369 \pm 0.036$.  Even though the SW baseline data was not taken for the carbon stars in this study, comparison of the NS-NW giant stars fits relative to the NS-NW-SW fits illustrate the robust quality of the NS-NW fits, even in the lack of the additional $\{u,v\}$ constraint provided by the SW baseline data.

\section{Effective Temperatures}\label{sec_TEFF}

\subsection{Bolometric Flux Estimates}\label{sec_SED_fitting}

\begin{figure}
\includegraphics[scale=0.6,angle=270]{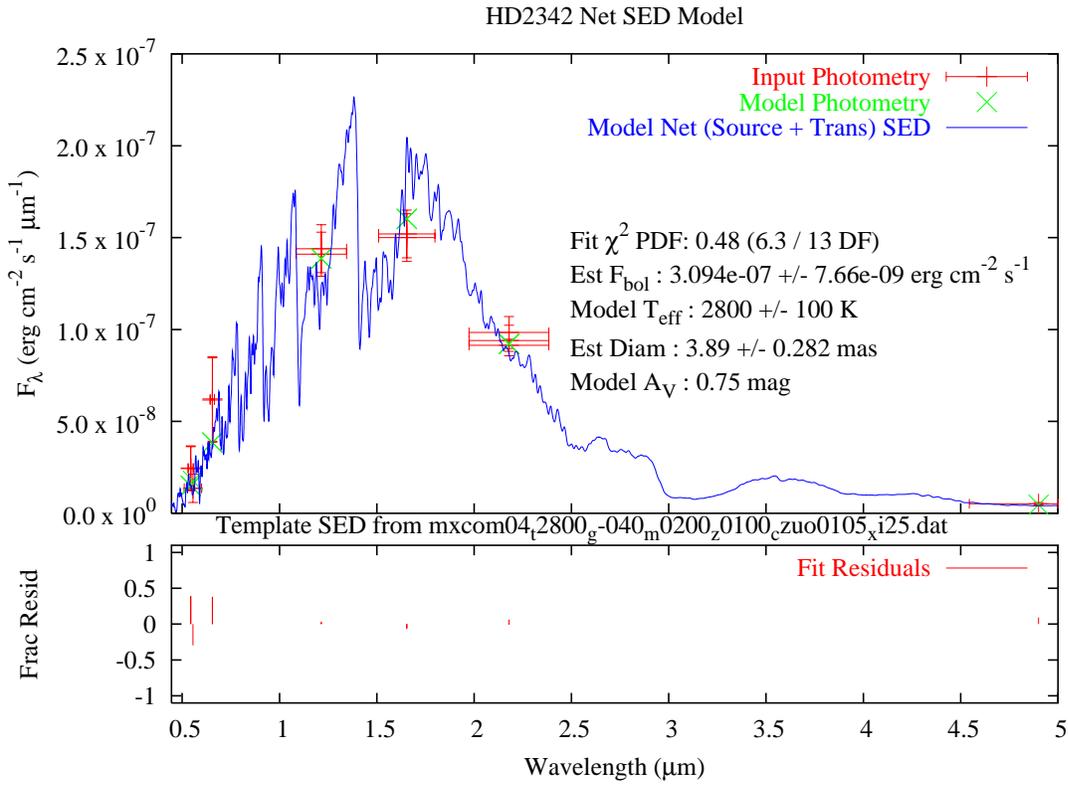}
\caption{\label{fig_spectra} Using a model atmosphere from \citet{Aringer2009A_A...503..913A}, a plot of a representative spectra for a limb-darkened model atmosphere with $T_{\rm EFF}=2800$K, $\log g=-0.40$, $M = 2 M_\odot$, fit to HD2342 photometry with $A_{\rm V}=0.75$.}
\end{figure}

For each of the target stars observed in this investigation, a bolometric flux $(F_{\rm BOL})$ estimate was established through spectral energy distribution (SED) fits.  This fit was accomplished using literature photometry values, with spectra from our carbon star models (\S \ref{sec_carbonmodels}; see example spectra in Figure \ref{fig_spectra}). The model spectra were adjusted to account for overall flux level, wavelength-dependent reddening.  Reddening corrections were based upon the empirical reddening determination described by \citet{Cardelli1989ApJ...345..245C}, which differs little from van de Hulst's theoretical reddening curve number 15 \citep{Johnson1968nim..book..167J,Dyck1996AJ....111.1705D}. Both narrowband and wideband photometry in the 0.5 $\mu$m to 25 $\mu$m were used as available, including Johnson $UBV$ \citep[see, for example,][]{Eggen1963AJ.....68..483E,Eggen1972ApJ...175..787E,Moreno1971A_A....12..442M},
Stromgren $ubvy\beta$ \citep{Piirola1976HelR....1....0P}, Geneva \citep{Rufener1976A_AS...26..275R}, 2Mass $JHK_s$ \citep{Cutri2003tmc..book.....C}, Vilnius $UPXYZS$ \citep{Zdanavicius1972VilOB..34....3Z}, $V$ and $I$ band data from the TASS Mark IV survey \citep{Droege2007yCat.2271....0D}, and the COBE DIRBE Point Source catalog \citep{Smith2004ApJS..154..673S}, which included additional near- to mid-infrared data from the MSX satellite as well. Zero-magnitude flux density calibrations were based upon the values given in \citet{Fukugita1995PASP..107..945F} and \citet{cox00}, or the system reference papers cited above.  The full list of photometry used for the SED fits is given in Table \ref{tab_SED_phot}.

One concern regarding the photometry used herein is that it was not taken contemporaneously with the interferometric angular sizes. However, this concern is mitigated by noting a few properties of these particular objects.  First, these are non-Mira carbon stars, and as such, have lower levels of photometric variability in the visible than those extreme variables; the visible levels of variability are characterized for our target stars in Table \ref{tab_Targets}.

Second, and more importantly, in the near- to mid-infrared, where the bulk of the bolometric flux is emitted, these objects show significantly less variability.  This can be seen informally for those stars in Table \ref{tab_SED_phot} with multiple measures over time --- e.g. HD 33016 has $m_{\rm K} = 2.29, 2.26,2.31$,  HD 173291 has $m_{\rm K} = 1.69,1.64,1.62$ --- where the epoch-to-epoch change in the near-infrared magnitudes is less than the measurement error (typically $\Delta m_{\rm K} = 0.1$).  The study of \citet{Whitelock2006MNRAS.369..751W} found that, for non-Mira carbon stars, $K$-band variability was, on average, $\Delta \overline{m_{\rm K}}=0.26$.  (The Mira carbons in the same study showed significant $K$-band variability, with $\Delta \overline{m_{\rm K}}=1.14$).

Since more than 50\% of a carbon star's flux is emitted between $J$ and $K$ (and over 90\% from $R$ through $L$, see Figure \ref{fig_spectra}), we are honing in on the significant foundation of the emitted flux for these objects.  For those cases were $V$ band photometry was available from the {\it General Catalog of Variable Stars} \citep{Samus2009yCat....102025S}, the $V$ band points reflect the average value between the minimum and maximum contained in the GCVS, with an uncertainty reflecting the range of $V$ band values.  Overall, these uncertainties due to variability should not affect the general trends derived for carbon stars in \S \ref{sec_subTEFF}.

As a sanity check on whether static model carbons star atmospheres are necessary for this evaluation, versus hydrostatic atmospheres, we can check the average effective temperatures derived in the next section, for those stars in the lower half of visible light variability, versus those in the upper half.  Examining Table \ref{tab_Targets}, we find our largest homogenous data set of visible light variability data is from the AFOEV (L'Association Française des Observateurs d'\'{E}toiles Variables)\footnote{http://cdsarc.u-strasbg.fr/afoev/}, with the separation occurring at $\Delta V=1.55$mag.  For the less variable stars, we have 17 objects\footnote{Omitting HD59643, for the reasons discussed in \S \ref{sec_yamashita} and \S \ref{sec_BergeatCompare}.} with $\overline{T_{\rm EFF}}=2950 \pm 270$; for the more variable stars, we have 12 objects with $\overline{T_{\rm EFF}}=2850 \pm 280$ -- an indication that, there is no significant difference based upon this metric.


For each star under consideration, the collected photometry was fit to a full grid of the model spectra, with an additional parameter of reddening being tested in steps of $\Delta A_V$=0.25 from $A_V$=0.5 to 2.5.  By selecting the result with the best reduced $\chi^2_\nu$, this fitting procedure resulted in a selection of the optimal model in terms of parameters $T_{\rm EFF}$ and $\log g$; the fitting itself also provided fit values for reddening $A_V$ and source bolometric flux $F_{\rm BOL}$.

Thus, in Table \ref{tab_SED_fits}, a value for $F_{\rm BOL}$ is presented for each star, based upon a best fit achieved through searching a grid of $\{ T_{\rm EFF}, \log g, A_{\rm V} \}$; the  $\chi^2_\nu$ value of that fit is also presented in the table.  The values for linear radius $R$ and bolometric magnitude $M_{\rm BOL}$ associated with the best-fit model are also noted in Table \ref{tab_SED_fits}.  These models follow the usual mass-radius-luminosity relationships, $L=4\pi R^2 \sigma T_{\rm EFF}^4=4\pi (GM/g) \sigma T_{\rm EFF}^4$; as such, for a constant $T_{\rm EFF}$ and $g$, $\Delta L$ will scale linearly with $\Delta M$, $\Delta R$ scales $\Delta M^{-1/2}$, and $\Delta M_{\rm BOL}$ scales as $2.5 \log(\Delta M)$.  As noted in \S \ref{sec_carbonmodels}, models with the C/O ratio of 1.05 were used; to simplify the search grid, available models with C/O=1.10, 1.40, and 2.00 were not.  However, we tested these models to ascertain the impact upon the derived $F_{\rm BOL}$ values as C/O varied; across a range of $T_{\rm EFF}$ values, for C/O=1.10 and 1.40, $\Delta F_{\rm BOL}$ was less than $<2\%$, well within the $\sim4\%$  $F_{\rm BOL}$ errors reported in Table \ref{tab_SED_fits}.  Figure \ref{fig_spectra} illustrates this well: although a changing C/O ratio will change the fine details of the SED, the gross nature of the SED will remain the same, and as such, the computed $F_{\rm BOL}$ from this procedure; Figure 3 of \citet{Aringer2009A_A...503..913A} illustrates this further.

Using the distances we derive in \S \ref{sec_distances}, the average rate of reddening for this ensemble is $A_{\rm V}=1.06$ mag/kpc.  In comparison to the `canonical' value of 0.7-1.0 mag/kpc \citep{Gottlieb1969ApJ...157..611G,Milne1980AJ.....85...17M,Lynga1982A_A...109..213L}, this value is on the high end of the expected envelope, though it is worth noting that most of these targets are near the galactic plane with an average absolute galactic latitude of $\bar{|b|}=10^o \pm 9$, where greater levels of extinction are expected.  Circumstellar reddening is a significant phenomenon for these stars, and is accounted for in the inclusion of mid-IR photometric data from IRAS and DIRBE/MSX \citep{Smith2004ApJS..154..673S} -- but is not significant factor (less than $<1\%$ of the total) when computing $F_{\rm BOL}$: again, the dominant bandpass when integrating under the SED curve for such a determination is 1-2$\mu$m (see Figure \ref{fig_spectra}).  As an independent check, $A_V$ was estimated using the 3D map of \citet{Gontcharov2012AstL...38...87G} (noting these stars are $\sim 1.5 \times$ more distant than the 1 kpc calibration of that study).  On average, our calculations of $A_V$ are slightly redder, but the comparison is framed by a considerable amount of scatter, with $\Delta A_V=0.22 \pm 0.83$.  Given that these objects emit the principal portion of their flux redwards of the $V$-band, in the 1.0-2.0$\mu$m regime (as illustrated in Figure \ref{fig_spectra}) where reddening drops markedly (eg. $A_K \simeq 0.11 A_V$), we find that this low value of $\Delta A_V$ indicates our $F_{\rm BOL}$ values should, on average, be accurate.

An important note regarding the errors reported on $F_{\rm BOL}$: the uncertainties reported with the $F_{\rm BOL}$ values are the {\it formal} values from the fitting routine for an individual spectra; however, these values probably under-represent the actual uncertainty, with an average uncertainty of $\sim 4\%$.  As such, a limit of 10\% uncertainty was taken as the actual $F_{\rm BOL}$ error for computation of the derived quantity of $T_{\rm EFF}$ in \S \ref{sec_subTEFF}.  Such a limit is consistent with derived values of $F_{\rm BOL}$ even in the presence of modest deviations in the model parameters (e.g. using a $T_{\rm EFF}=2800$K model when a $T_{\rm EFF}=3000$K model is appropriate, $\Delta \log g= \pm 0.20$, $\Delta A_{\rm V}= \pm 0.25$, etc.).  
Additionally, we expect this more conservative approach to $F_{\rm BOL}$ uncertainty is more consistent with the non-contemporaneous nature of the photometric data and the variability documented in \citet{Whitelock2006MNRAS.369..751W} for non-Mira carbon stars.

\begin{deluxetable}{ll|ccc|ccc|cccc|ccc}
\rotate
\tablecolumns{15}
\tabletypesize{\scriptsize}
\tablewidth{0pc}
\tablecaption{SED fits results and derived quantities.\label{tab_SED_fits}}
\tablehead{
Star & Yamashita & \multicolumn{3}{c}{Model Parameters} & \multicolumn{3}{c}{Fit Parameters} & \multicolumn{4}{c}{Derived} & \multicolumn{3}{c}{Previous}\\
ID  & Spectral & $[ T_{\rm EFF} / \log g]$ & $R$ & $M_{\rm BOL}$ & $A_{\rm V}$ & $f_{\rm BOL}$ & $\chi^2_{\rm RED}$ & $T_{\rm EFF}$ & $\Delta T$ & $d_{\rm R}$ & $d_{\rm BOL}$ & $T_{\rm EFF}$ & $\Delta T$ & Ref \\
 & Type\tablenotemark{e} &[ (K) / (dex)] & ($R_\odot$) & (mag) & (mag) & \tablenotemark{b} & & (K) & (K)\tablenotemark{c} & (pc) & (pc) & (K) & (K) &
}											
\startdata
HD225217 & C6,4 & [3000 / -0.6] & 467 & -5.78 & 0.75 & $17.7 \pm 0.60$ & 2.1 & $2989 \pm 102$ & -11 & 1684 & 1698 & $3057$ & -68 & OT96 \\
HD2342 & C5,4 & [2800 / -0.4] & 371 & -4.98 & 0.75 & $30.9 \pm 0.77$ & 0.5 & $2808 \pm 95$ & 8 & 892 & 887 & $2835\pm 191$ & -27 & OT96 T07 \\
HD19881 & C7,4 & [3300 / -0.2] & 295 & -5.19 & 2.25 & $22.3 \pm 0.75$ & 1.4 & $3314 \pm 113$ & 14 & 1162 & 1151 & $3247$ & 67 & OT96 \\
IRC+40067 & C6,3 & [2600 / -0.4] & 371 & -4.66 & 0.75 & $16.5 \pm 0.71$ & 2.2 & $2606 \pm 89$ & 6 & 1051 & 1047 & $2955$ & -349 & OT96 \\
HD30443 & C4,3 CH4: CN2: Ba4 & [4000 / 0.0] & 234 & -5.53 & 2.50 & $15.5 \pm 0.57$ & 0.1 & $3884 \pm 161$ & -116 & 1521 & 1615 & $3580$ & 304 & T91 \\
HD280188 & C8,1J & [3300 / -0.2] & 295 & -5.19 & 2.50 & $49.1 \pm 2.59$ & 2.2 & $3361 \pm 114$ & 61 & 806 & 776 &   &   &  \\
HD33016 & C5,4 & [2700 / -0.4] & 371 & -4.82 & 0.50 & $16.0 \pm 0.50$ & 0.8 & $2727 \pm 93$ & 27 & 1171 & 1147 & $3021\pm 112$ & -294 & OT96 T07 \\
HD34467 & C6,3 & [3000 / -0.2] & 295 & -4.78 & 1.00 & $13.6 \pm 0.68$ & 0.2 & $3003 \pm 102$ & 3 & 1224 & 1223 & $3207$ & -204 & OT96 \\
HIP25004 & C4:,4 & [2700 / -0.8] & 588 & -5.82 & 2.25 & $19.5 \pm 0.82$ & 0.7 & $2715 \pm 92$ & 15 & 1664 & 1646 &   &   &  \\
HD38218 & C5,4 & [2700 / -0.4] & 371 & -4.82 & 0.50 & $27.2 \pm 0.89$ & 0.4 & $2710 \pm 92$ & 10 & 885 & 878 & $2961\pm 55$ & -251 & OT96 T07 \\
HD247224 & C5,4 & [3100 / -0.4] & 371 & -5.42 & 1.00 & $7.5 \pm 0.34$ & 1.1 & $3056 \pm 104$ & -44 & 2152 & 2213 &   &   &  \\
HD38572 & C7,2 & [3500 / -0.2] & 295 & -5.45 & 2.25 & $34.8 \pm 1.08$ & 1.4 & $3483 \pm 118$ & -17 & 1028 & 1039 & $3099\pm 140$ & 384 & OT96 T07 \\
HD38521 & C4,4 & [2600 / -0.4] & 371 & -4.66 & 1.25 & $8.0 \pm 0.55$ & 2.1 & $2580 \pm 88$ & -20 & 1479 & 1504 &   &   &  \\
HIP29896 & C5,4e: & [2600 / -0.6] & 467 & -5.16 & 2.25 & $18.6 \pm 0.66$ & 0.6 & $2597 \pm 88$ & -3 & 1240 & 1245 & $2735$ & -138 & OT96 \\
HD45087 & C5,4 & [2700 / -0.6] & 467 & -5.32 & 2.00 & $13.9 \pm 0.61$ & 1.1 & $2723 \pm 92$ & 23 & 1575 & 1548 &   &   &  \\
HIP31349 & C8,3e & [3200 / -0.6] & 467 & -6.06 & 2.50 & $52.6 \pm 1.12$ & 0.7 & $3208 \pm 109$ & 8 & 1123 & 1118 &   &   &  \\
HD46321 & C4,5 & [2800 / -0.4] & 371 & -4.98 & 0.75 & $9.2 \pm 0.48$ & 0.2 & $2794 \pm 95$ & -6 & 1622 & 1630 &   &   &  \\
HD47883 & C5,4 Ba5 & [3100 / -0.2] & 295 & -4.92 & 0.75 & $15.0 \pm 0.73$ & 0.9 & $3118 \pm 106$ & 18 & 1252 & 1238 & $3236$ & -118 & OT96 \\
HD48664 & C4,5 & [2600 / -0.6] & 467 & -5.16 & 1.25 & $12.9 \pm 0.63$ & 0.6 & $2612 \pm 89$ & 12 & 1505 & 1494 & $2860$ & -248 & OT96 \\
HD51620 & C4,4 MS2 & [3300 / -0.2] & 295 & -5.19 & 0.75 & $45.7 \pm 2.09$ & 0.6 & $3290 \pm 112$ & -10 & 800 & 804 & $3330$ & -40 & OT96 \\
HD54361 & C6,3 & [2900 / 0.0] & 234 & -4.13 & 0.50 & $56.8 \pm 1.95$ & 1.0 & $2907 \pm 99$ & 7 & 445 & 443 & $3085\pm 290$ & -178 & L86 OT96 \\
IRC+10158 & C5,5 & [2500 / -0.6] & 467 & -4.99 & 1.25 & $12.4 \pm 0.61$ & 1.9 & $2561 \pm 87$ & 61 & 1473 & 1406 &   &   &  \\
HD57160 & C5,4J & [2900 / -0.2] & 295 & -4.63 & 0.50 & $11.1 \pm 0.45$ & 3.1 & $2871 \pm 98$ & -29 & 1237 & 1262 & $3105\pm 134$ & -234 & OT99 T07 \\
HD59643 & C6,2 CH3 & [3800 / 0.0] & 234 & -5.31 & 1.50 & $15.9 \pm 0.42$ & 1.5 & $3775 \pm 128$ & -25 & 1420 & 1442 & $3351\pm 71$ & 425 & OT96 Z09 \\
HD60826 & C5,5 & [2700 / -0.2] & 295 & -4.32 & 0.75 & $10.1 \pm 0.46$ & 3.5 & $2694 \pm 92$ & -6 & 1143 & 1147 &   &   &  \\
HD70072 & C6,4e & [2400 / -0.4] & 371 & -4.31 & 1.50 & $18.1 \pm 0.46$ & 1.9 & $2381 \pm 81$ & -19 & 838 & 852 & $2632\pm 45$ & -251 & OT96 T07 \\
HD144578 & C8,1e & [3500 / 0.0] & 234 & -4.95 & 1.25 & $11.8 \pm 1.09$ & 4.0 & $3450 \pm 117$ & -50 & 1375 & 1416 &   &   &  \\
HD173291 & C7,4 & [2800 / -0.4] & 371 & -4.98 & 0.75 & $31.5 \pm 1.01$ & 0.8 & $2793 \pm 95$ & -7 & 874 & 879 & $2866$ & -73 & OT96 \\
HIP92194 & C5,4 & [2800 / -0.6] & 467 & -5.48 & 2.25 & $24.4 \pm 0.96$ & 0.8 & $2803 \pm 95$ & 3 & 1259 & 1257 &   &   &  \\
HIP95024 & C4,5e & [2500 / -0.6] & 467 & -4.99 & 1.50 & $19.6 \pm 0.57$ & 3.6 & $2483 \pm 84$ & -17 & 1104 & 1121 & $2702$ & -219 & OT96 \\
HIP95777 & C4,5 & [2800 / -0.6] & 467 & -5.48 & 1.00 & $23.4 \pm 0.77$ & 1.0 & $2798 \pm 95$ & -2 & 1282 & 1284 & $2759$ & 39 & OT96 \\
HD186047 & {\it C6.5,y} & [2900 / 0.0] & 234 & -4.13 & 0.50 & $25.0 \pm 0.57$ & 1.9 & $2878 \pm 98$ & -22 & 658 & 668 & $3046$ & -168 & OT96 \\
HIP99336 & C6,3e & [3000 / -0.4] & 371 & -5.28 & 2.00 & $11.3 \pm 0.60$ & 1.3 & $2924 \pm 113$ & -76 & 1599 & 1685 &   &   &  \\
HD191783 & C6,4 & [2800 / -0.4] & 371 & -4.98 & 0.75 & $11.5 \pm 0.45$ & 1.0 & $2796 \pm 95$ & -4 & 1447 & 1453 &   &   &  \\
HIP102764 & C6,4 & [2700 / 0.0] & 234 & -3.82 & 0.75 & $15.0 \pm 0.56$ & 1.3 & $2705 \pm 92$ & 5 & 750 & 746 &   &   &  \\
HIP105539 & C6,3e & [2700 / -0.4] & 371 & -4.82 & 1.00 & $11.8 \pm 0.82$ & 1.2 & $2703 \pm 92$ & 3 & 1338 & 1335 &   &   &  \\
IRC+50399 & {\it C7,y} & [3000 / -0.6] & 467 & -5.78 & 2.25 & $11.6 \pm 0.48$ & 1.5 & $3006 \pm 103$ & 6 & 2097 & 2090 &   &   &  \\
HD208512 & C5,4e: & [2700 / -0.8] & 588 & -5.82 & 1.25 & $27.3 \pm 1.04$ & 1.8 & $2691 \pm 91$ & -9 & 1382 & 1390 & $2500$ & 191 & Z09 \\
HD208526 & C4,4J MS5 & [2800 / 0.0] & 234 & -3.98 & 0.75 & $18.4 \pm 0.67$ & 1.4 & $2804 \pm 95$ & 4 & 728 & 726 & $2890$ & -86 & OT99 \\
HIP113715 & C3,5J j6 MS6 & [2500 / -0.2] & 295 & -3.99 & 0.75 & $8.0 \pm 0.21$ & 0.9 & $2510 \pm 85$ & 10 & 1115 & 1108 &   &   &  \\
HD220870 & C7,3 & [2900 / -0.2] & 295 & -4.63 & 0.50 & $8.4 \pm 0.35$ & 0.9 & $2886 \pm 98$ & -14 & 1438 & 1451 &   &   &  \\
\enddata
\tablenotetext{a}{Additional non-variable model parameters were $M$=2$M_\odot$, $Z/Z_\odot=1$, and C/O $= 1.05$.}
\tablenotetext{b}{In units of $10^{-8}$ erg cm$^{-2}$ s$^{-1}$.  See discussion of $F_{\rm BOL}$ error in \ref{sec_SED_fitting}.}
\tablenotetext{c}{Difference between model $T_{\rm EFF}$ (column 2) and measured $T_{\rm EFF}$  (column 9) derived from $\theta$ (Table \ref{tab_Targets}) and $F_{\rm BOL}$ (column 7).}
\tablenotetext{d}{References: \citet[][L86]{Lambert1986ApJS...62..373L}, \citet[][T91]{Tsuji1991A&A...252L...1T}, \citet[][OT96]{Ohnaka1996A&A...310..933O}, \citet[][OT99]{Ohnaka1999A&A...345..233O}, \citet[][T07]{Tanaka2007PASJ...59..939T}, \citet[][Z09]{Zamora2009A&A...508..909Z}.}
\tablenotetext{e}{Values in italics predicted from measured $T_{\rm EFF}$ and Equation \ref{eqn_YamTeffFit2} for those stars (HD186047, IRC+50399) for which no Yamashita spectra type was given.}

\end{deluxetable}



\subsection{Effective Temperatures}\label{sec_subTEFF}

Stellar effective temperature $(T_{\rm EFF})$ is defined in terms of a star's luminosity ($L$) and linear radius ($R$) by $L=4 \pi \sigma R^2 T_{\rm EFF}^4$, which can be rewritten in terms of stellar angular diameter and source bolometric flux: $T_{\rm EFF} \propto (F_{\rm BOL} / \theta^2 )^{1/4}$.  Inherent in the expression for $F_{\rm BOL}$ is accounting for the effects of interstellar reddening; hence the attention paid to the subject in the derivation of $F_{\rm BOL}$ in \S \ref{sec_SED_fitting}.  As noted in \S \ref{sec_PA}, an angular size error floor of $\pm 4.6\%$ was used (regardless of the formal fitting error in Table \ref{tab_Targets}) due to uncharacterized stellar oblateness.

The derived values of $T_{\rm EFF}$ are presented to the left in Table \ref{tab_SED_fits}, along with the difference $(\Delta T)$ between those values and the input spectral model $T_{\rm EFF}^M$.  On average, the difference between those two temperatures is $0.25\sigma$, indicating a consistency in the overall approach
.  For these stars, the average error in determination of $T_{\rm EFF}$ is 100K --- a factor of $2.4 \times$ improvement over the previous large interferometric study in \citet{Dyck1996AJ....112..294D}.  Comparison to previous studies that have measured $T_{\rm EFF}$ by other methods for carbon stars \citep{Lambert1986ApJS...62..373L,Tsuji1991A&A...252L...1T,Ohnaka1996A&A...310..933O,Ohnaka1999A&A...345..233O,Tanaka2007PASJ...59..939T,Zamora2009A&A...508..909Z} show 23 stars in common (the rightmost columns of Table \ref{tab_SED_fits}), with an average agreement at the level of  of $\overline{\Delta T_{\rm EFF}} = -67\pm215$K.

Overall, the median temperature of this group is $T_{\rm EFF}=2803\pm337$.  If we exclude the `hot carbon' stars HD59643 and HD30443 (see the discussions in \S \ref{sec_yamashita} and \S \ref{sec_BergeatCompare}), the median remains largely the same but the spread drops considerably with $T_{\rm EFF}=2798\pm268$.  

We can also compare the four stars (HIP31349/CR Gem,
HD51620/RV Mon, HD173291/HK Lyr,
HIP92194/DR Ser) in common between this study and our previous study \citep{Paladini2011A_A...533A..27P}, which used a spectral analysis approach for determination of $T_{\rm EFF}$ (e.g. see \S 4.1 of that paper).  Overall, the agreement is at the $1-\sigma$ level, with the average $T_{\rm EFF}$ for those four stars being $3082\pm74$K from that paper, and $2989\pm51$K herein.

\subsection{Distances and Linear Radii}\label{sec_distances}

A distance determination was possible for each star in our sample via two separate methods.  First, through comparison of the measured angular size to the linear size associated with the best-fit model, a distance $(d_{\rm R})$ can be simply and directly extracted.  Second, each model has an absolute bolometric magnitude ($M_{\rm BOL}$) associated with it as well; by comparison of $M_{\rm BOL}$ with the unreddened apparent bolometric flux, a bolometric distance ($d_{\rm BOL}$) can also be estimated.  The values associated with these derivations can be found in Table \ref{tab_SED_fits}, with $d_{\rm R}$ and $d_{\rm BOL}$ on the right-hand columns.

These two separate (though not entirely independent) methods produce distances that agree well with each other --- on average, only a $1 \pm 7$\% difference between $d_{\rm R}$ and $d_{\rm BOL}$.  A more independent check is comparison to the distances found in \citet{Claussen1987ApJS...65..385C}: for the 35 objects in common, the agreement is $2 \pm 20$\% --- still very good on average, though with more scatter for individual objects.  This scatter is perhaps connected to the somewhat general approach of \citet{Claussen1987ApJS...65..385C} for distance determination; the assumption of that investigation was that all carbon stars had the same 2.2$\mu$m luminosity, with $M(K)=-8.1$, an average derived from 54 LMC/SMC carbon stars in \citet{Frogel1980ApJ...239..495F} and \citet{Cohen1981ApJ...249..481C} and assumed distance moduli of 18.6 and 19.1 for the LMC and SMC, respectively.

Although entries exist for 32 of these objects in the {\it Hipparcos} catalog, we chose to not use these data: the average fractional error $(|\pi| / \sigma_\pi)$ was 173\%.  The average parallax error was $\sigma_\pi=1.3$mas for these stars with $H_p=8.8 \pm 1.0$ and $B-V=2.5\pm0.5$ (parameters typical of carbon stars), markedly greater than the average of $\sigma_\pi=0.65$mas for stars of similar brightness, but with $B-V=0.5\pm0.5$ (typically A- to K-type main sequence stars).  As noted in \S 3.5 of \citet{vanBelle2002AJ....124.1706V} for the similar case of Mira variables, comparison of the diameter data of Table \ref{tab_Targets} and the distance information in Table \ref{tab_SED_fits} indicates that, on average, the angular sizes of these objects are $3.4 \times$ larger than their parallaxes, making such measurements particularly difficult (and, in {\it Hipparcos's} visible bandpass, susceptible to surface morphology, which should evolve on time scales comparable to the parallax measures.)

The median linear radii of these carbon stars is $R=360 \pm 100 R_\odot$, though it should be emphasized that the values for this parameter are based upon the models and not the angular size data (since distances to carbon stars remain quite uncertain).

\section{Discussion}\label{sec_discussion}

This large, homogenous data set allows us a number of interesting avenues to explore the implications for carbon stars.  First, we can compare to previous studes \citep{Yamashita1972AnTok..13..169Y,Yamashita1975AnTok..15...47Y,Bergeat2001A_A...369..178B,Bergeat2002A&A...390..987B,Bergeat2002A_A...390..967B}. Second, we can use the new data to calibrate our expectations for angular size prediction for these stars.  The indication that some carbon stars are oblate is then examined in some detail, including connections to the stellar angular momentum history, and mass loss.  Finally, the proposition that most carbon stars could be oblate is considered.

\subsection{Comparison to \citet{Yamashita1972AnTok..13..169Y,Yamashita1975AnTok..15...47Y} Spectral Types}\label{sec_yamashita}

\begin{figure}
\includegraphics[scale=0.90,angle=0]{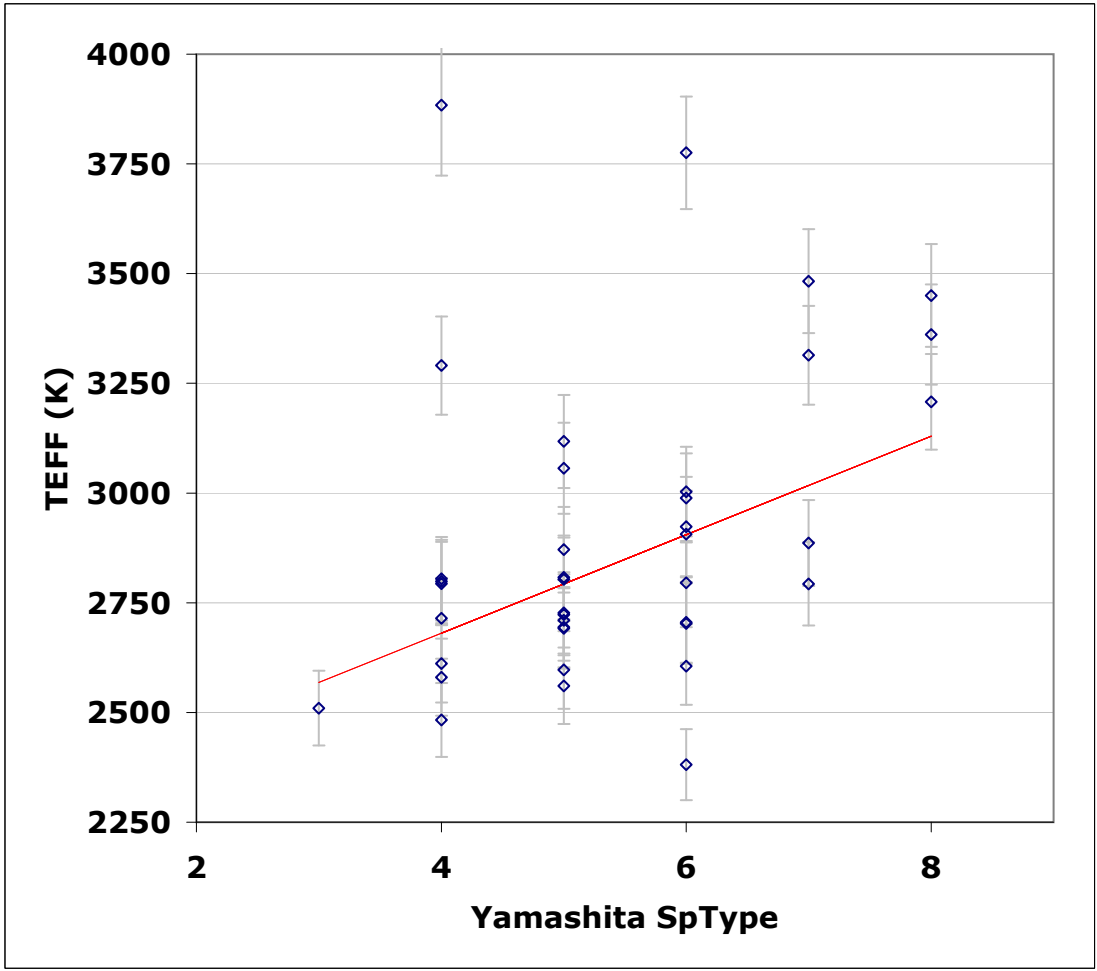}
\caption{\label{fig_yamashita_compare} From \citet{Yamashita1972AnTok..13..169Y,Yamashita1975AnTok..15...47Y}, a comparison of temperature versus spectra types for the common sample of 40 stars.  The fit line excludes the two CH star outliers HD30443 and HD59643 above 3750K, and is described in Equation \ref{eqn_YamTeffFit2}.}
\end{figure}

\citet{Yamashita1972AnTok..13..169Y,Yamashita1975AnTok..15...47Y} spectra types are written as $Cx,y$, where the criteria for sub-types $x$ and $y$ are described in detail in those investigations. The $x$ index quantifies the D-lines of Na I at 5890\AA~and 5896\AA; $y$ is indicative of the strength of the Swan bands of $C_2$ at 5635\AA~and 6191\AA.  The classification scheme of Yamashita is largely based upon (and intended to retain the principal features of) \citet{Keenan1941ApJ....94..501K}; the $x$ index was intended to establish a temperature sequence, where $T_{\rm EFF}$ was expected to decrease with increasing $x$.  However, an anti-correlation of $T_{\rm EFF}$ with the $x$ parameter is expected and has tentatively been detected in previous studies \citep{Tsuji1981JApA....2...95T,Dyck1996AJ....112..294D}.  For our target stars, these carbon spectral types can be found in Table \ref{tab_SED_fits}, and a plot of the $T_{\rm EFF}$ versus subtype $x$ can be seen in Figure \ref{fig_yamashita_compare}.

A straight fit to our spectral type data gives the relationship between subtype $x$ and $T_{\rm EFF}$
\begin{equation}
T_{\rm EFF} = (112 \pm 13.2) x + (2231 \pm 71.6)
\end{equation}
with $\chi^2_\nu=7.71$  and an average temperature offset of $\overline{|\Delta T|}=203$K. If we exclude the obvious outliers HD30443 and HD59643 above $T_{\rm EFF}>3750$ K --- justified by the presence of the CH marker in their Yamashita spectral types --- there is little change in the parameters of the fit
\begin{equation}\label{eqn_YamTeffFit2}
T_{\rm EFF} = (117 \pm 13.3) x + (2183 \pm 72.2)
\end{equation}
although the statistics of the fit improve markedly, with $\chi^2_\nu=5.16$ and $\overline{|\Delta T|}=152$K.  CH stars are thought to be a distinct group of high-velocity carbon stars \citep[][and references therein]{Yamashita1975PASJ...27..325Y}, which justifies this exclusion.

As evidenced by the poor $\chi^2_\nu$ value of the fit in Equation \ref{eqn_YamTeffFit2}, and illustrated in Figure \ref{fig_yamashita_compare}, the usefulness of $T_{EFF}$ derived from the spectral index is less than optimal.
A summary is presented in Table \ref{tab_YamashitaDataSummary} including the $T_{\rm EFF}$ weighted average by spectral type, and the $T_{\rm EFF}$ line fit value from Equation \ref{eqn_YamTeffFit2}; a comparison of these summary values to the previous values in Table 4 of \citet{Dyck1996AJ....112..294D} shows favorable agreement between these two studies.\footnote{As suggested by our referee, it would be an interesting task to investigate the relationship of the spectral index~$x$ versus $T_{\rm EFF}$ based on codes such as MOOG \citep{Sneden1973ApJ...184..839S} or SYNTHE \citep{Kurucz1981SAOSR.391.....K}. However, the sodium lines used to determine the $x$ index are very intense - thus, they are partly formed in the outer regions of the stellar atmosphere. This makes them very sensitive to deviations from hydrostatic equilibrium - especially to the appearance of even weak stellar winds - and to non-LTE effects. Consequently, it is quite problematic to model them based on the hydrostatic LTE atmospheres utilized in this investigation, or the MOOG/SYNTHE codes which also have underlying LTE assumptions. Additionally, we caution the interested investigator that their behavior will not relate to the parameters of the star in a simple way, since what will be measured is a combination of effective temperature, shock intensities, mass loss rates, atmospheric extension, etc.}

\begin{deluxetable}{lccc}




\tablecaption{Summary of $T_{\rm EFF}$ expectations for Yamashita spectral types.\label{tab_YamashitaDataSummary}}


\tablehead{
Yamashita & $N$ & $T_{\rm EFF,wtd}$\tablenotemark{a} & $T_{\rm EFF,fit}$\\
Sp Type $x$ & & (K)& (K)
}
\startdata
3 & 1 & $2510$ & 2534 \\
4 & 8 & $2726 \pm 247$ & 2650 \\
5 & 12 & $2763 \pm 168$ & 2767 \\
6 & 9 & $2746 \pm 207$ & 2883 \\
7 & 4 & $3067 \pm 337$ & 3000 \\
8 & 3 & $3334 \pm 123$ & 3116
\enddata
\tablenotetext{a}{Weighted average of $T_{\rm EFF}$ of those stars with corresponding $x$-index in $Cx,y$ Yamashita spectral type.}




\end{deluxetable}

\subsection{Comparison to \citet{Bergeat2001A_A...369..178B,Bergeat2002A&A...390..987B}}\label{sec_BergeatCompare}

\begin{figure}
\includegraphics[scale=0.90,angle=0]{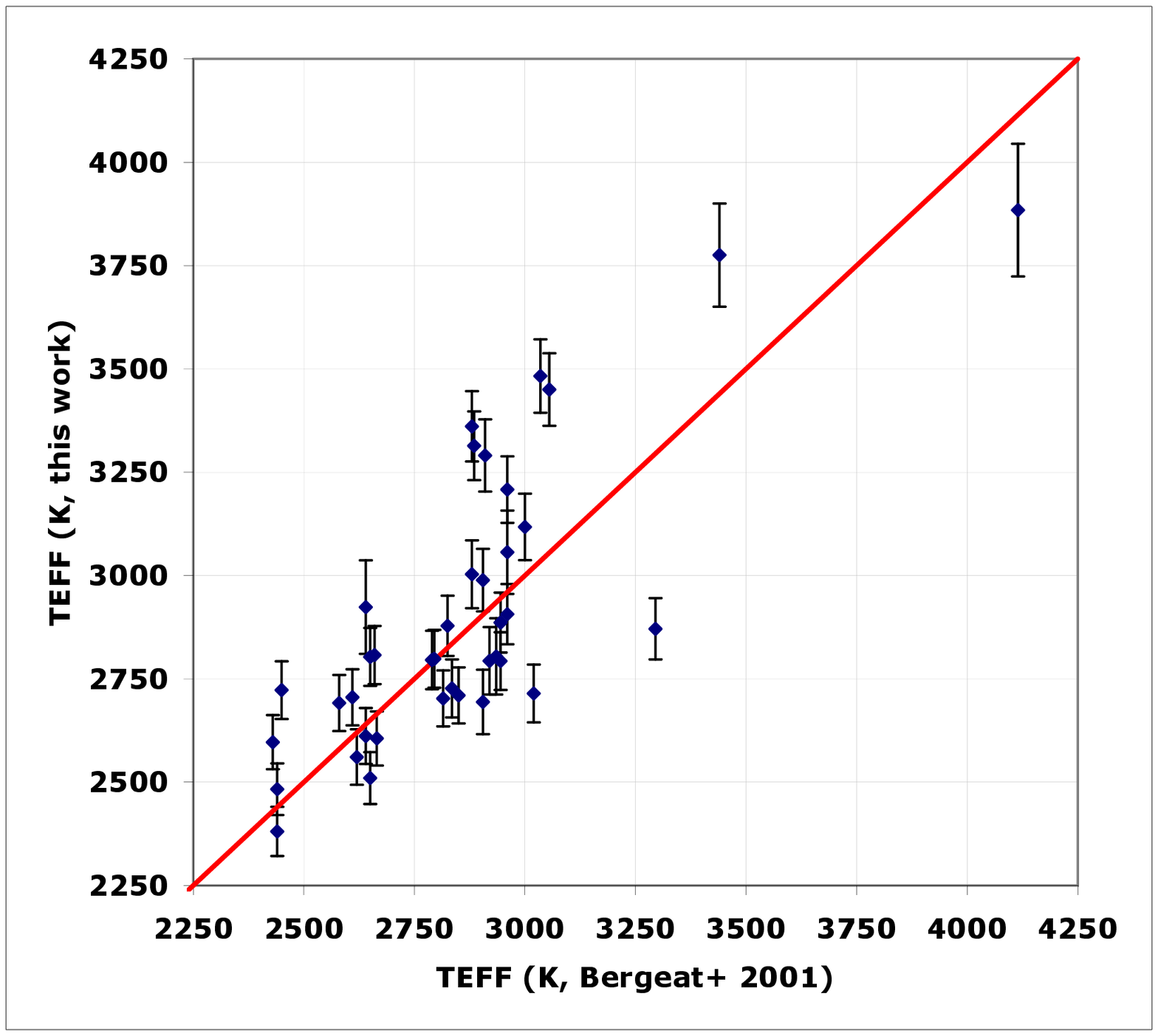}
\caption{\label{fig_bergeat_compare} From \citet{Bergeat2002A&A...390..987B}, a comparison of effective temperatures for the common sample of 38 stars; the red line is the 1:1 relationship.}
\end{figure}

We can compare our $T_{\rm EFF}$ determinations to the SED determinations found in \citet{Bergeat2002A_A...390..967B}, whose sample overlaps with 38 of our stars. As seen in Figure \ref{fig_bergeat_compare}, the agreement is quite good up to 3,000K, with the average difference being $0.21\sigma$ --- our ensemble indicating temperatures on average only 17K lower, well within the margin of error; the average absolute temperature difference in this range is 131K.  However, for our temperatures above 3,000K, our stars indicate temperatures on average 92K hotter, possibly indicating a preference in the \citet{Bergeat2002A_A...390..967B} process for $T_{\rm EFF}$ results below 3,000K.  It is difficult to isolate the source of this discrepancy, however; tracing the calibration of that study to \citet{Bergeat2001A_A...369..178B}, we find their SED calibrations are grounded in angular size determinations.  Indeed, it appears 17 of the 54 diameter measurements in \citet{Bergeat2001A_A...369..178B} were preliminary PTI diameters from a conference proceeding of ours \citep{vanBelle1999AAS...195.4501V}, and while the quality of the measurements has improved --- angular size errors are $\sim 3 \times$ smaller --- the indicated angular sizes have, on average, not changed significantly.  Overall, though, the average difference is only 15K, with an average absolute temperature difference of 153K.

\citet{Bergeat2001A_A...369..178B} utilize the photometric classification scheme proposed in \citet{Knapik1997A&A...321..236K}  to classify carbon stars into six groups, CV1--CV6, and extended this classification to `hot' carbon stars groups HC1--HC5 in \citet{Bergeat1999A&A...342..773B}.  These groups are each represented by characteristic unreddened $B-V$, $CI_{\rm B}=[0.78]-[1.08]$, and $J-K$ colors \citep[see Table 1 of][]{Knapik1997A&A...321..236K}.

As seen in Table 7 of \citet{Bergeat2001A_A...369..178B}, $T_{\rm EFF}$ determinations for these groups, from angular sizes, are listed alongside similar such determinations from SEDs and calibrated color indices.  The latter two $T_{\rm EFF}$ determinations are systematically lower in temperature for a given group than the former, particularly for temperatures above 3,000K; the final adopted $T_{\rm EFF}$ scale reflects this lower scale as well.

On the highest temperature end ($T_{\rm EFF} >$3,400K), it is significant to note that the two objects visible there in Figure \ref{fig_bergeat_compare}, HD30443 and HD59643, are HC4 and HC5 members, respectively, of the `hot carbon'(HC) group of \citet{Bergeat2001A_A...369..178B}, who noted that no angular size measurements of such group members were available at that time.  (As noted already in \S \ref{sec_yamashita}, these two stars are also identified as members of the unique CH subgroup of carbon stars.)  As such, all Bergeat et al. $T_{\rm EFF}$ calibrations for HCs relied solely upon the indirect approaches of SEDs and calibrated color indices.

However, there is overall good agreement with the CV group temperatures, as seen in Table \ref{tab_CVgroups} --- the only notable discrepancy is for CV6, which in our temperature calibration shows a smoother transition from CV5 than the large $\Delta T=335$K drop expected from \citet{Bergeat2001A_A...369..178B}.

\begin{deluxetable}{lccc}											
\tablecolumns{4}
\tablewidth{0pc}
\tablecaption{$T_{\rm EFF}$ calibration for the CV groups of \citet{Bergeat2001A_A...369..178B}.\label{tab_CVgroups}}
\tablehead{
Group & N & $T_{\rm EFF}$ & $T_{\rm EFF}$ \\
 & & (This work) & (Bergeat+ 2001)
}											
\startdata
CV2 & 5 & $3145 \pm 315$ & $3130 \pm 70$ \\
CV3 & 12 & $2916 \pm 206$ & $2940 \pm 80$ \\
CV4 & 5 & $2778 \pm 69$ & $2790 \pm 130$ \\
CV5 & 9 & $2638 \pm 154$ & $2720 \pm 135$ \\
CV6 & 4 & $2585 \pm 158$ & $2385 \pm 110$
\enddata
\end{deluxetable}



\subsection{Reference Angular Diameters}

\begin{figure}
\includegraphics[scale=1,angle=0]{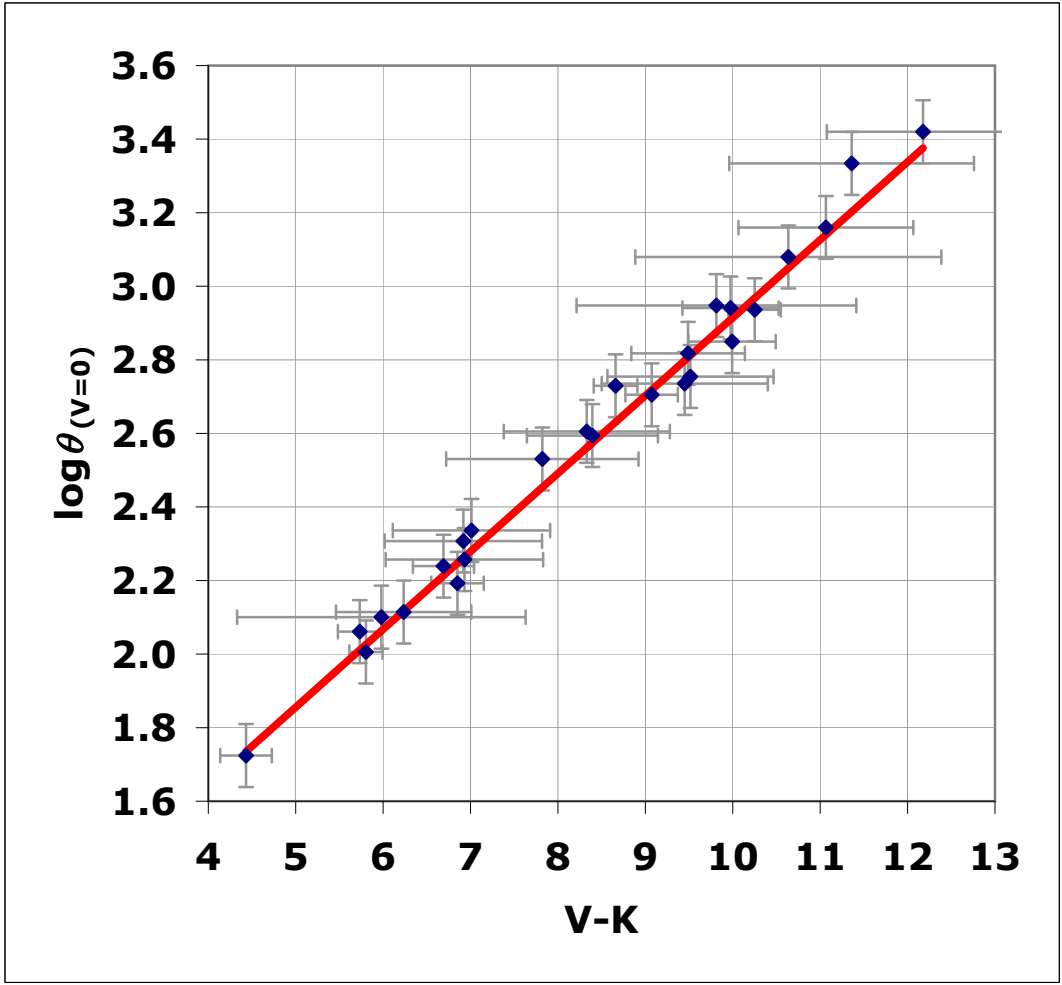}
\caption{\label{fig_ref_sizes} Reference angular sizes $\theta_{V=0}$ versus $V-K$ color, and associated fit (Equation \ref{ref_ang_sizes}).}
\end{figure}

Using these data we can refine parameters for the reliable prediction of angular sizes of these stars, as presented in \citet{vanBelle1999PASP..111.1515V}.  As in that paper, we can use $V$ band data to scale angular sizes to a common scale, $\theta_{V=0} = \theta \times 10^{V/5}$, and then fit that reference $\theta_{V=0}$ to $V-K$ color:
\begin{equation}\label{ref_ang_sizes}
\log_{10} \theta_{V=0} = (0.796 \pm 0.107) + (0.212 \pm 0.014) \times (V-K)
\end{equation}
with a rms error of 12\%; the $V-K$ uncertainty in this fit is dominated by the $V$ variability (Figure \ref{fig_ref_sizes}).  The more general fit to semiregular variables, Mira variables, and carbon stars in the previous study of \citet{vanBelle1999PASP..111.1515V} (which ignored $V-K$ uncertainty) had a slope of $0.218\pm0.014$ and intercept of $0.789\pm0.119$ and a rms error of 26\%.  Overall, this represents a significant improvement in the ability to predict carbon star angular sizes, solely upon $V$ and $K$ photometry.

\subsection{Dependency of $T_{\rm EFF}$ on $V_0-K_0$}\label{sec_TEFF_vs_V0K0}

\begin{figure}
\includegraphics[scale=0.90,angle=0]{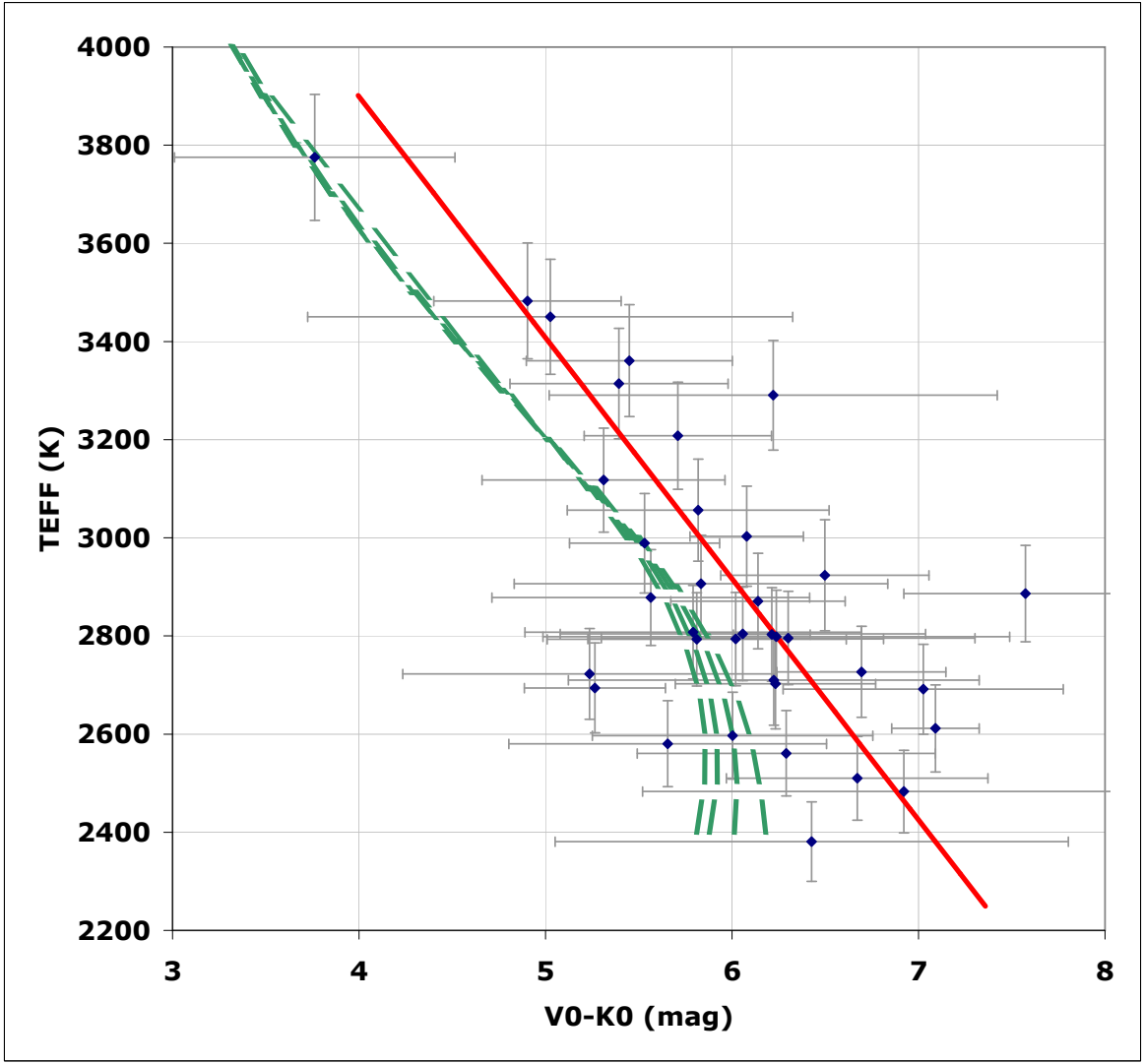}
\caption{\label{fig_TEFF_vs_V0K0} $T_{\rm EFF}$ as a function of dereddened color $V_0-K_0$.  The red solid line is the linear fit to our data, as discussed in \S \ref{sec_TEFF_vs_V0K0}.  The green dotted lines are the $\{ T_{\rm EFF} , V_0-K_0 \}$ tracks from \citet[][eg. Fig. 16]{Aringer2009A_A...503..913A} for models with $M=2 M_\odot, Z=1 Z_\odot, \log(g)=0.0,$ and $C/O=2.00, 1.40, 1.10, 1.05$ (left-to-right).}
\end{figure}

We can examine the effective temperature dependency of our measurements upon a commonly utilized color index, $V-K$, particularly in comparison to the synthetic photometry of \citet{Aringer2009A_A...503..913A}.  However, to effectively execute this comparison, we have to deredden the observed $V$ and $K$ photometry; fortunately, from the values for $A_V$ calculated in \S \ref{sec_SED_fitting}, this is rather straightforward.  $A_K$ was computed from the standard relationship of $A_K = 0.108 \times A_V$ \citep[][assuming $R_V=3.1$; other reasonable values for $R_V$ do not alter $A_K$ in a significant way]{Cardelli1989ApJ...345..245C}.  A plot of $T_{\rm EFF}$ versus $V_0-K_0$ can be seen in Figure \ref{fig_TEFF_vs_V0K0}.  The principle uncertainty in the color is the $V$-band variability, the range of which (as was the case in \S \ref{sec_SED_fitting}) was taken to be the uncertainty.

Referencing a simple linear fit line to the central $V_0-K_0$ color of 6, we derived the following fit presented in Figure \ref{fig_TEFF_vs_V0K0}:
\begin{equation}
T_{\rm EFF} = [(V_0-K_0)-6]\times (491\pm109) -  (2917 \pm 53)
\end{equation}
with a standard linear fit with errors in both dimensions \citep{Press92}; the reduced $\chi^2_\nu$ was 0.698.  Omitting the data point from HD59643, previously discussed as an outlier in \S \ref{sec_BergeatCompare}, alters the slope only slightly to $-525\pm141$ with $\chi^2_\nu=0.704$, and does not change the intercept significantly (the other high-velocity carbon star, HD30443, did not have $V$ data and was not considered in either fit).

Also plotted on Figure \ref{fig_TEFF_vs_V0K0} are the corresponding model points from \citet[][see Figure 16 of that article, albeit with transposed axes]{Aringer2009A_A...503..913A}.  Selecting the models with $M=2 M_\odot, Z=1 Z_\odot, \log(g)=0.0,$ and a range of carbon abundances $C/O=1.05, 1.10, 1.40, 2.00$, we can compare the predicted dependency of $T_{\rm EFF}$ upon $V_0-K_0$.  Our observations agree in the slope of the relationship in the range of $V_0-K_0=\{4,5.5\}$, although there is an offset of $\Delta T_{\rm EFF} = +200K$; redwards of $V_0-K_0>5.5$, the models drop in $T_{\rm EFF}$ at a constant color of $V_0-K_0\sim6$, a feature not captured by our simple linear fit.  However, in a qualitative sense, our observational data do exhibit an increasing spread in $V_0-K_0$ at a given $T_{\rm EFF}$ for temperatures below 3000K, which perhaps is indeed indicative of a spread in $C/O$ as suggested by the \citet{Aringer2009A_A...503..913A} synthetic photometry.  The blueward shift of the models relative to observations at a given $T_{\rm EFF}$ was already seen in this previous investigation, which linked this phenomenon to concerns about the applied corrections of interstellar reddening and the limitations of non-simultaneous photometry.  The former concern is perhaps mitigated by our treatment of the subject in this investigation; the latter remains an area where improvements can be made, both from the specific standpoint of simultaneous measurement of $V$ and $K$ photometry, and the general application of $F_{\rm BOL}$ determination from (spectro-)photometry.  Broadly speaking, this could also an area where the limitations of hydrostatic models as applied to even mildly variable objects are highlighted.

\subsection{Departures from Spherical Symmetry}\label{sec_departures_from_spherical}

As shown in \S \ref{sec_PA}, a simple uniform disk fit of the visibility data does not present itself with a constant value as a function of on-sky position angle, for those stars for which we have sufficient data.  It appears that, regardless of the cause (which will be discussed below), it is possible that {\it all} carbon stars examined in sufficient detail this way will exhibit similar departures from circular symmetry.

\begin{figure}
\includegraphics[scale=1,angle=0]{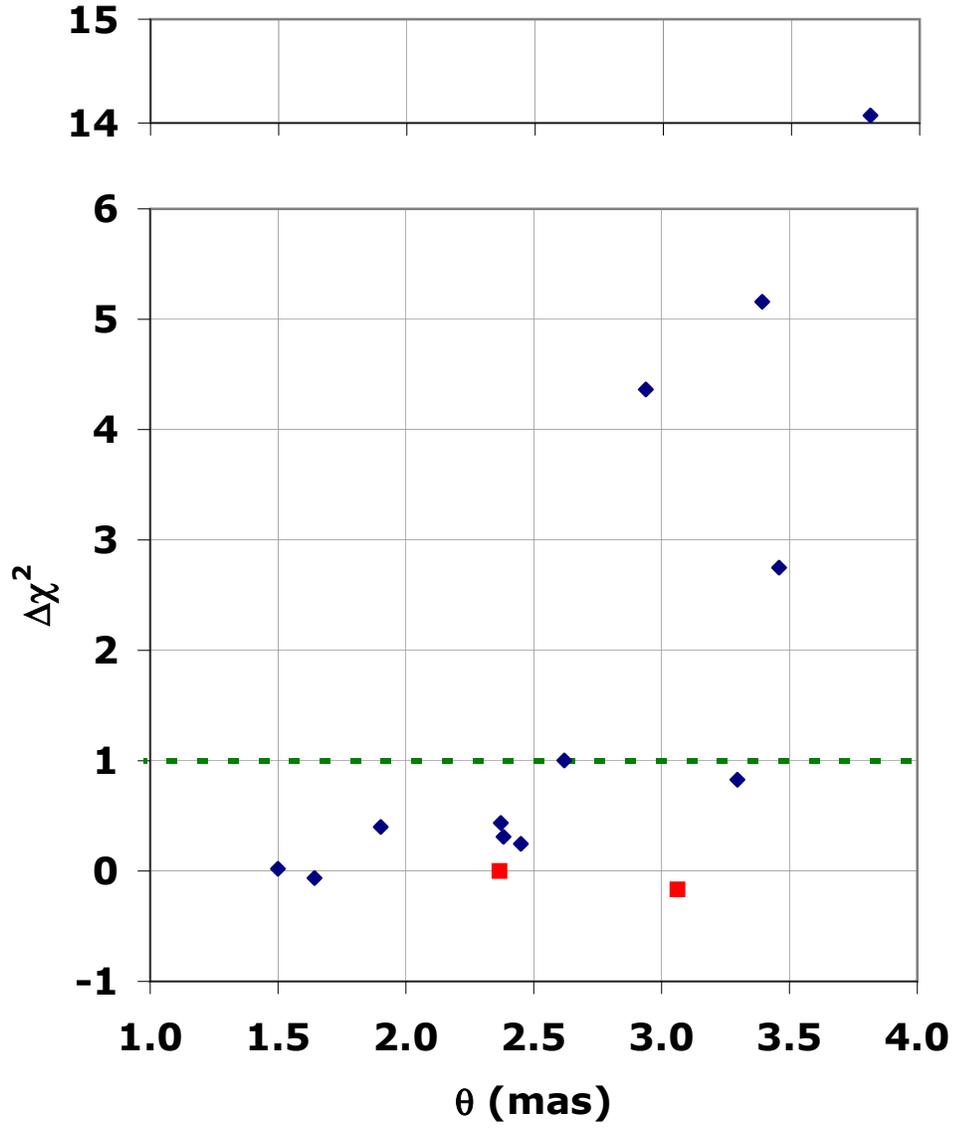}
\caption{\label{fig_delta-chi2} Difference in reduced $\chi^2_\nu$ fit values for elliptical versus circular on-sky photosphere fits; blue points are carbon stars, red points are the giant check stars.}
\end{figure}

In this study, `large' stars with $\theta > 2.5$mas observed by PTI have the average difference between the reduced chi-squared value for elliptical versus circular on-sky photospheric fits is $\overline{\Delta \chi^2_\nu}=4.70$, indicative of significant improvement of the elliptical fits over the circular fits (Figure \ref{fig_delta-chi2}).  With the exception of IRC+40067, all of the stars in this range have $\Delta \chi^2_\nu \gtrsim 1$ (and typically $\gg 1$); as noted in \S \ref{sec_PA}, only IRC+40067 has $\Delta \chi^2_\nu$ slightly less than one, at $0.83$. For `intermediate' sized stars with $1.75 < \theta < 2.5$mas, $\overline{\Delta \chi^2_\nu}=0.35$; Figure \ref{fig_delta-chi2} further shows that while the improvement does not rise to the significance level of $\Delta \chi^2_\nu \geq 1$, it is still $\Delta \chi^2_\nu > 0$ for all stars in this size range.  Only for the two smallest stars, with $\theta<1.75$mas, is the comparison between circular and elliptical photospheric fits moot, with $\overline{\Delta \chi^2_\nu} \simeq 0$ (and comparable to the giants).  By comparison, for our giant check stars, $\overline{\Delta \chi^2_\nu}=-0.01$, even though one is in the `intermediate' size range and one is in the `large' size range.

One possible explanation for the trend of increasing $\Delta \chi^2_\nu$ with increasing $\theta$ is that, for the stars with larger apparent size, the data is of sufficiently high signal-to-noise-ratio (SNR) to detect departures from sphericity  --- departures that are present in essentially {\it all} carbon stars, that merely remain undetected for the smaller stars in our sample.  (The proposition that, for these sizes in question, increasing $\theta$ means increasing SNR is considered in detail in the Appendix.)  An earlier Infrared-Optical Telescope Array (IOTA) study of evolved stars by \citet{Ragland2006ApJ...652..650R} included 10 semi-regular (non-Mira) carbon stars, for which 4 of these objects indicated detectable asymmetries in their closure phase data.  Within the context of SNR-dependent detections, it is unsurprising that the asymmetric objects are, on average, $\sim$30-50\% larger than those objects for which no asymmetries were detected.

There are three likely scenarios for these observed departures from spherical symmetry: the presence of a pronounced stellar spot (region of increased flux) on a circularly symmetric stellar photosphere; the presence of a disk around a circularly symmetric stellar photosphere; or, a truly elliptical (in appearance) photosphere.  We will examine the first two possibilities in \S \ref{sec_surface_inhomogeneities}, and the third in \S \ref{sec_mass_loss_and_rotation}.

\subsubsection{Surface Inhomogeneities: Spots, Disks}\label{sec_surface_inhomogeneities}

\begin{figure}
\includegraphics[scale=1,angle=0]{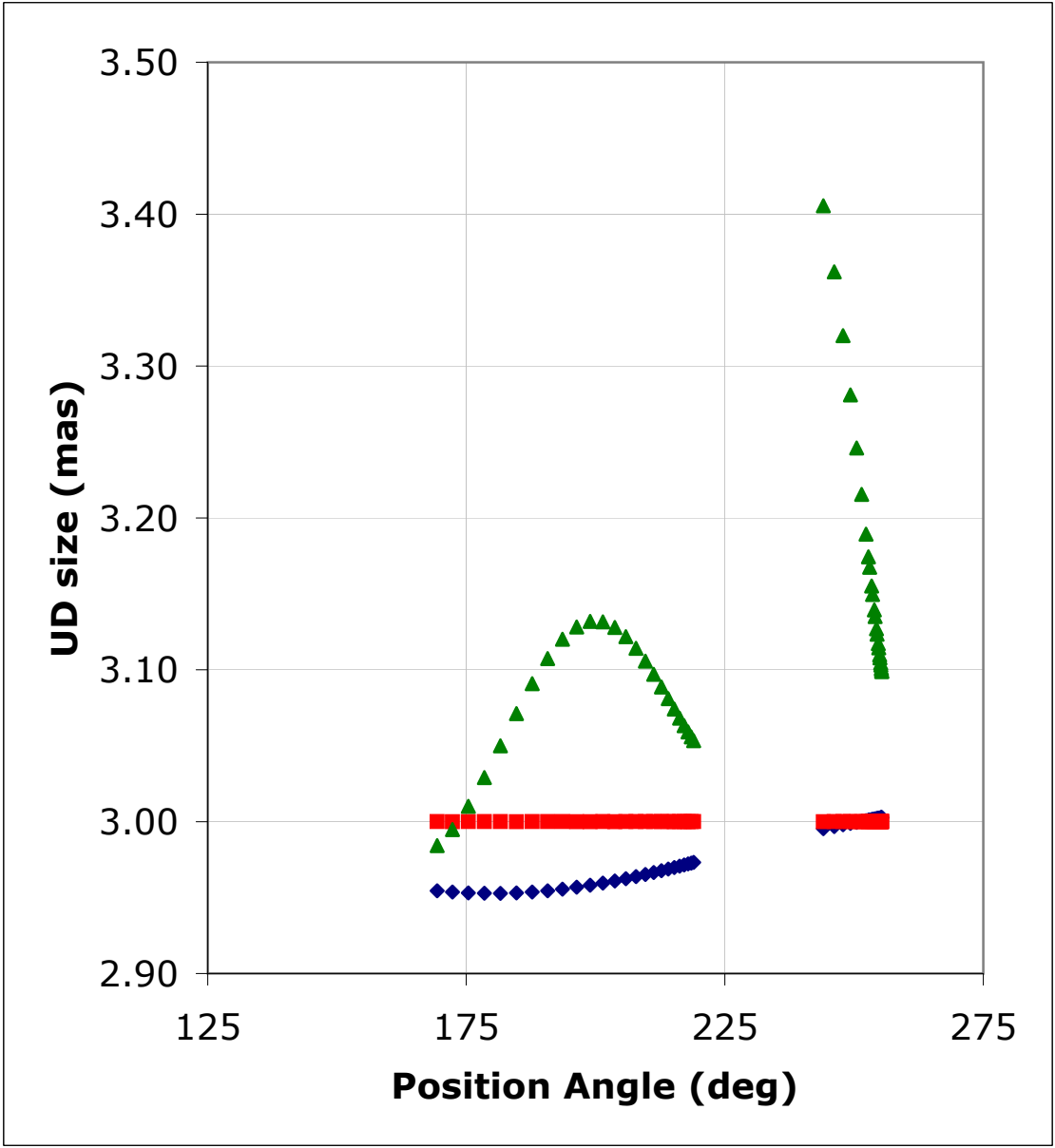}
\caption{\label{fig_surfinhomo} Indicated uniform disk angular size expected from PTI NS, NW single-baseline observations of three toy models at $\delta=+0^o$: (a) a plain disk, 3.0 mas in diameter (red squares), (b) adding a 1.0 mas diameter spot to (a), offset from the disk center by 0.75 mas, with 25\% greater flux emission than the rest of the disk (blue diamonds); (c) adding a circumstellar ring to (a), projected to be $12\times2$ mas with an inner hole of $6\times1$ mas, and 3\% of the flux of the central star (green triangles).  The left-hand grouping of points is consistent with PTI's 109-m NS baseline; the right-hand, with PTI's 85-m SW baseline.}
\end{figure}

When viewed by a single-baseline interferometer such as PTI, a circular, but non-uniform, stellar photosphere could produce data sets that give on-sky position-angle dependent uniform disk sizes.  This is illustrated in Figure \ref{fig_surfinhomo} with 3 toy models.  The first model is merely a simple uniform disk of 3 mas in diameter, observed by PTI's NS and NW baselines (as is the case for the real stars in this study), at a declination of $\delta=+0^o$ for hour angles between $HA=\{-3^h,+3^h\}$.  Using NExScI's Visibility Modeling Tool\footnote{http://nexsciweb.ipac.caltech.edu/vmt/vmtWeb/}, a simple set of $V^2$ values were synthesized, and uniform disk values fit to the individual $V^2$ points.  For the simple uniform disk, each $V^2(HA,\rm{baseline})$ point produced an apparent size of 3 mas.

To illustrate the case of a stellar hot spot, we added a 1 mas spot in our VMT toy model, which had 25\% greater flux emitted per unit area than the rest of the star, offset from the center of the toy model by 0.75mas.  A flux differential of this magnitude corresponds to a $\Delta T$ of $\sim$175K for our $\sim$3000K stars.  As seen in Figure \ref{fig_surfinhomo}, this introduces a gentle variation of 2\% in apparent UD diameter, from 2.95 to 3.0 mas, as the on-sky position angle varies with $HA$ and baseline.

A second iteration on our toy model was to add not a spot but a circumstellar disk, projected such that outer dimensions were $12\times2$ mas on-sky, with inner dimensions of $6\times1$ mas, with a major axis position angle of $45^o$.  What is striking about the disk toy model --- and this is representative of all other disk toy models we tried, at a variety of inclinations --- is that, while there is a gentle variation in fitted UD sizes for $V^2$ data from a given baseline, there is a strong discontinuity in UD sizes between baselines.    This is due to the combination of differing projection lengths for the two PTI baselines on a complex overresolved structure, such as a disk. In general, any putative disk-like structures that surround these objects will be greater in angular extent than the host stars.  Since the host stars are in the `sweet spot' of angular resolution for PTI (2-3mas), which corresponds to the mid-point of the central lobe of the visibility function, structures that are $\gtrsim1.5\times$ greater in size will be overresolved and the visibility contribution will be nearly zero.  While it is perhaps possible that a specific set of parameters for disk toy models could result in `smoother' variation in UD sizes in those specific cases, the complete lack of such discontinuities found in our observational data set lead to expect that disks are an unlikely explanation for our observations.

Additionally, such a dusty disk would need to be optically thick even in the $K$-band.  (Eg. see the illustration in Figure 13(b) of \citet{Johnson1991AJ....101.1735J}, and the related discussion in \S 5 of that paper.)  However, our expectation is that such a dense dust disk is inconsistent with the overall reddening levels for these stars - our SED results of $A_V \simeq 0.75-2.50$ (\S \ref{sec_SED_fitting}) imply $A_K \simeq 0.08-0.28$, of which most if not all of this effect at these distances should be interstellar and not circumstellar.

Certainly for objects represented by the extreme case of IRC+10216, with 15-24\% polarization features in the $H$-band and a mass loss rate of $\sim 2 \times 10^{-5} M_\odot$ yr$^{-1}$ \citep{Murakawa2005A&A...436..601M}, such considerations could not be discounted.  However, the carbon star in our sample, the polarization levels are closer to the 1\% level \citep{Lopez2011AJ....142...11L,Lopez2011RMxAA..47...63L}, and the mass loss rates are $<10^{-6.25} M_\odot$ yr$^{-1}$.

However, the possibility of significant hot spots remains quite likely.  This seems specifically supported by stellar models: \citet{Freytag2008A&A...483..571F} find that spatial inhomogeneities, induced by huge convection cells, should be significant surface features on AGB stars.  \citet{Chiavassa2010sf2a.conf..339C,Chiavassa2011ASPC..445..169C} find that for red supergiants such as Betelgeuse, significant photocentric excursions (up to $\sim$7.5\% of the stellar radius) are expected in the visible due to smaller-scale convection cells; in the $H$-band, \cite{Chiavassa2011A&A...528A.120C} show that model has larger `conspicuous spots' due to subphotospheric convective cells, which have lives of several years.

These sorts of structures --- if present in the related but non-identical carbon stars --- could lead to the `hot spot' observable we have treated here with our toy model.  A specific numeric solution is not provided for each of the stars in our sample, because the parameters for even a single simple hot spot (eg. size, location, $\Delta T$) are too complex (and therefore, degenerate) to be uniquely solvable with our data.

\subsubsection{Oblateness and Mass Loss}\label{sec_mass_loss_and_rotation}

A plain spheroid of gas will deform into an oblate spheroid when subjected to rotation, which in turn will present itself upon the sky (roughly) as an ellipsoidal photosphere.  Furthermore, oblateness due to rotation may be intrinsically connected to the phenomenon of mass loss for carbon stars.  The previous studies of \citet{Kahn1985MNRAS.212..837K} and \citet{Johnson1991AJ....101.1735J} established that mass loss for these stars must be aspherical, without positing the underlying cause.  The latter study based its empirical result upon polarization observations, which have been expanded upon in \citet{Lopez2011AJ....142...11L,Lopez2011RMxAA..47...63L}.   Much of the attention on carbon star mass loss has concentrated on chemistry and dust grain size \citep[e.g.][]{Mattsson2011A&A...533A..42M}.  However, there has been at least some discussion in the literature exploring the relationship between rotation and mass loss rates - e.g. \citep{Dorfi1996A&A...313..605D} discuss non-spherical winds due to `slowly' rotation C-rich AGB stars, up to $v \sin i \sim 7$ km s$^{-1}$.

For the carbon star V Hya, a spectroscopic rotation velocity of $v \sin i=13.5$ km s$^{-1}$ is acknowledged as `rapid' by \citet{Barnbaum1995ApJ...450..862B}, who note in their conclusions that the star would `not be spherically symmetric'. Such rotation is consistent with a large fraction ($>50$\%) of planetary nebulae being aspherical \citep{Zuckerman1986ApJ...301..772Z}, and this fact being potentially connected to central star rotation \citep{Mufson1975ApJ...202..183M,Phillips1977A&A....59...91P,Pascoli1987A&A...180..191P}.

Following Equation (A1) of \citet{vanbelle2006ApJ...637..494V}, the range of oblatenesses found \S \ref{sec_PA} roughly correspond (for $\overline{R}  \sim 360 R_\odot$, from \S \ref{sec_distances}) to $v \sin i$ values of 9-17 km s$^{-1}$ for  $\overline{M} \sim 2 M_\odot$, and 7-13.5 km s$^{-1}$ for  $\overline{M} \sim 1.2 M_\odot$.  This range of putative $v \sin i$ values actually bridges the region between the somewhat qualitatively defined regimes of `slow' rotation as defined by \citep{Dorfi1996A&A...313..605D} and the `rapid' rotation of \citet{Barnbaum1995ApJ...450..862B}.  The values exceed the macroturbulent velocities reported by \citet{Lambert1986ApJS...62..373L} $(4 \lesssim V_{\rm macro} \lesssim 7 {\rm ~km~s}^{-1})$, but we are cautious about being too concerned with this conflict, noting that that investigation relied upon plane-parallel models; their assertion that the effects of spherical extension were found to be unimportant based upon the method of \citet{Nordlund1984mrt..book..211N} has been contradicted somewhat in subsequent studies.  For example, \citet{Aringer2009A_A...503..913A} shows $\sim20\%$ reductions in SiO equivalent widths at a given $T_{\rm EFF}$ when plane-parallel model atmospheres are replaced by spherical ones.



\begin{figure}
\includegraphics[scale=0.9,angle=0]{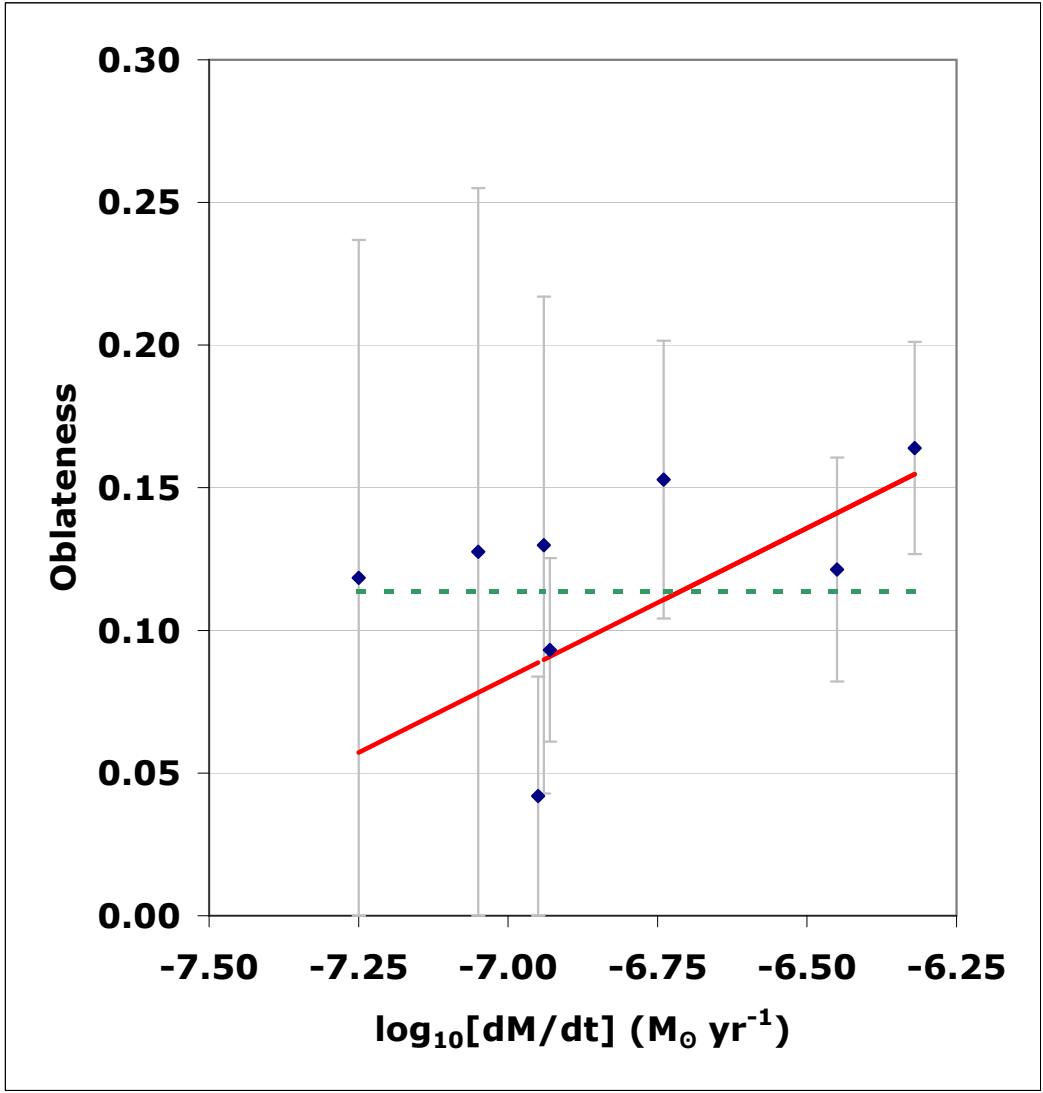}
\caption{\label{fig_oblateness_vs_mass_loss} Our values for oblateness $o_{\rm ab}=a/b-1$ versus mass loss rates from \citet{Claussen1987ApJS...65..385C}, with a solid fit line described in Equation \ref{eqn_mass_loss_rotation} with  $\chi^2_\nu=0.49$; a dotted flat line is also shown fit to the data, with intercept $o_{ab}=0.12$ and $\chi^2_\nu=0.99$.}
\end{figure}

We examined the stars for which we could derive oblate fits in \S \ref{sec_PA} against the mass-loss data of \citet{Claussen1987ApJS...65..385C}.  Other mass loss references were examined \citep{Schoeier2001A&A...368..969S,Groenewegen2002A&A...390..511G,Guandalini2006A&A...445.1069G} but insufficient (if any) overlap was found with our carbon sample.  A statistically weighted fit of oblateness $(o_{ab} = a/b-1)$ for those stars in our sample with robust, or even marginal, $o_{ab}$ data versus the mass loss rates of \citet{Claussen1987ApJS...65..385C} gives:
\begin{equation}\label{eqn_mass_loss_rotation}
o_{ab} = (0.11 \pm 0.06)\times\log_{10} \left( {dM \over dt}\right)  + (0.82 \pm 0.41)
\end{equation}
with a reduced chi-squared of $\chi^2_\nu=0.49$, and can be seen in Figure \ref{fig_oblateness_vs_mass_loss}.  This is {\it not} statistically significant --- a flat line of intercept 0.12 has $\chi^2_\nu=0.99$, so the level of improvement is not quite $\Delta \chi^2_\nu=1$ --- but is a tantalizing suggestion that oblateness is a significant factor when considering carbon star mass loss rates.

Spectral measurements of carbon star rotational velocities are challenging, given the complicated spectra of these objects that are characterized principally by a dense set of vibration-rotation transitions of CN and C$_2$ \citep{Barnbaum1995ApJ...450..862B}; direct measurements of equivalent widths are not possible without using spectral synthesis models.

Although \citet{Pereira2006A&A...452..571P} characterize rapid rotation among AGB stars as `very rare', this statement is made without substantiation.  Indeed, \citet{deMedeiros2004} point out that in their evolved star rotation catalog \citep{deMedeiros1999A&AS..139..433D}, `a growing list' of evolved stars present moderate to fast rotation: 200 of the 1540 non-carbon FGK evolved stars in their catalog have $v \sin i> 10$ km s$^{-1}$, though the investigation of \citet{Carlberg2011ApJ...732...39C} specifically directed at characterizing the frequency of K giant rapid rotation (again, $v \sin i> 10$ km s$^{-1}$) was 2.2\%.  It simply may be the case that, for carbon stars, rotation rates have not been comprehensively measured due to the difficulty of the task.  

For the oblateness explanation, the parent population could plausibly be evolved rapid rotators, or stars spun up from exoplanet ingestion \citep{Soker2000MNRAS.317..861S,Soker2001MNRAS.324..699S,Carlberg2009ApJ...700..832C,Carlberg2011AIPC.1331...33C}.  Indeed, \citet{Soker2001MNRAS.324..699S} argues that fairly small exoplanets ($M \sim 0.01-0.1 M_{\rm Jupiter}$) within 2AU could be sufficient to spin-up stars as they reach the AGB.  If correct, this hypothesis is significantly boosted by recent Kepler results that indicate $\sim 23\%$ of FGK stars have 1-3 $M_{\rm Earth}$ (0.003-0.01 $M_{\rm Jupiter}$) planets within 0.25AU \citep{Fressin2013arXiv1301.0842F}.  That result is further extended by indications that the number of planets as a function of orbital period ($\delta N_{\rm planets} / \delta \log P$) remains constant \citep{Dong2012arXiv1212.4853D} out to 500 days, which implies $\sim 45\%$ of FGK stars have 1-3 $M_{\rm Earth}$ planets out to 1.2AU (with additional planets yet to be found in the range 1.2-2.0AU).

\section{Conclusion}\label{sec_conclusion}

Angular size measurements are presented for a homogenous dataset of 41 carbon stars, the largest such single study to date.  Our results compare favorably with the previous studies of \citet{Dyck1996AJ....112..294D} and \citet{Bergeat2001A_A...369..178B,Bergeat2002A&A...390..987B,Bergeat2002A_A...390..967B}, and provide new calibrations of $T_{\rm EFF}$ versus Yamashita spectral type, and carbon star angular predictions.  As a group these stars have a median $T_{\rm EFF}=2800\pm270$K, though this temperature shows a decrease with increasing Yamashita spectral type. Notably, our data indicate that some carbon stars clearly present non-spherical solutions, due to either hot spots or rotational oblateness, and suggest that potentially all carbon stars will present such solutions when examined in sufficient detail.

Follow-up observations have the potential to separate the hypotheses of hot spots versus rotational oblateness, in two ways: First, repeating the current set stars with similar detail will present a long time baseline ($\sim$ years) between observations, and presumably any putative hot spots will migrate during that interval.  If each carbon star presents repeatable degrees of oblateness and position angles upon the sky, the notion of hot spots due to stochastically varying convection seems less likely.  Second, increasing the sample size, particularly as it can populate Figure \ref{fig_oblateness_vs_mass_loss} to the point of statistically significance in the $o_{ab}$ versus mass loss relationship, would increase the confidence in or motivate discarding the rotation hypothesis.

Unfortunately, continuing this investigation with PTI is not possible, since the facility was decommissioned in 2009.  However, newer interferometric imaging facilities such as the CHARA Array and VLTI, particularly with their near- to mid-IR multi-baseline instruments such as MIRC, PIONIER, and (in the future) MATISSE and GRAVITY, have the potential to carry on this line of inquiry.
If non-spherical surface morphology is indeed a general characteristic of carbon stars, the nature of that morphology as it relates to the parameters of the main sequence parent population, the overall angular momentum history (including possible exoplanet ingestion), and structure of evolved stars atmospheres becomes quite interesting to consider.

\acknowledgements

{\it Acknowledgements.} We thank the staff of the Palomar Observatory --- in particular, Kevin Rykoski --- for support during our observations and the PTI Collaboration for a generous allotment of time that made this research possible.  We have made extensive use of the SIMBAD database and the VizieR catalogue access tool, operated by the CDS in Strasbourg, France \citep{Ochsenbein2000A&AS..143...23O}.  This research has made use of the AFOEV database, operated at CDS, France. This research has made use of NASA's Astrophysics Data System.  Funding for PTI was provided to the Jet Propulsion Laboratory under its TOPS (Towards Other Planetary Systems), ASEPS (Astronomical Studies of Extrasolar Planetary Systems), and Origins programs and from the JPL Director's Discretionary Fund. Portions of this work were performed at the Jet Propulsion Laboratory, California Institute of Technology under contract with the National Aeronautics and Space Administration.

GvB would like to thank H. M. Dyck for the original inspiration and direction to do interferometric observations of both carbon stars and rotationally oblate stars, and to thank R. Canterna for generous support during the (very) early phases of this work.  Helpful input for this article resulted from discussions with and feedback from Travis Barman, Kaspar von Braun, Jay Farihi, Terry Jones, Phil Massey, and Greg Sloan.  Portions of this research were conducted while GvB was in residence at the European Southern Observatory; funding for this research for GvB has been generously provided in part by Lowell Observatory.  This material is based upon work supported by the National Science Foundation under Grant No. AST-1212203. CP, BA, and JH were supported by the Projects P23006 of the Austrian Science Fund (FWF).  The research leading to these results has received funding from the European Community’s Seventh Framework Programme under Grant Agreement 226604.


\bibliography{journal-references}


\appendix

\begin{figure}
\includegraphics[scale=1,angle=0]{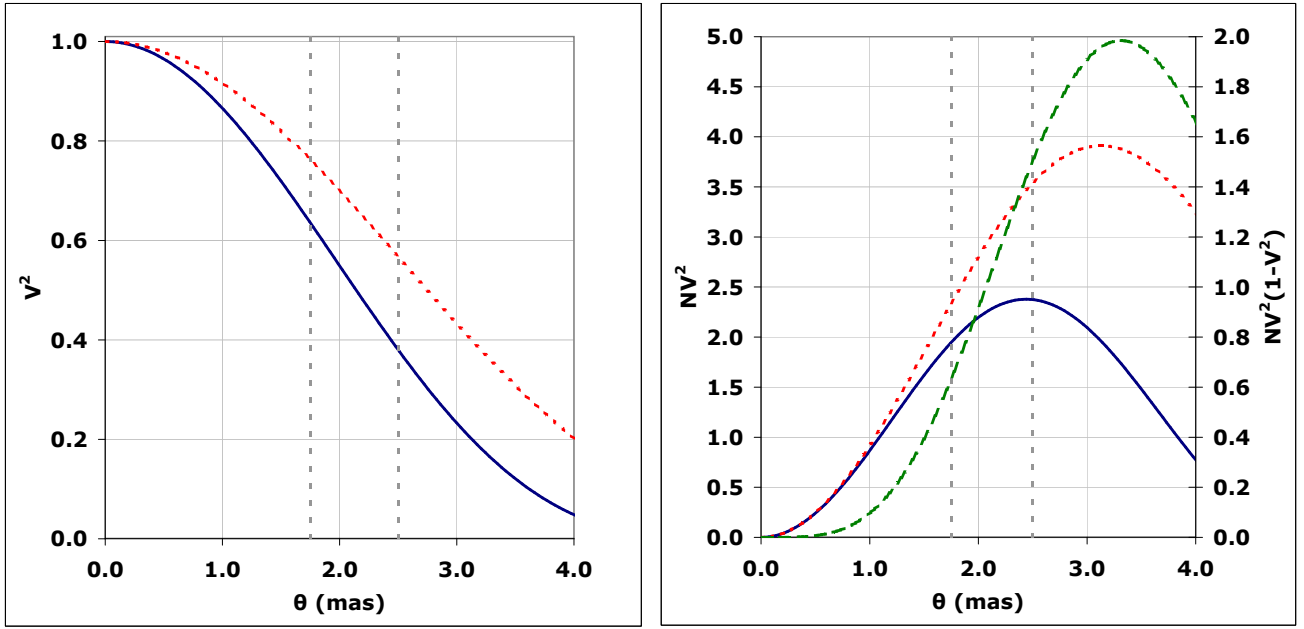}
\caption{\label{fig_NV2_vs_theta} (Left panel) Visibility plots for the 109-m NS PTI baseline (blue solid line) and 85-m NW baseline (red dotted line), at $2.2\mu$m, as a function of angular size $\theta$.  Grey dotted vertical lines separate objects into the `small', `intermediate', and `large' angular size ranges as discussed in \S \ref{sec_departures_from_spherical}.  (Right panel, left axis) Relative signal-to-noise $NV^2$ for visibility points for the two baselines as a function of $\theta$ (arbitrary units).  (Right axis) Expected signal-to-noise for $\theta$ (green dashed line) as a function of $\theta$.}
\end{figure}

In \S \ref{sec_departures_from_spherical} we suggest that, for our carbon star sample, data quality improves directly with increasing $\theta$.  This improvement in data quality is illustrated qualitatively in Figure \ref{fig_NV2_vs_theta}.  In the left panel, notional $2.2\mu$m visibility curves for both the 109-m NS and 85-m NW PTI baselines are shown, with the three size ranges partitioned off with grey dotted vertical lines.  In the right panel, we illustrate the relative signal-to-noise for a fringe-tracking interferometer such as PTI, which scales as the flux and the visibility squared, $NV^2$ \citep{Colavita1999PASP..111..111C}.  For the flux $N$, since all of these stars are fairly similar in temperature (\S\ref{sec_TEFF}), we shall assume it simply scales with apparent angular size $\theta$, $N \propto \theta^2$.  As such, $NV^2$ increases rapidly with $\theta$ initially --- until the visibility has dropped sufficiently low that these SNR curves turn over and drop.

For measurement of angular sizes, however, one may consider as a primary influence on those measures not the quantity $V^2$, but the distance from unit visibility $(1-V^2)$.  Consider how a solution for the visibility curve --- and the resulting measure of $\theta$ --- is constrained by two visibility points of differing amplitudes but identical uncertainties.  The low-level point will robustly fix the curve in place, while the high-level point has a large family of visibility curve solutions that fit through it.  As such, we have plotted the proxy product of visibility point SNR $NV^2$, and angular size `goodness' $(1-V^2)$, averaged for both baselines, on the right-hand vertical axis of Figure \ref{fig_NV2_vs_theta}'s lower panel.  Using the 3 size regimes of \S \ref{sec_departures_from_spherical}, the angular size regimes are partitioned on the plot.  Using this proxy, we can see how angular sizes in the `large' range have roughly $\sim 2 \times$ the SNR of those in the `intermediate' range, and $\sim 4 \times$ the `small' range; the `intermediate' points beat the `small' points SNR by $>2\times$.

\begin{deluxetable}{lcccl}											
\tablecolumns{6}											
\tabletypesize{\scriptsize}											
\tablewidth{0pc}											
\tablecaption{Target photometry used in SED fits.\label{tab_SED_phot}}											
\tablehead{											
Star & System /  & Bandpass /  & Value &  \\											
 ID &  Wavelength &  Bandwidth & [mag/Jy]\tablenotemark{a}  & Reference											
}											
\startdata
HD225217 & Johnson & V & $8.25 \pm 0.25$ & \citet{Samus2009yCat....102025S} \\
HD225217 & Johnson & J & $3.71 \pm 0.10$ & \citet{1981PASJ...33..373N} \\
HD225217 & Johnson & H & $2.78 \pm 0.10$ & \citet{1981PASJ...33..373N} \\
HD225217 & Johnson & K & $2.33 \pm 0.08$ & \citet{1969NASSP3047.....N} \\
HD225217 & Johnson & K & $2.40 \pm 0.10$ & \citet{1976A_A....52..227B} \\
HD225217 & Johnson & K & $2.49 \pm 0.10$ & \citet{1980A_A....87..139B} \\
HD225217 & Johnson & K & $2.34 \pm 0.10$ & \citet{1981PASJ...33..373N} \\
HD225217 & 4900 & 712 & $18.70 \pm 6.40$ & \citet{Smith2004ApJS..154..673S} \\
HD225217 & 12000 & 6384 & $14.30 \pm 35.00$ & \citet{Smith2004ApJS..154..673S} \\
HD2342 & Vilnius & V & $7.96 \pm 0.75$ & \citet{Straizys1989BICDS..37..179S} \\
HD2342 & Vilnius & V & $7.98 \pm 0.75$ & \citet{Straizys1989BICDS..37..179S} \\
HD2342 & Johnson & V & $10.85 \pm 0.95$ & \citet{Samus2009yCat....102025S} \\
HD2342 & Vilnius & S & $6.16 \pm 0.50$ & \citet{Straizys1989BICDS..37..179S} \\
HD2342 & Vilnius & S & $6.17 \pm 0.50$ & \citet{Straizys1989BICDS..37..179S} \\
HD2342 & Johnson & J & $3.43 \pm 0.10$ & \citet{1976A_A....52..227B} \\
HD2342 & Johnson & J & $3.40 \pm 0.10$ & \citet{1980A_A....87..139B} \\
HD2342 & Johnson & H & $2.20 \pm 0.10$ & \citet{1976A_A....52..227B} \\
HD2342 & Johnson & H & $2.21 \pm 0.10$ & \citet{1980A_A....87..139B} \\
HD2342 & Johnson & K & $1.64 \pm 0.04$ & \citet{Neugebauer1969tmss.book.....N} \\
HD2342 & Johnson & K & $1.56 \pm 0.10$ & \citet{1976A_A....52..227B} \\
HD2342 & Johnson & K & $1.61 \pm 0.10$ & \citet{1980A_A....87..139B} \\
HD2342 & 4900 & 712 & $41.40 \pm 5.60$ & \citet{Smith2004ApJS..154..673S} \\
HD2342 & 12000 & 6384 & $20.50 \pm 19.20$ & \citet{Smith2004ApJS..154..673S} \\
HD19881 & Vilnius & Z & $10.24 \pm 0.20$ & \citet{Straizys1989BICDS..37..179S} \\
HD19881 & Vilnius & V & $9.41 \pm 0.20$ & \citet{Straizys1989BICDS..37..179S} \\
HD19881 & Vilnius & S & $7.55 \pm 0.20$ & \citet{Straizys1989BICDS..37..179S} \\
HD19881 & Cousins & Ic & $6.33 \pm 0.10$ & \citet{Droege2007yCat.2271....0D} \\
HD19881 & Johnson & J & $3.97 \pm 0.10$ & \citet{1981PASJ...33..373N} \\
HD19881 & Johnson & H & $3.01 \pm 0.10$ & \citet{1981PASJ...33..373N} \\
HD19881 & Johnson & K & $2.48 \pm 0.07$ & \citet{Neugebauer1969tmss.book.....N} \\
HD19881 & Johnson & K & $2.52 \pm 0.10$ & \citet{1981PASJ...33..373N} \\
HD19881 & 4900 & 712 & $33.70 \pm 7.00$ & \citet{Smith2004ApJS..154..673S} \\
HD19881 & 12000 & 6384 & $6.00 \pm 384.00$ & \citet{Smith2004ApJS..154..673S} \\
IRC+40067 & Johnson & V & $12.1 \pm 0.30$ & \citet{Samus2009yCat....102025S} \\
IRC+40067 & Cousins & Ic & $6.42 \pm 0.15$ & \citet{Droege2007yCat.2271....0D} \\
IRC+40067 & Johnson & J & $3.86 \pm 0.10$ & \citet{1981PASJ...33..373N} \\
IRC+40067 & Johnson & H & $2.89 \pm 0.10$ & \citet{1981PASJ...33..373N} \\
IRC+40067 & Johnson & K & $2.38 \pm 0.12$ & \citet{Neugebauer1969tmss.book.....N} \\
IRC+40067 & Johnson & K & $2.17 \pm 0.10$ & \citet{1981PASJ...33..373N} \\
IRC+40067 & 4900 & 712 & $31.90 \pm 10.30$ & \citet{Smith2004ApJS..154..673S} \\
HD30443 & Vilnius & Z & $9.68 \pm 0.20$ & \citet{Straizys1989BICDS..37..179S} \\
HD30443 & Vilnius & Z & $9.63 \pm 0.20$ & \citet{Straizys1989BICDS..37..179S} \\
HD30443 & Vilnius & V & $8.97 \pm 0.20$ & \citet{Straizys1989BICDS..37..179S} \\
HD30443 & Vilnius & V & $8.97 \pm 0.20$ & \citet{Straizys1989BICDS..37..179S} \\
HD30443 & Vilnius & S & $7.52 \pm 0.20$ & \citet{Straizys1989BICDS..37..179S} \\
HD30443 & Vilnius & S & $7.49 \pm 0.20$ & \citet{Straizys1989BICDS..37..179S} \\
HD30443 & Johnson & H & $3.62 \pm 0.10$ & \citet{1997A_A...319..244U} \\
HD30443 & Johnson & K & $3.34 \pm 0.10$ & \citet{1997A_A...319..244U} \\
HD30443 & 4900 & 712 & $9.80 \pm 5.60$ & \citet{Smith2004ApJS..154..673S} \\
HD30443 & 12000 & 6384 & $6.40 \pm 24.40$ & \citet{Smith2004ApJS..154..673S} \\
HD280188 & Vilnius & V & $9.04 \pm 0.20$ & \citet{Straizys1989BICDS..37..179S} \\
HD280188 & Vilnius & S & $7.09 \pm 0.20$ & \citet{Straizys1989BICDS..37..179S} \\
HD280188 & Johnson & V & $12.1 \pm 0.90$ & \citet{Samus2009yCat....102025S} \\
HD280188 & 2Mass & J & $3.10 \pm 0.18$ & \citet{Cutri2003tmc..book.....C} \\
HD280188 & 2Mass & H & $1.82 \pm 0.16$ & \citet{Cutri2003tmc..book.....C} \\
HD280188 & 2Mass & Ks & $1.40 \pm 0.17$ & \citet{Cutri2003tmc..book.....C} \\
HD280188 & Johnson & K & $1.57 \pm 0.10$ & \citet{1969NASSP3047.....N} \\
HD280188 & 4900 & 712 & $32.20 \pm 6.70$ & \citet{Smith2004ApJS..154..673S} \\
HD280188 & 12000 & 6384 & $-36.10 \pm 23.10$ & \citet{Smith2004ApJS..154..673S} \\
HD33016 & Vilnius & V & $9.16 \pm 0.90$ & \citet{Straizys1989BICDS..37..179S} \\
HD33016 & Johnson & V & $8.85 \pm 0.35$ & \citet{Samus2009yCat....102025S} \\
HD33016 & Vilnius & S & $7.29 \pm 0.90$ & \citet{Straizys1989BICDS..37..179S} \\
HD33016 & Cousins & Ic & $6.07 \pm 0.07$ & \citet{Droege2007yCat.2271....0D} \\
HD33016 & Johnson & J & $3.84 \pm 0.10$ & \citet{1981PASJ...33..373N} \\
HD33016 & 2Mass & J & $3.93 \pm 0.22$ & \citet{Cutri2003tmc..book.....C} \\
HD33016 & Johnson & H & $2.95 \pm 0.10$ & \citet{1981PASJ...33..373N} \\
HD33016 & 2Mass & H & $2.84 \pm 0.20$ & \citet{Cutri2003tmc..book.....C} \\
HD33016 & 2Mass & Ks & $2.16 \pm 0.28$ & \citet{Cutri2003tmc..book.....C} \\
HD33016 & Johnson & K & $2.29 \pm 0.10$ & \citet{1969NASSP3047.....N} \\
HD33016 & Johnson & K & $2.26 \pm 0.10$ & \citet{1981PASJ...33..373N} \\
HD33016 & Johnson & K & $2.31 \pm 0.10$ & \citet{1989ApJ...336..924J} \\
HD33016 & 4900 & 712 & $15.80 \pm 6.80$ & \citet{Smith2004ApJS..154..673S} \\
HD33016 & 12000 & 6384 & $1.40 \pm 22.50$ & \citet{Smith2004ApJS..154..673S} \\
HD34467 & Vilnius & V & $9.41 \pm 0.20$ & \citet{Straizys1989BICDS..37..179S} \\
HD34467 & Vilnius & S & $7.60 \pm 0.20$ & \citet{Straizys1989BICDS..37..179S} \\
HD34467 & Johnson & V & $12.5 \pm 0.50$ & \citet{Samus2009yCat....102025S} \\
HD34467 & Cousins & Ic & $6.43 \pm 0.26$ & \citet{Droege2007yCat.2271....0D} \\
HD34467 & 2Mass & J & $4.38 \pm 0.18$ & \citet{Cutri2003tmc..book.....C} \\
HD34467 & Johnson & H & $3.07 \pm 0.10$ & \citet{1981PASJ...33..373N} \\
HD34467 & Johnson & K & $2.63 \pm 0.10$ & \citet{1989ApJ...336..924J} \\
HIP25004 & Cousins & Ic & $7.07 \pm 0.15$ & \citet{Droege2007yCat.2271....0D} \\
HIP25004 & 2Mass & J & $4.30 \pm 0.32$ & \citet{Cutri2003tmc..book.....C} \\
HIP25004 & 1250 & 310 & $32.10 \pm 7.80$ & \citet{Smith2004ApJS..154..673S} \\
HIP25004 & 2Mass & H & $2.96 \pm 0.22$ & \citet{Cutri2003tmc..book.....C} \\
HIP25004 & 2Mass & Ks & $2.19 \pm 0.27$ & \citet{Cutri2003tmc..book.....C} \\
HIP25004 & Johnson & K & $2.26 \pm 0.06$ & \citet{1969NASSP3047.....N} \\
HIP25004 & 2200 & 361 & $83.60 \pm 9.30$ & \citet{Smith2004ApJS..154..673S} \\
HIP25004 & 4900 & 712 & $21.80 \pm 6.40$ & \citet{Smith2004ApJS..154..673S} \\
HIP25004 & 12000 & 6384 & $107.20 \pm 47.40$ & \citet{Smith2004ApJS..154..673S} \\
HD38218 & Johnson & V & $11.8 \pm 0.70$ & \citet{Samus2009yCat....102025S} \\
HD38218 & Johnson & J & $3.50 \pm 0.10$ & \citet{1990ApJS...73..841A} \\
HD38218 & Johnson & H & $2.12 \pm 0.10$ & \citet{1980A_A....87..139B} \\
HD38218 & Johnson & H & $2.22 \pm 0.10$ & \citet{1981PASJ...33..373N} \\
HD38218 & Johnson & H & $2.34 \pm 0.10$ & \citet{1990ApJS...73..841A} \\
HD38218 & Johnson & K & $1.66 \pm 0.10$ & \citet{1989ApJ...336..924J} \\
HD38218 & Johnson & K & $1.72 \pm 0.10$ & \citet{1990ApJS...73..841A} \\
HD38218 & Johnson & K & $1.76 \pm 0.10$ & \citet{1990ApJS...73..841A} \\
HD38218 & 4900 & 712 & $36.20 \pm 6.10$ & \citet{Smith2004ApJS..154..673S} \\
HD38218 & 12000 & 6384 & $14.30 \pm 42.90$ & \citet{Smith2004ApJS..154..673S} \\
HD247224 & Vilnius & S & $7.85 \pm 0.20$ & \citet{Straizys1989BICDS..37..179S} \\
HD247224 & Johnson & V & $12.75 \pm 0.95$ & \citet{Samus2009yCat....102025S} \\
HD247224 & Cousins & Ic & $7.20 \pm 0.18$ & \citet{Droege2007yCat.2271....0D} \\
HD247224 & Johnson & J & $5.00 \pm 0.10$ & \citet{1993A_AS...99...31G} \\
HD247224 & Johnson & H & $3.81 \pm 0.10$ & \citet{1993A_AS...99...31G} \\
HD247224 & Johnson & K & $3.39 \pm 0.10$ & \citet{1993A_AS...99...31G} \\
HD247224 & 3800 & 760 & $2.73 \pm 0.05$ & \citet{1993A_AS...99...31G} \\
HD247224 & 4900 & 712 & $12.90 \pm 5.50$ & \citet{Smith2004ApJS..154..673S} \\
HD247224 & 12000 & 6384 & $-3.50 \pm 25.00$ & \citet{Smith2004ApJS..154..673S} \\
HD38572 & Johnson & V & $11.8 \pm 0.4$ & \citet{Samus2009yCat....102025S} \\
HD38572 & Vilnius & S & $6.92 \pm 0.05$ & \citet{Straizys1989BICDS..37..179S} \\
HD38572 & 1000 & 200 & $4.04 \pm 0.05$ & \citet{1981PASJ...33..373N} \\
HD38572 & Johnson & J & $3.51 \pm 0.10$ & \citet{1990ApJS...73..841A} \\
HD38572 & Johnson & H & $2.44 \pm 0.10$ & \citet{1990ApJS...73..841A} \\
HD38572 & Johnson & K & $2.14 \pm 0.10$ & \citet{1989ApJ...336..924J} \\
HD38572 & Johnson & K & $2.08 \pm 0.10$ & \citet{1990ApJS...73..841A} \\
HD38572 & 3120 & 624 & $2.00 \pm 0.05$ & \citet{1981PASJ...33..373N} \\
HD38572 & 4900 & 712 & $30.90 \pm 5.60$ & \citet{Smith2004ApJS..154..673S} \\
HD38572 & 12000 & 6384 & $14.70 \pm 28.40$ & \citet{Smith2004ApJS..154..673S} \\
HD38521 & Johnson & V & $11.55 \pm 0.25$ & \citet{Samus2009yCat....102025S} \\
HD38521 & Cousins & Ic & $7.04 \pm 0.19$ & \citet{Droege2007yCat.2271....0D} \\
HD38521 & 2Mass & J & $5.27 \pm 0.21$ & \citet{Cutri2003tmc..book.....C} \\
HD38521 & 2Mass & H & $3.89 \pm 0.22$ & \citet{Cutri2003tmc..book.....C} \\
HD38521 & 2Mass & Ks & $3.09 \pm 0.30$ & \citet{Cutri2003tmc..book.....C} \\
HD38521 & Johnson & K & $2.88 \pm 0.11$ & \citet{Neugebauer1969tmss.book.....N} \\
HD38521 & 4900 & 712 & $6.70 \pm 5.70$ & \citet{Smith2004ApJS..154..673S} \\
HD38521 & 12000 & 6384 & $5.60 \pm 20.50$ & \citet{Smith2004ApJS..154..673S} \\
HIP29896 & Johnson & V & $10.25 \pm 0.75$ & \citet{Samus2009yCat....102025S} \\
HIP29896 & Johnson & J & $4.37 \pm 0.10$ & \citet{1990A_A...227...82E} \\
HIP29896 & Johnson & H & $2.96 \pm 0.10$ & \citet{1981ApJ...249..481C} \\
HIP29896 & Johnson & H & $2.99 \pm 0.10$ & \citet{1981PASJ...33..373N} \\
HIP29896 & Johnson & H & $3.08 \pm 0.10$ & \citet{1990A_A...227...82E} \\
HIP29896 & Johnson & K & $2.24 \pm 0.10$ & \citet{1981ApJ...249..481C} \\
HIP29896 & Johnson & K & $2.28 \pm 0.10$ & \citet{1990A_A...227...82E} \\
HIP29896 & 4900 & 712 & $17.00 \pm 5.90$ & \citet{Smith2004ApJS..154..673S} \\
HIP29896 & 12000 & 6384 & $0.40 \pm 24.30$ & \citet{Smith2004ApJS..154..673S} \\
HD45087 & Johnson & V & $15.4 \pm 0.6$ & \citet{Samus2009yCat....102025S} \\
HD45087 & Johnson & J & $4.60 \pm 0.10$ & \citet{1990A_A...227...82E} \\
HD45087 & Johnson & H & $3.37 \pm 0.10$ & \citet{1990A_A...227...82E} \\
HD45087 & Johnson & K & $2.55 \pm 0.10$ & \citet{1969NASSP3047.....N} \\
HD45087 & Johnson & K & $2.60 \pm 0.10$ & \citet{1990A_A...227...82E} \\
HD45087 & 4900 & 712 & $27.00 \pm 14.60$ & \citet{Smith2004ApJS..154..673S} \\
HD45087 & 12000 & 6384 & $-0.60 \pm 25.40$ & \citet{Smith2004ApJS..154..673S} \\
HIP31349 & Johnson & V & $11.35 \pm 0.45$ & \citet{Samus2009yCat....102025S} \\
HIP31349 & Johnson & J & $3.36 \pm 0.05$ & \citet{1990A_A...227...82E} \\
HIP31349 & 2Mass & J & $3.58 \pm 0.30$ & \citet{Cutri2003tmc..book.....C} \\
HIP31349 & Johnson & H & $2.07 \pm 0.05$ & \citet{1990A_A...227...82E} \\
HIP31349 & 2Mass & H & $2.03 \pm 0.25$ & \citet{Cutri2003tmc..book.....C} \\
HIP31349 & 2Mass & Ks & $1.54 \pm 0.24$ & \citet{Cutri2003tmc..book.....C} \\
HIP31349 & Johnson & K & $1.45 \pm 0.05$ & \citet{1969NASSP3047.....N} \\
HIP31349 & Johnson & K & $1.46 \pm 0.05$ & \citet{1990A_A...227...82E} \\
HIP31349 & 3400 & 680 & $0.88 \pm 0.05$ & \citet{1990A_A...227...82E} \\
HIP31349 & 4900 & 712 & $55.40 \pm 9.50$ & \citet{Smith2004ApJS..154..673S} \\
HIP31349 & Johnson & M & $1.10 \pm 0.10$ & \citet{1990A_A...227...82E} \\
HIP31349 & 12000 & 6384 & $51.70 \pm 25.90$ & \citet{Smith2004ApJS..154..673S} \\
HD46321 & Johnson & V & $12.45 \pm 0.65$ & \citet{Samus2009yCat....102025S} \\
HD46321 & 2Mass & J & $4.70 \pm 0.16$ & \citet{Cutri2003tmc..book.....C} \\
HD46321 & 2Mass & H & $3.57 \pm 0.21$ & \citet{Cutri2003tmc..book.....C} \\
HD46321 & 2Mass & Ks & $2.96 \pm 0.26$ & \citet{Cutri2003tmc..book.....C} \\
HD46321 & Johnson & K & $2.94 \pm 0.08$ & \citet{Knapp2003A_A...403..993K} \\
HD46321 & 2200 & 361 & $45.10 \pm 5.70$ & \citet{Smith2004ApJS..154..673S} \\
HD46321 & 4900 & 712 & $13.00 \pm 5.50$ & \citet{Smith2004ApJS..154..673S} \\
HD46321 & 12000 & 6384 & $8.70 \pm 22.70$ & \citet{Smith2004ApJS..154..673S} \\
HD47883 & Johnson & V & $8.3 \pm 0.2$ & \citet{Samus2009yCat....102025S} \\
HD47883 & Vilnius & Z & $9.35 \pm 0.20$ & \citet{Straizys1989BICDS..37..179S} \\
HD47883 & Vilnius & V & $8.65 \pm 0.20$ & \citet{Straizys1989BICDS..37..179S} \\
HD47883 & Johnson & V & $8.33 \pm 0.19$ & \citet{Watson2006SASS...25...47W} \\
HD47883 & Vilnius & S & $6.94 \pm 0.20$ & \citet{Straizys1989BICDS..37..179S} \\
HD47883 & Johnson & J & $4.14 \pm 0.10$ & \citet{1980A_A....87..139B} \\
HD47883 & 2Mass & H & $3.12 \pm 0.24$ & \citet{Cutri2003tmc..book.....C} \\
HD47883 & Johnson & K & $2.69 \pm 0.10$ & \citet{1969NASSP3047.....N} \\
HD47883 & 4900 & 712 & $11.00 \pm 6.30$ & \citet{Smith2004ApJS..154..673S} \\
HD47883 & 12000 & 6384 & $1.90 \pm 28.50$ & \citet{Smith2004ApJS..154..673S} \\
HD48664 & Johnson & V & $13 \pm 0.5$ & \citet{Samus2009yCat....102025S} \\
HD48664 & Cousins & Ic & $6.92 \pm 0.24$ & \citet{Droege2007yCat.2271....0D} \\
HD48664 & Johnson & J & $4.43 \pm 0.10$ & \citet{1981PASJ...33..373N} \\
HD48664 & Johnson & H & $3.28 \pm 0.10$ & \citet{1981PASJ...33..373N} \\
HD48664 & Johnson & K & $2.55 \pm 0.10$ & \citet{1981PASJ...33..373N} \\
HD48664 & 4900 & 712 & $20.40 \pm 7.40$ & \citet{Smith2004ApJS..154..673S} \\
HD48664 & 12000 & 6384 & $1.40 \pm 25.30$ & \citet{Smith2004ApJS..154..673S} \\
HD51620 & Vilnius & V & $7.03 \pm 0.50$ & \citet{Straizys1989BICDS..37..179S} \\
HD51620 & Vilnius & V & $7.05 \pm 0.50$ & \citet{Straizys1989BICDS..37..179S} \\
HD51620 & Johnson & V & $11.1 \pm 0.8$ & \citet{Samus2009yCat....102025S} \\
HD51620 & Vilnius & S & $5.41 \pm 0.50$ & \citet{Straizys1989BICDS..37..179S} \\
HD51620 & Vilnius & S & $5.43 \pm 0.50$ & \citet{Straizys1989BICDS..37..179S} \\
HD51620 & Johnson & J & $3.06 \pm 0.10$ & \citet{1990A_A...227...82E} \\
HD51620 & Johnson & H & $1.96 \pm 0.10$ & \citet{1990A_A...227...82E} \\
HD51620 & Johnson & K & $1.43 \pm 0.10$ & \citet{1990A_A...227...82E} \\
HD51620 & 4900 & 712 & $55.20 \pm 6.90$ & \citet{Smith2004ApJS..154..673S} \\
HD51620 & 12000 & 6384 & $19.20 \pm 23.20$ & \citet{Smith2004ApJS..154..673S} \\
HD54361 & Vilnius & V & $7.00 \pm 0.20$ & \citet{Straizys1989BICDS..37..179S} \\
HD54361 & Johnson & V & $9.2 \pm 0.5$ & \citet{Samus2009yCat....102025S} \\
HD54361 & Vilnius & S & $5.34 \pm 0.20$ & \citet{Straizys1989BICDS..37..179S} \\
HD54361 & Johnson & J & $2.68 \pm 0.10$ & \citet{1993ApJS...87..305O} \\
HD54361 & Johnson & J & $2.59 \pm 0.10$ & \citet{1996A_AS..118..397K} \\
HD54361 & Johnson & H & $1.53 \pm 0.10$ & \citet{1993ApJS...87..305O} \\
HD54361 & Johnson & H & $1.52 \pm 0.10$ & \citet{1996A_AS..118..397K} \\
HD54361 & Johnson & K & $1.02 \pm 0.10$ & \citet{1993ApJS...87..305O} \\
HD54361 & Johnson & K & $1.01 \pm 0.10$ & \citet{1996A_AS..118..397K} \\
IRC+10158 & Cousins & Ic & $7.08 \pm 0.19$ & \citet{Droege2007yCat.2271....0D} \\
IRC+10158 & Johnson & J & $4.19 \pm 0.10$ & \citet{Claussen1987ApJS...65..385C} \\
IRC+10158 & Johnson & H & $3.22 \pm 0.10$ & \citet{Claussen1987ApJS...65..385C} \\
IRC+10158 & Johnson & K & $2.57 \pm 0.10$ & \citet{Neugebauer1969tmss.book.....N} \\
HD57160 & Vilnius & Z & $9.34 \pm 0.20$ & \citet{Straizys1989BICDS..37..179S} \\
HD57160 & Vilnius & V & $8.62 \pm 0.20$ & \citet{Straizys1989BICDS..37..179S} \\
HD57160 & Johnson & V & $11.8 \pm 0.3$ & \citet{Samus2009yCat....102025S} \\
HD57160 & Vilnius & S & $7.03 \pm 0.20$ & \citet{Straizys1989BICDS..37..179S} \\
HD57160 & Johnson & J & $4.28 \pm 0.10$ & \citet{1981PASJ...33..373N} \\
HD57160 & Johnson & H & $3.32 \pm 0.10$ & \citet{1981PASJ...33..373N} \\
HD57160 & Johnson & K & $2.95 \pm 0.10$ & \citet{1969NASSP3047.....N} \\
HD57160 & Johnson & K & $2.75 \pm 0.10$ & \citet{1981PASJ...33..373N} \\
HD57160 & 4900 & 712 & $13.10 \pm 6.80$ & \citet{Smith2004ApJS..154..673S} \\
HD57160 & 12000 & 6384 & $35.60 \pm 30.30$ & \citet{Smith2004ApJS..154..673S} \\
HD59643 & Vilnius & U & $13.19 \pm 0.75$ & \citet{Straizys1989BICDS..37..179S} \\
HD59643 & Vilnius & U & $13.85 \pm 0.75$ & \citet{Straizys1989BICDS..37..179S} \\
HD59643 & Vilnius & P & $12.83 \pm 0.75$ & \citet{Straizys1989BICDS..37..179S} \\
HD59643 & Vilnius & P & $13.39 \pm 0.75$ & \citet{Straizys1989BICDS..37..179S} \\
HD59643 & Vilnius & X & $12.15 \pm 0.50$ & \citet{Straizys1989BICDS..37..179S} \\
HD59643 & Vilnius & X & $12.19 \pm 0.50$ & \citet{Straizys1989BICDS..37..179S} \\
HD59643 & Vilnius & Y & $9.75 \pm 0.10$ & \citet{Straizys1989BICDS..37..179S} \\
HD59643 & Vilnius & Y & $9.65 \pm 0.10$ & \citet{Straizys1989BICDS..37..179S} \\
HD59643 & Vilnius & Z & $8.90 \pm 0.10$ & \citet{Straizys1989BICDS..37..179S} \\
HD59643 & Vilnius & Z & $8.77 \pm 0.10$ & \citet{Straizys1989BICDS..37..179S} \\
HD59643 & Vilnius & V & $8.19 \pm 0.10$ & \citet{Straizys1989BICDS..37..179S} \\
HD59643 & Vilnius & V & $8.06 \pm 0.10$ & \citet{Straizys1989BICDS..37..179S} \\
HD59643 & Johnson & V & $7.70 \pm 0.30$ & \citet{Watson2006SASS...25...47W} \\
HIP36623 & Johnson & V & $7.70 \pm 0.30$ & \citet{Watson2006SASS...25...47W} \\
HD59643 & Vilnius & S & $6.91 \pm 0.10$ & \citet{Straizys1989BICDS..37..179S} \\
HD59643 & Vilnius & S & $6.84 \pm 0.10$ & \citet{Straizys1989BICDS..37..179S} \\
HD59643 & Johnson & J & $4.05 \pm 0.10$ & \citet{1981PASJ...33..373N} \\
HIP36623 & Johnson & J & $4.05 \pm 0.10$ & \citet{1981PASJ...33..373N} \\
HD59643 & Johnson & H & $3.33 \pm 0.10$ & \citet{1981PASJ...33..373N} \\
HIP36623 & Johnson & H & $3.33 \pm 0.10$ & \citet{1981PASJ...33..373N} \\
HD59643 & Johnson & K & $2.95 \pm 0.10$ & \citet{1981PASJ...33..373N} \\
HIP36623 & Johnson & K & $2.95 \pm 0.10$ & \citet{1981PASJ...33..373N} \\
HD59643 & 3120 & 624 & $2.87 \pm 0.10$ & \citet{1981PASJ...33..373N} \\
HIP36623 & 3120 & 624 & $2.87 \pm 0.05$ & \citet{1981PASJ...33..373N} \\
HIP36623 & Johnson & L & $2.93 \pm 0.05$ & \citet{1981PASJ...33..373N} \\
HD59643 & 3700 & 740 & $2.93 \pm 0.10$ & \citet{1981PASJ...33..373N} \\
HD60826 & Vilnius & V & $9.42 \pm 0.20$ & \citet{Straizys1989BICDS..37..179S} \\
HD60826 & Vilnius & S & $7.60 \pm 0.20$ & \citet{Straizys1989BICDS..37..179S} \\
HD60826 & Cousins & Rc & $7.49 \pm 0.20$ & \citet{Stritzinger2005PASP..117..810S} \\
HD60826 & Cousins & Rc & $7.43 \pm 0.20$ & \citet{Stritzinger2005PASP..117..810S} \\
HD60826 & Cousins & Ic & $6.31 \pm 0.20$ & \citet{Stritzinger2005PASP..117..810S} \\
HD60826 & Cousins & Ic & $6.23 \pm 0.20$ & \citet{Stritzinger2005PASP..117..810S} \\
HD60826 & Cousins & Ic & $6.68 \pm 0.19$ & \citet{Droege2007yCat.2271....0D} \\
HD60826 & 2Mass & J & $5.16 \pm 0.27$ & \citet{Cutri2003tmc..book.....C} \\
HD60826 & 2Mass & H & $3.69 \pm 0.24$ & \citet{Cutri2003tmc..book.....C} \\
HD60826 & 2Mass & Ks & $3.09 \pm 0.26$ & \citet{Cutri2003tmc..book.....C} \\
HD60826 & Johnson & K & $2.83 \pm 0.08$ & \citet{1969NASSP3047.....N} \\
HD60826 & 4900 & 712 & $11.80 \pm 5.50$ & \citet{Smith2004ApJS..154..673S} \\
HD60826 & 12000 & 6384 & $7.40 \pm 18.60$ & \citet{Smith2004ApJS..154..673S} \\
HD60826 & Cousins & V & $8.98 \pm 0.20$ & \citet{Landolt1983AJ.....88..439L} \\
HD70072 & Vilnius & Z & $10.82 \pm 1.00$ & \citet{Straizys1989BICDS..37..179S} \\
HD70072 & Vilnius & V & $9.68 \pm 1.00$ & \citet{Straizys1989BICDS..37..179S} \\
HD70072 & Johnson & V & $13.6 \pm 1.4$ & \citet{Samus2009yCat....102025S} \\
HD70072 & Vilnius & S & $7.45 \pm 0.75$ & \citet{Straizys1989BICDS..37..179S} \\
HD70072 & Johnson & J & $4.06 \pm 0.10$ & \citet{1980A_A....87..139B} \\
HD70072 & Johnson & J & $4.14 \pm 0.10$ & \citet{1981PASJ...33..373N} \\
HD70072 & Johnson & J & $4.10 \pm 0.10$ & \citet{1990A_A...227...82E} \\
HD70072 & 2Mass & J & $4.61 \pm 0.24$ & \citet{Cutri2003tmc..book.....C} \\
HD70072 & 1250 & 310 & $45.80 \pm 6.60$ & \citet{Smith2004ApJS..154..673S} \\
HD70072 & Johnson & H & $2.93 \pm 0.10$ & \citet{1980A_A....87..139B} \\
HD70072 & Johnson & H & $3.15 \pm 0.10$ & \citet{1981PASJ...33..373N} \\
HD70072 & Johnson & H & $2.90 \pm 0.10$ & \citet{1990A_A...227...82E} \\
HD70072 & 2Mass & H & $3.18 \pm 0.23$ & \citet{Cutri2003tmc..book.....C} \\
HD70072 & 2Mass & Ks & $2.46 \pm 0.23$ & \citet{Cutri2003tmc..book.....C} \\
HD70072 & Johnson & K & $2.14 \pm 0.10$ & \citet{1969NASSP3047.....N} \\
HD70072 & Johnson & K & $2.32 \pm 0.10$ & \citet{1980A_A....87..139B} \\
HD70072 & Johnson & K & $2.35 \pm 0.10$ & \citet{1981PASJ...33..373N} \\
HD70072 & Johnson & K & $2.16 \pm 0.10$ & \citet{1990A_A...227...82E} \\
HD70072 & 2200 & 361 & $88.10 \pm 8.30$ & \citet{Smith2004ApJS..154..673S} \\
HD70072 & 4900 & 712 & $34.40 \pm 4.70$ & \citet{Smith2004ApJS..154..673S} \\
HD70072 & 12000 & 6384 & $20.20 \pm 20.30$ & \citet{Smith2004ApJS..154..673S} \\
HD144578 & Vilnius & Z & $8.88 \pm 0.20$ & \citet{Straizys1989BICDS..37..179S} \\
HD144578 & Vilnius & V & $8.07 \pm 0.20$ & \citet{Straizys1989BICDS..37..179S} \\
HD144578 & Johnson & V & $11.15 \pm 2.35$ & \citet{Samus2009yCat....102025S} \\
HD144578 & Vilnius & S & $6.69 \pm 0.20$ & \citet{Straizys1989BICDS..37..179S} \\
HD144578 & 2Mass & J & $5.20 \pm 0.33$ & \citet{Cutri2003tmc..book.....C} \\
HD144578 & 2Mass & H & $3.96 \pm 0.24$ & \citet{Cutri2003tmc..book.....C} \\
HD144578 & 2Mass & Ks & $3.44 \pm 0.47$ & \citet{Cutri2003tmc..book.....C} \\
HD144578 & 4900 & 712 & $8.20 \pm 4.90$ & \citet{Smith2004ApJS..154..673S} \\
HD144578 & 12000 & 6384 & $3.10 \pm 16.50$ & \citet{Smith2004ApJS..154..673S} \\
HD173291 & Vilnius & V & $8.81 \pm 1.00$ & \citet{Straizys1989BICDS..37..179S} \\
HD173291 & Vilnius & V & $8.47 \pm 1.00$ & \citet{Straizys1989BICDS..37..179S} \\
HD173291 & Vilnius & V & $8.50 \pm 1.00$ & \citet{Straizys1989BICDS..37..179S} \\
HD173291 & Johnson & V & $10.55 \pm 1.05$ & \citet{Samus2009yCat....102025S} \\
HD173291 & Vilnius & S & $7.18 \pm 1.00$ & \citet{Straizys1989BICDS..37..179S} \\
HD173291 & Vilnius & S & $6.40 \pm 1.00$ & \citet{Straizys1989BICDS..37..179S} \\
HD173291 & Vilnius & S & $6.44 \pm 1.00$ & \citet{Straizys1989BICDS..37..179S} \\
HD173291 & Johnson & J & $3.26 \pm 0.10$ & \citet{1981PASJ...33..373N} \\
HD173291 & Johnson & J & $3.23 \pm 0.10$ & \citet{1996A_AS..118..397K} \\
HD173291 & Johnson & H & $2.21 \pm 0.10$ & \citet{1981PASJ...33..373N} \\
HD173291 & Johnson & H & $2.15 \pm 0.10$ & \citet{1996A_AS..118..397K} \\
HD173291 & Johnson & K & $1.69 \pm 0.10$ & \citet{1969NASSP3047.....N} \\
HD173291 & Johnson & K & $1.64 \pm 0.10$ & \citet{1981PASJ...33..373N} \\
HD173291 & Johnson & K & $1.62 \pm 0.10$ & \citet{1996A_AS..118..397K} \\
HD173291 & 4900 & 712 & $41.10 \pm 5.30$ & \citet{Smith2004ApJS..154..673S} \\
HD173291 & 12000 & 6384 & $16.50 \pm 15.50$ & \citet{Smith2004ApJS..154..673S} \\
HIP92194 & Johnson & V & $11.45 \pm 1.05$ & \citet{Samus2009yCat....102025S} \\
HIP92194 & Cousins & Ic & $6.75 \pm 0.41$ & \citet{Droege2007yCat.2271....0D} \\
HIP92194 & Johnson & J & $4.12 \pm 0.10$ & \citet{1993A_AS...99...31G} \\
HIP92194 & Johnson & H & $2.79 \pm 0.10$ & \citet{1993A_AS...99...31G} \\
HIP92194 & Johnson & K & $1.99 \pm 0.07$ & \citet{Neugebauer1969tmss.book.....N} \\
HIP92194 & Johnson & K & $2.08 \pm 0.10$ & \citet{1993A_AS...99...31G} \\
HIP95024 & Cousins & Ic & $7.11 \pm 0.38$ & \citet{Droege2007yCat.2271....0D} \\
HIP95024 & Johnson & J & $3.88 \pm 0.10$ & \citet{1981PASJ...33..373N} \\
HIP95024 & Johnson & H & $2.95 \pm 0.10$ & \citet{1981PASJ...33..373N} \\
HIP95024 & Johnson & K & $2.20 \pm 0.04$ & \citet{Neugebauer1969tmss.book.....N} \\
HIP95024 & Johnson & K & $2.14 \pm 0.10$ & \citet{1981PASJ...33..373N} \\
HIP95024 & 3500 & 898 & $69.30 \pm 14.30$ & \citet{Smith2004ApJS..154..673S} \\
HIP95024 & 4900 & 712 & $44.30 \pm 7.50$ & \citet{Smith2004ApJS..154..673S} \\
HIP95024 & 12000 & 6384 & $20.40 \pm 16.90$ & \citet{Smith2004ApJS..154..673S} \\
HIP95777 & Johnson & V & $12.75 \pm 1.75$ & \citet{Samus2009yCat....102025S} \\
HIP95777 & Cousins & Ic & $5.80 \pm 1.00$ & \citet{Droege2007yCat.2271....0D} \\
HIP95777 & Johnson & J & $3.62 \pm 0.10$ & \citet{1981PASJ...33..373N} \\
HIP95777 & Johnson & J & $3.77 \pm 0.10$ & \citet{1994A_AS..106..397K} \\
HIP95777 & Johnson & H & $2.58 \pm 0.10$ & \citet{1981PASJ...33..373N} \\
HIP95777 & Johnson & H & $2.61 \pm 0.10$ & \citet{1994A_AS..106..397K} \\
HIP95777 & Johnson & K & $2.02 \pm 0.10$ & \citet{1969NASSP3047.....N} \\
HIP95777 & Johnson & K & $1.92 \pm 0.10$ & \citet{1994A_AS..106..397K} \\
HIP95777 & Johnson & K & $1.92 \pm 0.10$ & \citet{1994A_AS..106..397K} \\
HIP95777 & 12000 & 6384 & $38.30 \pm 30.90$ & \citet{Smith2004ApJS..154..673S} \\
HD186047 & Vilnius & Y & $9.85 \pm 0.20$ & \citet{Straizys1989BICDS..37..179S} \\
HD186047 & Vilnius & Y & $9.86 \pm 0.20$ & \citet{Straizys1989BICDS..37..179S} \\
HD186047 & Vilnius & Y & $9.88 \pm 0.20$ & \citet{Straizys1989BICDS..37..179S} \\
HD186047 & Vilnius & Z & $8.50 \pm 0.20$ & \citet{Straizys1989BICDS..37..179S} \\
HD186047 & Vilnius & Z & $8.50 \pm 0.20$ & \citet{Straizys1989BICDS..37..179S} \\
HD186047 & Vilnius & Z & $8.53 \pm 0.20$ & \citet{Straizys1989BICDS..37..179S} \\
HD186047 & Vilnius & V & $7.93 \pm 0.20$ & \citet{Straizys1989BICDS..37..179S} \\
HD186047 & Vilnius & V & $7.96 \pm 0.20$ & \citet{Straizys1989BICDS..37..179S} \\
HD186047 & Vilnius & V & $7.96 \pm 0.20$ & \citet{Straizys1989BICDS..37..179S} \\
HD186047 & Johnson & V & $11.35 \pm 0.55$ & \citet{Samus2009yCat....102025S} \\
HD186047 & Vilnius & S & $6.17 \pm 0.20$ & \citet{Straizys1989BICDS..37..179S} \\
HD186047 & Vilnius & S & $6.20 \pm 0.20$ & \citet{Straizys1989BICDS..37..179S} \\
HD186047 & Vilnius & S & $6.21 \pm 0.20$ & \citet{Straizys1989BICDS..37..179S} \\
HD186047 & Johnson & J & $3.43 \pm 0.10$ & \citet{1981PASJ...33..373N} \\
HD186047 & Johnson & J & $3.59 \pm 0.10$ & \citet{1994A_AS..106..397K} \\
HD186047 & Johnson & H & $2.43 \pm 0.10$ & \citet{1981PASJ...33..373N} \\
HD186047 & Johnson & H & $2.46 \pm 0.10$ & \citet{1994A_AS..106..397K} \\
HD186047 & Johnson & K & $1.93 \pm 0.10$ & \citet{1969NASSP3047.....N} \\
HD186047 & Johnson & K & $1.94 \pm 0.04$ & \citet{Neugebauer1969tmss.book.....N} \\
HD186047 & Johnson & K & $1.90 \pm 0.10$ & \citet{1981PASJ...33..373N} \\
HIP99336 & Johnson & V & $12.1 \pm 0.40$ & \citet{Samus2009yCat....102025S} \\
HIP99336 & Cousins & Ic & $7.17 \pm 0.10$ & \citet{Droege2007yCat.2271....0D} \\
HIP99336 & 2Mass & J & $5.02 \pm 0.30$ & \citet{Cutri2003tmc..book.....C} \\
HIP99336 & 2Mass & H & $3.68 \pm 0.21$ & \citet{Cutri2003tmc..book.....C} \\
HIP99336 & 2Mass & Ks & $2.82 \pm 0.39$ & \citet{Cutri2003tmc..book.....C} \\
HIP99336 & Johnson & K & $2.92 \pm 0.09$ & \citet{Neugebauer1969tmss.book.....N} \\
HIP99336 & 3500 & 898 & $44.80 \pm 29.30$ & \citet{Smith2004ApJS..154..673S} \\
HIP99336 & 4900 & 712 & $31.20 \pm 10.00$ & \citet{Smith2004ApJS..154..673S} \\
HIP99336 & 12000 & 6384 & $-102.20 \pm 149.50$ & \citet{Smith2004ApJS..154..673S} \\
HD191783 & Johnson & V & $9.25 \pm 0.75$ & \citet{Samus2009yCat....102025S} \\
HD191783 & Cousins & Ic & $6.48 \pm 0.07$ & \citet{Droege2007yCat.2271....0D} \\
HD191783 & Johnson & J & $4.45 \pm 0.10$ & \citet{1990ApJS...73..841A} \\
HD191783 & Johnson & H & $3.32 \pm 0.10$ & \citet{1990ApJS...73..841A} \\
HD191783 & Johnson & K & $2.75 \pm 0.10$ & \citet{1990ApJS...73..841A} \\
HIP102764 & Cousins & Ic & $6.22 \pm 0.11$ & \citet{Droege2007yCat.2271....0D} \\
HIP102764 & Johnson & J & $4.00 \pm 0.10$ & \citet{1981PASJ...33..373N} \\
HIP102764 & Johnson & H & $2.95 \pm 0.10$ & \citet{1981PASJ...33..373N} \\
HIP102764 & 2Mass & Ks & $2.58 \pm 0.33$ & \citet{Cutri2003tmc..book.....C} \\
HIP102764 & Johnson & K & $2.41 \pm 0.10$ & \citet{1981PASJ...33..373N} \\
HIP105539 & Johnson & V & $12.65 \pm 0.55$ & \citet{Samus2009yCat....102025S} \\
HIP105539 & Cousins & Ic & $6.52 \pm 0.24$ & \citet{Droege2007yCat.2271....0D} \\
HIP105539 & 2Mass & J & $4.79 \pm 0.28$ & \citet{Cutri2003tmc..book.....C} \\
HIP105539 & 2Mass & H & $3.39 \pm 0.29$ & \citet{Cutri2003tmc..book.....C} \\
HIP105539 & 2Mass & Ks & $2.68 \pm 0.36$ & \citet{Cutri2003tmc..book.....C} \\
HIP105539 & Johnson & K & $2.61 \pm 0.10$ & \citet{1969NASSP3047.....N} \\
HIP105539 & 4900 & 712 & $38.10 \pm 14.50$ & \citet{Smith2004ApJS..154..673S} \\
HIP105539 & 12000 & 6384 & $22.80 \pm 29.50$ & \citet{Smith2004ApJS..154..673S} \\
IRC+50399 & Johnson & V & $14.0 \pm 0.5$ & \citet{Samus2009yCat....102025S} \\
IRC+50399 & Johnson & J & $4.74 \pm 0.10$ & \citet{1990ApJS...73..841A} \\
IRC+50399 & Johnson & H & $3.59 \pm 0.10$ & \citet{1990ApJS...73..841A} \\
IRC+50399 & Johnson & K & $2.94 \pm 0.10$ & \citet{1969NASSP3047.....N} \\
IRC+50399 & Johnson & K & $3.06 \pm 0.10$ & \citet{1990ApJS...73..841A} \\
IRC+50399 & 4900 & 712 & $26.90 \pm 17.10$ & \citet{Smith2004ApJS..154..673S} \\
HD208512 & Vilnius & V & $9.44 \pm 0.20$ & \citet{Straizys1989BICDS..37..179S} \\
HD208512 & Vilnius & V & $9.46 \pm 0.20$ & \citet{Straizys1989BICDS..37..179S} \\
HD208512 & Vilnius & V & $9.47 \pm 0.20$ & \citet{Straizys1989BICDS..37..179S} \\
HD208512 & Johnson & V & $13.4 \pm 1.1$ & \citet{Samus2009yCat....102025S} \\
HD208512 & Vilnius & S & $7.21 \pm 0.20$ & \citet{Straizys1989BICDS..37..179S} \\
HD208512 & Vilnius & S & $7.23 \pm 0.20$ & \citet{Straizys1989BICDS..37..179S} \\
HD208512 & Vilnius & S & $7.25 \pm 0.20$ & \citet{Straizys1989BICDS..37..179S} \\
HD208512 & Cousins & Ic & $5.83 \pm 0.14$ & \citet{Droege2007yCat.2271....0D} \\
HD208512 & 2Mass & J & $4.11 \pm 0.26$ & \citet{Cutri2003tmc..book.....C} \\
HD208512 & 2Mass & H & $2.75 \pm 0.21$ & \citet{Cutri2003tmc..book.....C} \\
HD208512 & Johnson & K & $1.71 \pm 0.06$ & \citet{1969NASSP3047.....N} \\
HD208512 & 4900 & 712 & $42.90 \pm 11.60$ & \citet{Smith2004ApJS..154..673S} \\
HD208512 & 12000 & 6384 & $2.50 \pm 32.90$ & \citet{Smith2004ApJS..154..673S} \\
HD208526 & Vilnius & Y & $10.58 \pm 0.50$ & \citet{Straizys1989BICDS..37..179S} \\
HD208526 & Vilnius & Y & $10.51 \pm 0.50$ & \citet{Straizys1989BICDS..37..179S} \\
HD208526 & Vilnius & Y & $10.53 \pm 0.50$ & \citet{Straizys1989BICDS..37..179S} \\
HD208526 & Vilnius & Z & $9.09 \pm 0.50$ & \citet{Straizys1989BICDS..37..179S} \\
HD208526 & Vilnius & Z & $9.07 \pm 0.50$ & \citet{Straizys1989BICDS..37..179S} \\
HD208526 & Vilnius & Z & $9.08 \pm 0.50$ & \citet{Straizys1989BICDS..37..179S} \\
HD208526 & Vilnius & V & $8.51 \pm 0.50$ & \citet{Straizys1989BICDS..37..179S} \\
HD208526 & Vilnius & V & $8.53 \pm 0.50$ & \citet{Straizys1989BICDS..37..179S} \\
HD208526 & Vilnius & V & $8.54 \pm 0.50$ & \citet{Straizys1989BICDS..37..179S} \\
HD208526 & Johnson & V & $10.65 \pm 0.95$ & \citet{Samus2009yCat....102025S} \\
HD208526 & Johnson & V & $8.23 \pm 0.45$ & \citet{Kharchenko2007AN....328..889K} \\
HD208526 & Vilnius & S & $6.77 \pm 0.20$ & \citet{Straizys1989BICDS..37..179S} \\
HD208526 & Vilnius & S & $6.79 \pm 0.20$ & \citet{Straizys1989BICDS..37..179S} \\
HD208526 & Vilnius & S & $6.79 \pm 0.20$ & \citet{Straizys1989BICDS..37..179S} \\
HD208526 & 2Mass & J & $4.12 \pm 0.24$ & \citet{Cutri2003tmc..book.....C} \\
HD208526 & 2Mass & H & $2.97 \pm 0.19$ & \citet{Cutri2003tmc..book.....C} \\
HD208526 & 2Mass & Ks & $2.32 \pm 0.24$ & \citet{Cutri2003tmc..book.....C} \\
HD208526 & Johnson & K & $2.23 \pm 0.05$ & \citet{1969NASSP3047.....N} \\
HD208526 & 4900 & 712 & $19.60 \pm 5.50$ & \citet{Smith2004ApJS..154..673S} \\
HD208526 & 12000 & 6384 & $2.60 \pm 19.60$ & \citet{Smith2004ApJS..154..673S} \\
HIP113715 & Johnson & V & $10.5 \pm 0.9$ & \citet{Samus2009yCat....102025S} \\
HIP113715 & Cousins & Ic & $6.94 \pm 0.10$ & \citet{Droege2007yCat.2271....0D} \\
HIP113715 & Johnson & J & $4.83 \pm 0.10$ & \citet{1981PASJ...33..373N} \\
HIP113715 & Johnson & J & $4.83 \pm 0.10$ & \citet{1981PASJ...33..373N} \\
HIP113715 & Johnson & J & $4.83 \pm 0.10$ & \citet{1981PASJ...33..373N} \\
HIP113715 & Johnson & H & $3.61 \pm 0.10$ & \citet{1980A_A....87..139B} \\
HIP113715 & Johnson & H & $3.80 \pm 0.10$ & \citet{1981PASJ...33..373N} \\
HIP113715 & Johnson & H & $3.81 \pm 0.10$ & \citet{1981PASJ...33..373N} \\
HIP113715 & Johnson & K & $2.98 \pm 0.10$ & \citet{1969NASSP3047.....N} \\
HIP113715 & Johnson & K & $3.05 \pm 0.10$ & \citet{1980A_A....87..139B} \\
HIP113715 & Johnson & K & $3.06 \pm 0.10$ & \citet{1980A_A....87..139B} \\
HIP113715 & Johnson & K & $2.99 \pm 0.10$ & \citet{1981PASJ...33..373N} \\
HIP113715 & 4900 & 712 & $10.90 \pm 6.40$ & \citet{Smith2004ApJS..154..673S} \\
HIP113715 & 12000 & 6384 & $12.60 \pm 17.90$ & \citet{Smith2004ApJS..154..673S} \\
HD220870 & Johnson & V & $11.3 \pm 0.5$ & \citet{Samus2009yCat....102025S} \\
HD220870 & Cousins & Ic & $6.53 \pm 0.11$ & \citet{Droege2007yCat.2271....0D} \\
HD220870 & Johnson & J & $4.68 \pm 0.10$ & \citet{1980A_A....87..139B} \\
HD220870 & 2Mass & J & $4.84 \pm 0.20$ & \citet{Cutri2003tmc..book.....C} \\
HD220870 & Johnson & H & $3.57 \pm 0.10$ & \citet{1980A_A....87..139B} \\
HD220870 & 2Mass & H & $3.68 \pm 0.22$ & \citet{Cutri2003tmc..book.....C} \\
HD220870 & 2Mass & Ks & $2.99 \pm 0.29$ & \citet{Cutri2003tmc..book.....C} \\
HD220870 & Johnson & K & $3.13 \pm 0.10$ & \citet{1980A_A....87..139B} \\
HD220870 & 4900 & 712 & $12.20 \pm 5.20$ & \citet{Smith2004ApJS..154..673S} \\
HD220870 & 12000 & 6384 & $-2.10 \pm 17.80$ & \citet{Smith2004ApJS..154..673S} \\

\enddata	
\tablenotetext{a}{Values for system-bandpass data are in magnitudes; values for wavelength-bandwidth data are in Janskys.}
										
\tablecomments{The collections of photometry used in the SED fitting routine for all objects. Refer to \S Section~\ref{sec_SED_fitting} for details.}											
											
\end{deluxetable}


											

\end{document}